\documentclass[a4paper,fleqn,usenatbib]{mnras}

\usepackage{newtxtext,newtxmath}

\usepackage[T1]{fontenc}
\usepackage{ae,aecompl}

\usepackage{graphicx}	
\usepackage{amsmath} 
\usepackage{amssymb} 
\usepackage{tabularx}
\usepackage{subfig}
\usepackage{siunitx}
\DeclareSIUnit\smass{M\ensuremath{_\odot}}
\DeclareSIUnit\zmet{Z\ensuremath{_\odot}}

\newcommand{\Ha}{\rm H\alpha }
\newcommand{\Hb}{\rm H\beta }
\newcommand{\kms}{km\,s^{-1}}

\def\la{\raise.5ex\hbox{$<$}\kern-.8em\lower 1mm\hbox{$\sim$}}
\def\ma{\raise.5ex\hbox{$>$}\kern-.8em\lower 1mm\hbox{$\sim$}}

\def\cm3{$\rm cm^{-3}$}

\def\Vs{$\rm V_{s}$~}
\def\n0{$\rm n_{0}$}
\def\B0{$\rm B_{0}$}

\def\erg{$\rm erg\, cm^{-2}\, s^{-1}$}

\def\L12{L$_{12\mu m}$~}
\def\F12{F$_{12\mu m}$~}

\title[High resolution spectroscopy of the ENLR of AGN]{High resolution spectroscopy of the extended narrow-line region of IC\,5063 and NGC\,7212}

\author[E.~Congiu et al.]{E.~Congiu$^{1,2}$,\thanks{E-mail: enrico.congiu@phd.unipd.it} M.~Contini$^3$, S.~Ciroi$^{1}$, V.~Cracco$^{1}$,\newauthor{M.~Berton$^{1,2}$, F.~Di Mille$^4$, M.~Frezzato$^{1}$, G.~La Mura$^{1}$,  P.~Rafanelli$^{1}$}
\\
$^{1}$Dipartimento di Fisica e Astronomia "G. Galilei", Universit\`a di Padova, Vicolo dell'Osservatorio 3, I-35122 Padova, Italy;\\
$^2$INAF-Osservatorio Astronomico di Brera, via E. Bianchi 46, I-23807, Merate (LC), Italy;\\
$^3$School of Physics and Astronomy, Tel Aviv University, Tel Aviv 69978, Israel;\\
$^4$Las Campanas Observatory - Carnegie Institution of Washington, Colina el Pino Casilla 601, La Serena, Chile.
}

\date{Accepted XXX. Received YYY; in original form ZZZ}

\pubyear{2017}

\begin{document}
\label{firstpage}
\pagerange{\pageref{firstpage}--\pageref{lastpage}}
\maketitle

\begin{abstract}
We studied the properties of the gas of the extended narrow line region (ENLR) of two Seyfert 2 galaxies: IC\,5063 and NGC\,7212.
We analysed high resolution spectra to investigate how the main properties of this region depend on the gas velocity.
We divided the emission lines in velocity bins and we calculated several line ratios.
Diagnostic diagrams and SUMA composite models (photo-ionization $+$ shocks), show that in both galaxies there might be evidence of shocks significantly contributing in the gas ionization at high $\lvert V \rvert$, even though photo-ionization from the active nucleus remains the main ionization mechanism.
In IC\,5063 the ionization parameter depends on $V$ and its trend might be explained assuming an hollow bi-conical shape for the ENLR, with one of the edges aligned with the galaxy disk.
On the other hand, NGC\,7212 does not show any kind of dependence.
The models show that solar O/H relative abundances reproduce the observed spectra in all the analysed regions.
They also revealed an high fragmentation of the gas clouds, suggesting that the complex kinematics observed in these two objects might be caused by interaction between the ISM and high velocity components, such as jets.

\end{abstract}




\section{Introduction}

Active galactic nuclei (AGN) are among the most luminous objects of the Universe and they have been intensively studied, in the last decades, because they are characterized by many important astrophysical processes.
Some nearby AGN, typically Seyfert 2 galaxies \citep[$\sim 50$ up to $z\sim 0.05$,][]{Netzer15}, are characterized by conical or bi-conical structures of highly ionized gas, whose apexes point the galaxy nucleus, the \emph{ionization cones}.
The extension of optical emission (which is usually traced with the the [\ion{O}{III}]$\lambda5007$ line) is of the order of kiloparsecs and in the most extended ionization cones it can be traced up to $15$--$20\,\si{kpc}$ \citep{Mulchaey96,Schmitt03}.
The presence of these structures was already predicted by the Unified Model \citep{Antonucci85, Antonucci93} which affirms that the core of the AGN is surrounded by a dusty torus that absorbs part of the radiation coming from the nucleus.
However, the radiation emitted along the torus axis can escape and ionize the surrounding gas, forming the narrow-line region (NLR).
When the host galaxy contains enough gas and the NLR is not absorbing completely the ionizing photons, it is also possible to observed the ionization cones \citep{Evans93}.
Due to its extension, this region of ionized gas is also called extended narrow-line region (ENLR)\footnote{We consider the ENLR as a natural extension of the NLR beyond $0.8$--$1\,\si{kpc}$}.

Both NLR and ENLR are characterized by spectra with narrow permitted and forbidden emission lines, with a typical full width at half maximum (FWHM) between $300$ and $800\,\si{\kms}$.
The presence of several forbidden lines indicates that the electron density of the regions is low, typically $n_{\rm e} \sim 10^2$ -- $10^4\,\si{cm^{-3}}$.
High resolution images of nearby galaxies with ionization cones \citep[e.g.][]{Schmitt03} reveal the presence of substructures, such as gaseous clouds and filaments.
Medium and high resolution spectra show line profiles characterized by asymmetries, bumps and multiple peaks \citep[e.g.][]{Dietrich98,Ozaki09}, indicating very complex kinematics.
These internal gas motion are expected to produce shock-waves, which can further influence the properties of the NLR/ENLR \citep[see e.g.][]{Contini12}.

There are several possible causes of the complex kinematics of these structures.
For example, the interaction between jets and the interstellar medium (ISM) of the galaxy, as suggested by the fact that the cone axis and the radio-jet axis, if present, are often aligned \citep{Unger87, Wilson94, Nagar99, Schmitt03}.
Another possibility is that the gas of the ENLR is the result of an episode of merging which brought new material toward the center of the host galaxy \citep{Veilleux99, Ciroi05, DiMille07, Cracco11}.
The third possibility is the presence of fast outflows ($400$ -- $600\,\si{\kms}$) which involve different kinds of gas, from the cold molecular one to the warm ionized one \citep{Baldwin87,Morganti15,Dasyra15}.
It is worth noting that the radio-jet -- ISM interaction could cause these outflows \citep{Tadhunter14,Morganti15,Dasyra15}. 

\begin{figure*}
\centering
\includegraphics[width=0.30\textwidth]{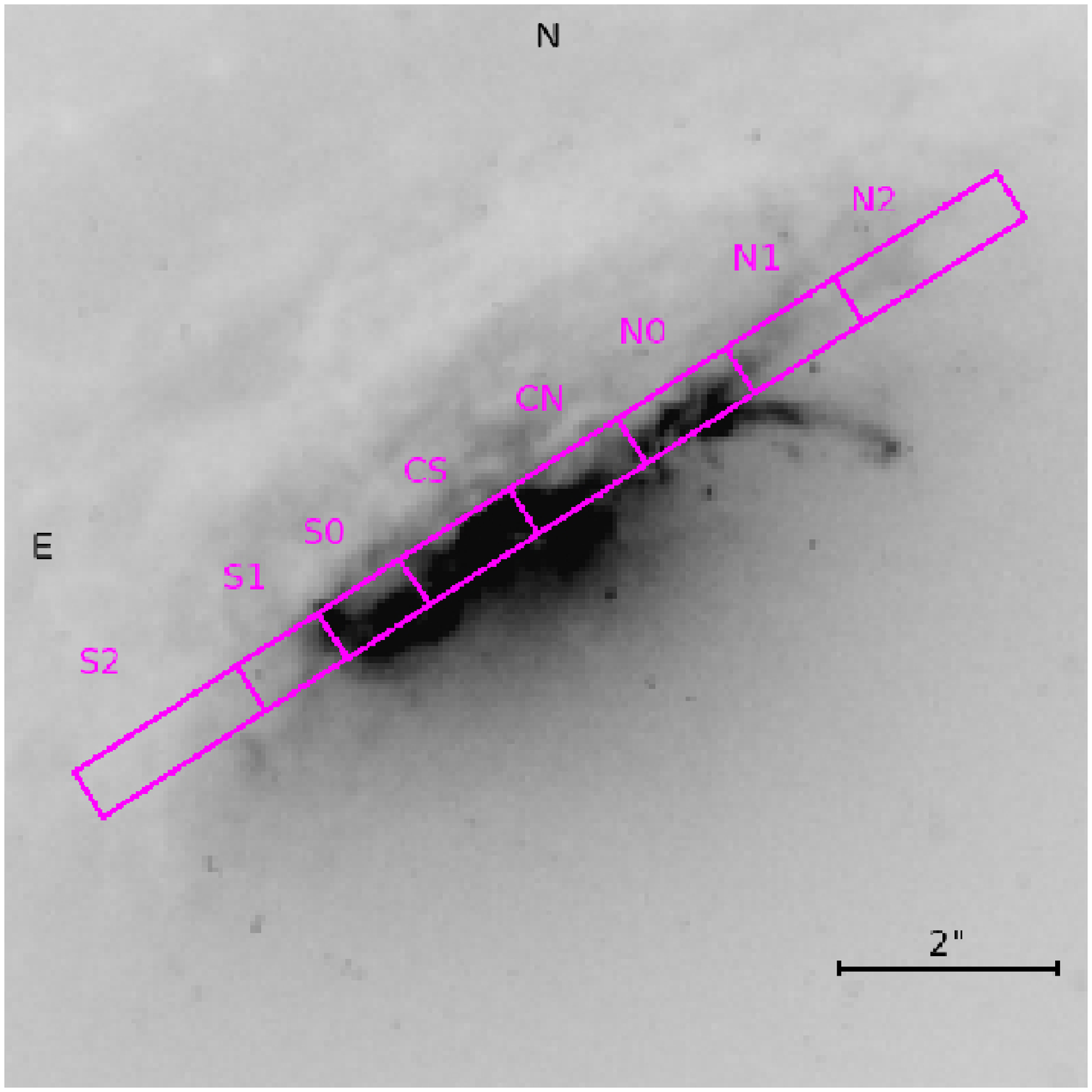} 
\quad
\includegraphics[width=0.30\textwidth]{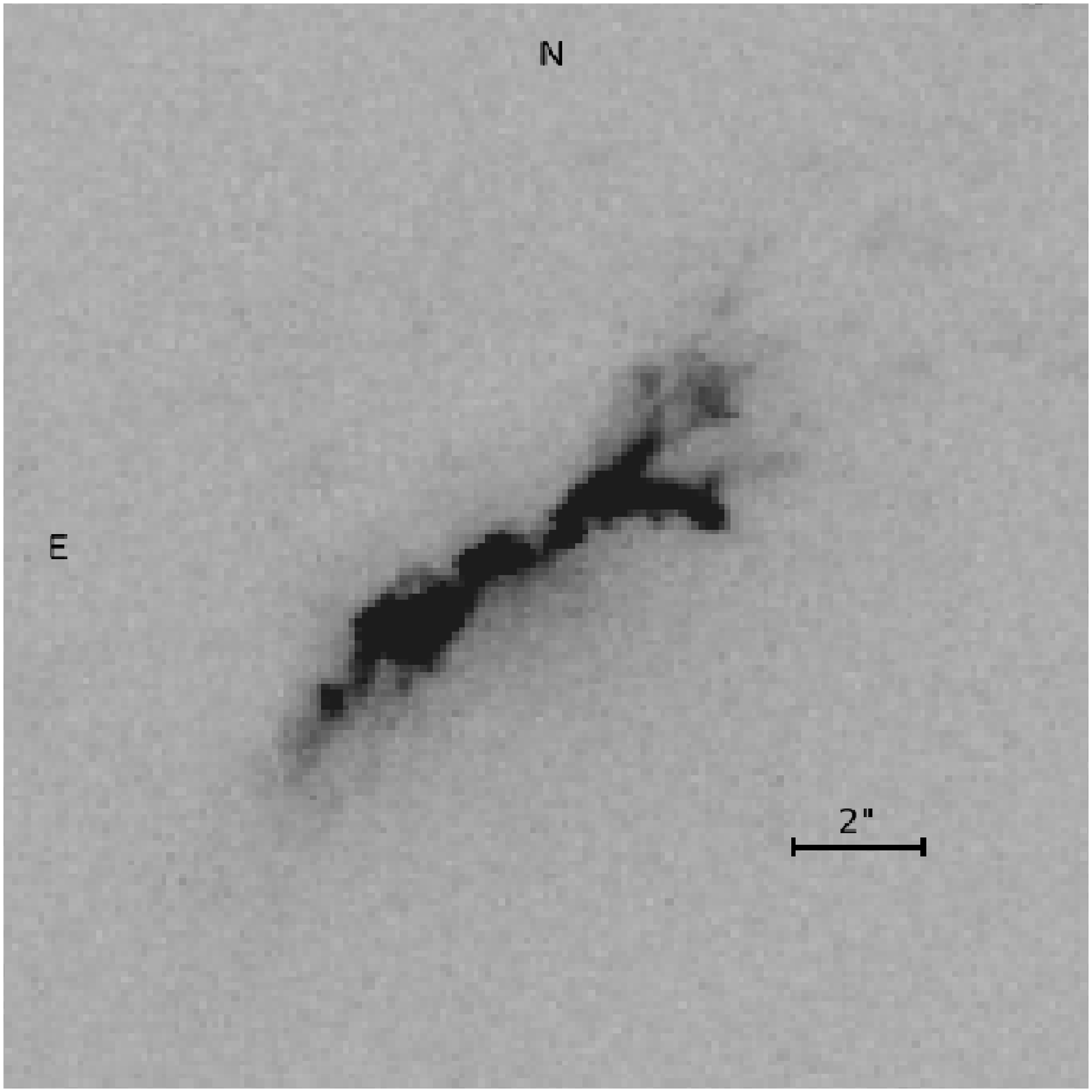} 
\quad
\includegraphics[width=0.30\textwidth]{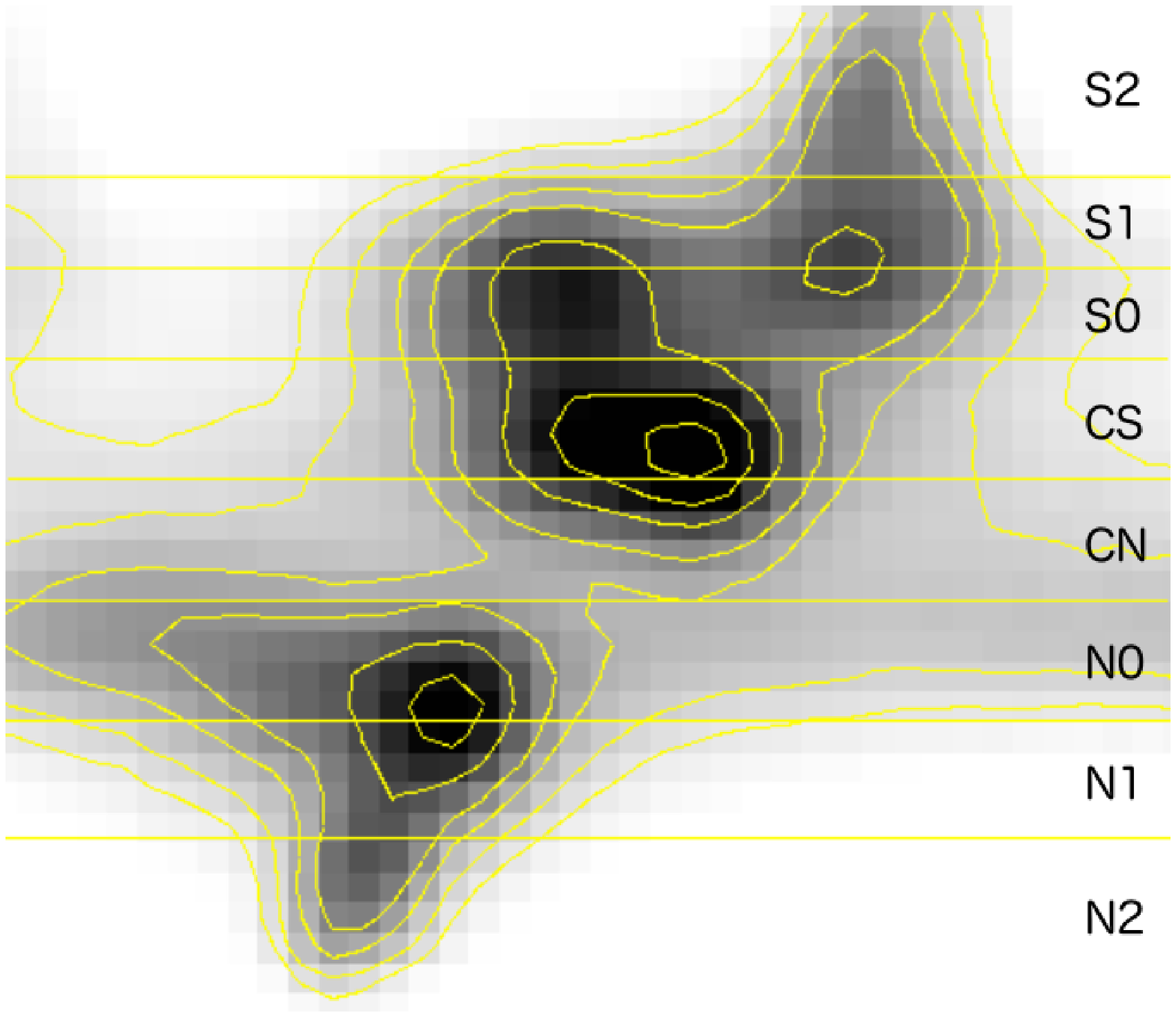} 
\caption[]{\textbf{Left:} Image of the Seyfert galaxy IC\,5063 taken with the F606W filter of the Wide Field and Planetary Camera 2 (WFPC2) of the Hubble Space Telescope (HST). In purple are shown the position of the slit and the regions from which we extracted the one-dimensional spectra. \textbf{Centre:} Image taken with the FR533N filter of the WFPC2. The images were recovered from the Hubble Legacy Archive (HLA).
\textbf{Right:} two-dimensional spectrum of the $\Ha$ region with contours. The figure also show the division of the spectrum in regions.}
\label{fig:IC5063}
\end{figure*}

Trying to distinguish among these causes it is clearly not an easy task. 
Data indicate that most AGN hold a NLR, while, up to $z\sim0.05$, very few of them show an ENLR  \citep{Netzer15}. 
The morphology of Seyfert galaxies up to $z\sim1$ is almost unperturbed, suggesting that the gravitational interactions, when occur, must be minor merger events \citep{Cisternas11}. 
Besides, most of Seyfert galaxies are classified as radio-quiet \citep{Singh15} with compact radio emission, even though recent studies found that kiloparsec scale emission might be more common than expected \citep[e.g.][]{Gallimore06, Singh15}. 
In spite of that, the kinematics of both NLR and ENLR is often characterized by radial motions \citep[e.g.][]{Morganti07,Ozaki09,Cracco11,Netzer15}	, suggesting that the ionized gas is not simply driven by the gravitational potential of the host galaxy.

Within this picture, we chose to carry out a pilot study of the physical properties of the NLR/ENLR gas as a function of velocity by comparing the intensities and profiles of emission lines analysed in medium and/or high resolution spectra.
We are well aware that this is a relatively unexplored field due to the need of observing nearby AGN with large diameter telescopes in order to have simultaneously good spectral and spatial resolution and high signal-to-noise ratio (SNR).
Nevertheless, our first tests showed us that with a resolution $R \sim  10000$ it is possible to highlight structures in the line profiles, such as multiple peaks and asymmetries, which are usually missed in low resolution spectra. 
Moreover, the lines seem to be intrinsically broad, therefore a further increase of the resolution will probably not improve the separation between the gaseous components.

In this work we improved the method developed by \citet{Ozaki09} who obtained and analysed high resolution spectra of the NLR of NGC 1068, and we applied it to medium resolution echelle spectra of two nearby Seyfert 2 galaxies with ENLR: IC\,5063 and NGC\,7212.
Both of them show extended ionization cones and are observable from the southern hemisphere.
We used the strongest optical lines to directly measure the physical properties of the gas such as density, extinction, ionization degree. 
Then, we calculated the gas physical conditions and element abundances through detailed modelling of the line ratios. 
We adopted the code SUMA \citep{Contini12} which accounts for the coupled effect of photo-ionization and shocks (see Sec.\,\ref{sec:models}).

In Sec.\,\ref{sec:sample} we present the observations and the sample, in Sec.\,\ref{sec:datared} we describe the data reduction, in Sec.\,\ref{sec:data_an} we show how we analysed the data and the results obtained and in Sec.\,\ref{sec:conclusions} we summarize our work.

In this paper we adopt the following cosmological parameters:  $H_0 = 70\,\si{\kms\,Mpc^{-1}}$, $\Omega_{m,0}=0.3$ and $\Omega_{\Lambda,0}=0.7$ \citep{Komatsu11}.

\section{Sample and observations}
\label{sec:sample}

In this work we studied two nearby Seyfert galaxies with known ENLR: IC\,5063 and NGC\,7212.
They were observed with the MagE (Magellan Echellette) spectrograph mounted at the Nasmyth focus of the Clay telescope, one of the two 6.5m Magellan telescopes of the Las Campanas Observatory (Chile).
MagE is able to cover simultaneously the complete visible spectrum ($3100$ -- $10000$\,\AA) with a maximum resolution $\rm R=8000$ \citep{Marshall08}.

Both galaxies were observed with multiple exposures of $1200\,\si{s}$ each.
Additional images were also acquired to perform data reduction (night-sky and calibration lamp spectra).
The spectrophotometric standard HR5501 was observed each night to carry out the flux calibration.
The details of our observations are reported in Table\,\ref{tab:obs}.

\begin{table*}
\caption[]{Principal properties of the observed galaxies, taken form NED, and characteristics of the observations.}
\label{tab:obs}
\centering
\begin{tabular}{lcccccccc}
\hline
Name&RA&Dec& Exp. time (s)& PA (deg)& Redshift& B (mag)& A$(V)$ (mag)& Morph.\\
\hline
IC\,5063& $20^{\rm h}$ $52^{\rm m}$ $02^{\rm s}$&$-57^{\rm d}$ $04^{\rm m}$ $08^{\rm s}$&$2400$&$123$&$0.01135$&$12.89$ &$0.165$&S0\\
NGC\,7212&$22^{\rm h}$ $07^{\rm m}$ $01^{\rm s}$&$10^{\rm d}$ $13^{\rm m}$ $52^{\rm s}$&$6000$&$16$&$0.02663$& $14.78^1$& $0.195$&Sab\\
\hline
\multicolumn{2}{l}{$^1$\footnotesize\cite{Munoz07}\normalsize}\\
\hline
\end{tabular}
\end{table*}

\subsection{IC\,5063}
\label{sec:IC5063}

IC\,5063 (Fig.\,\ref{fig:IC5063}) ($z=0.01135$) is a lenticular galaxy classified as a Seyfert 2 \citep{Morganti07}.
Table\,\ref{tab:obs} lists some of the principal properties about the object as reported by the NASA/IPAC Extragalactic Database (NED).
It is one of the most radio-loud Seyfert 2 galaxies known, its radio luminosity is 2 orders of magnitude larger than that of typical Seyfert galaxies \citep{Morganti98}.
IC\,5063 is characterized by a complex system of dust lanes aligned with the major axis of the galaxy.
It also shows a strong IR emission \citep{Hough87} and a very broad component of $\Ha$ in polarized light \citep{Inglis93}, likely a sign of an obscured broad line region (BLR).

The ENLR was discovered by \citet{Colina91}, who found a very extended region of ionized gas ($\sim 22\,\si{kpc}$, $H_0 =50\,\si{\kms.Mpc^{-1}}$) at position angle $PA = \ang{123}$, with a peculiar X shape and an opening angle of about $\ang{50}$. 
They supposed that the ENLR is the result of a quite recent merging between an elliptical galaxy and a small gas-rich spiral galaxy. 
\citet{Morganti07} traced [\ion{O}{III}]$\lambda\lambda4959, 5007$ and $\Ha$ emission lines up to a distance of about $15$--$16\,\si{arcsec}$ ($3.5$--$3.8\,\si{kpc}$) from the nucleus in a direction very close to the two resolved radio jets, $PA = \ang{117}$. 
The authors found signs of fast outflows, with velocities of about $600\,\si{\kms}$ in all the gas phases.

Similar velocities were found several years earlier by \citet{Morganti98} who discovered fast outflows in the neutral gas of the galaxy studying the \ion{H}{I} line at $21\,\si{cm}$.
These outflows were the first of their kind discovered in a galaxy and they were deeply studied in the following years \citep{Morganti07,Tadhunter14,Morganti15}.
These most recent papers explained them as gas accelerated by fast shocks caused by the radio jets plasma expanding in the ISM.

\subsection{NGC\,7212}
\label{sec:NGC7212}

\begin{figure*}
\centering
\includegraphics[width=0.30\textwidth]{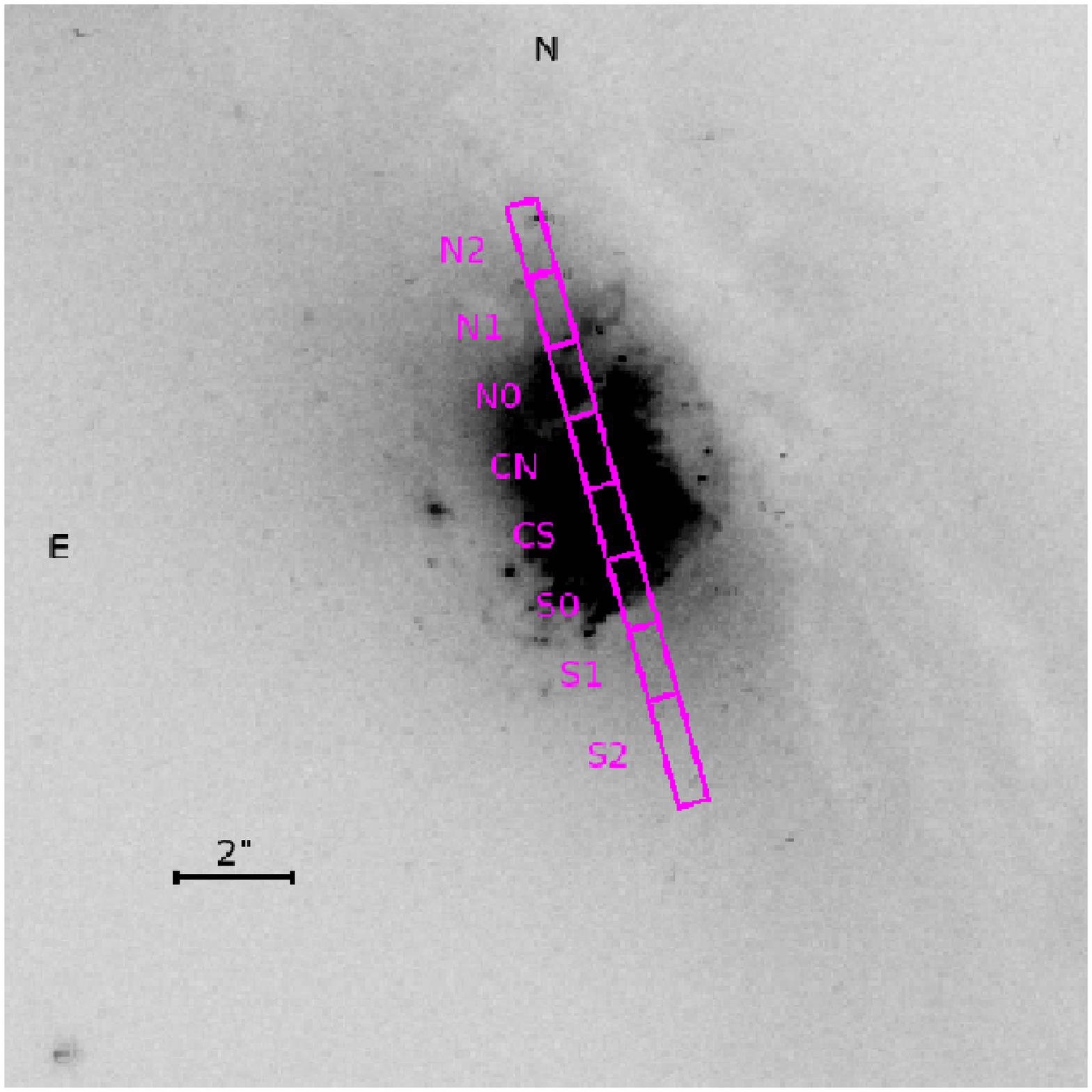} 
\quad
\includegraphics[width=0.30\textwidth]{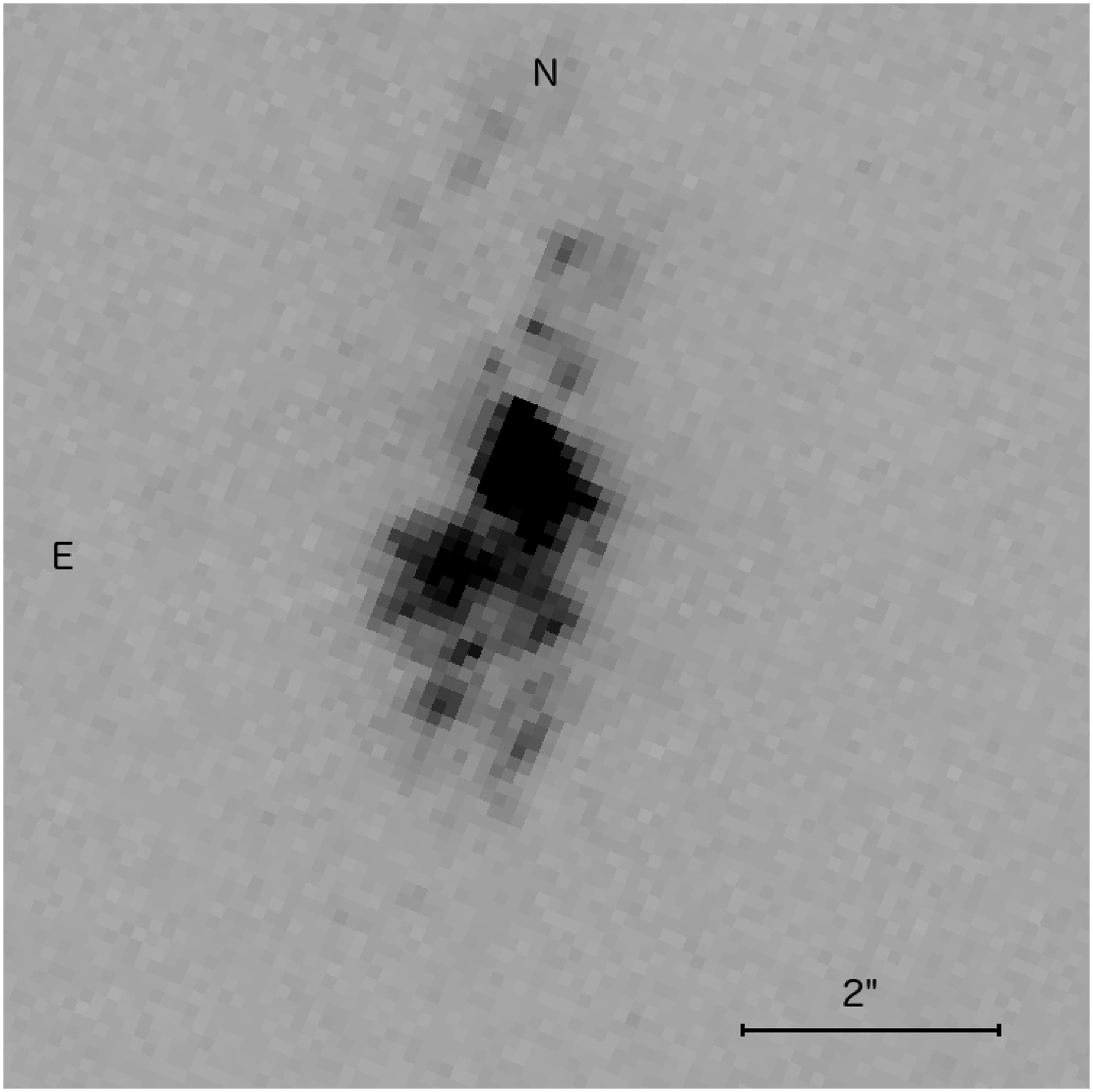}
\quad
\includegraphics[width=0.30\textwidth]{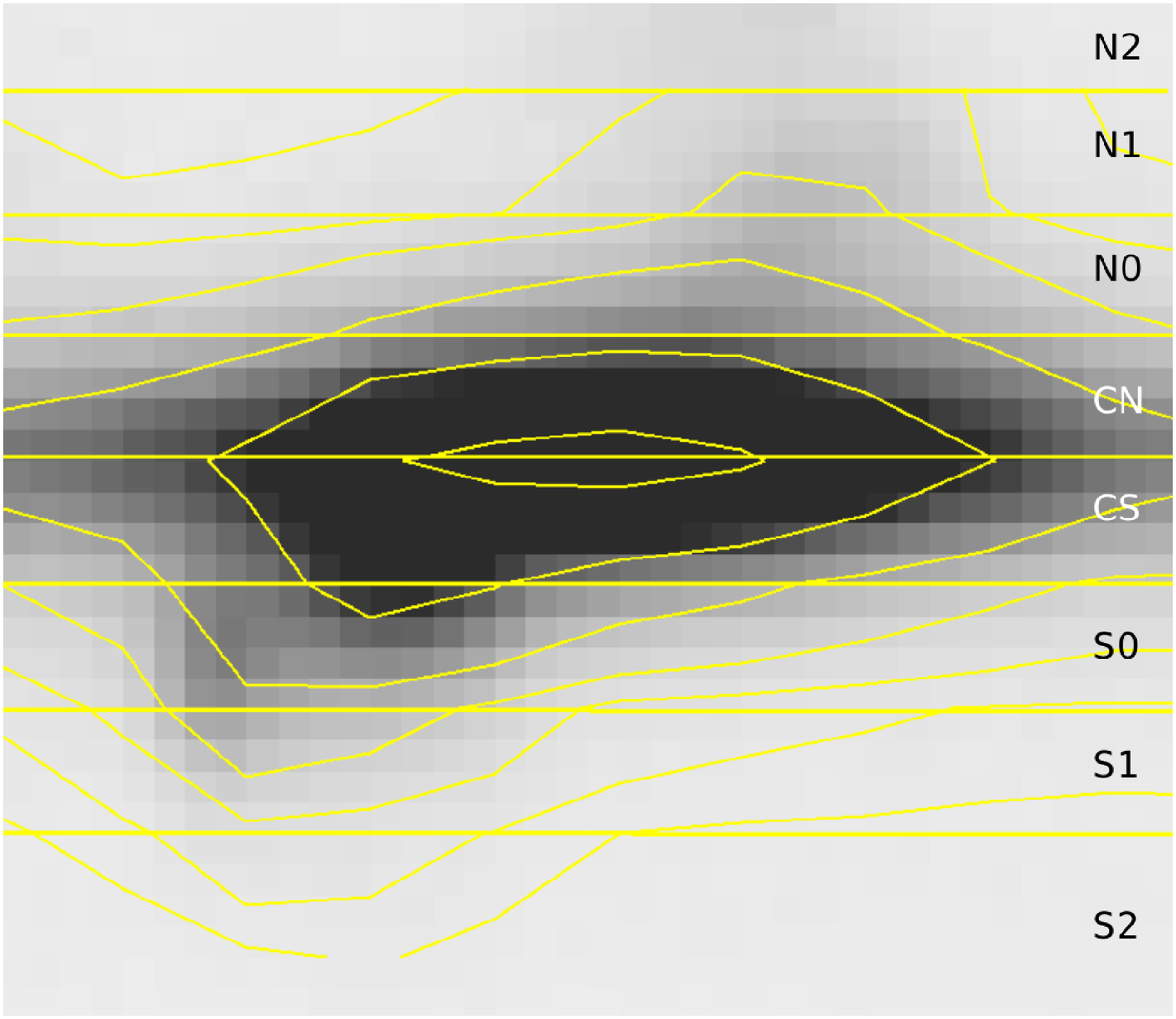}
\caption[]{\textbf{Left:} Image of the Seyfert galaxy NGC\,7212 taken with the F606W filter of the WFPC2 of the HST. In purple are shown the position of the slit and the regions from which we extracted the one-dimensional spectra. \textbf{Centre:} Image taken with the FR533N filter of the WFPC2. The images were recovered from the HLA.
\textbf{Right:} two-dimensional spectrum of the $\Ha$ region with contours. The figure also show the division of the spectrum in regions.}
\label{fig:NGC7212}
\end{figure*}

NGC\,7212 (Fig.\,\ref{fig:NGC7212}) is a nearby spiral galaxy ($z=0.02663$) belonging to a system composed by three interacting objects.
Some of its characteristic are reported in Table\,\ref{tab:obs}.
It was classified as Seyfert 2 by \citet{Wasilewski81}.
Later \citet{Tran95}, who studied some Seyfert 2 galaxies in polarized light found an elongated NLR in NGC\,7212, with the axis at $\rm PA=\ang{170}$, aligned with a structure of highly ionized gas.
Then, \citet{Falcke98} discovered the presence of a diffuse and extended NLR also in non-polarized light, in a compatible direction with the previously found elongated structure, but without the well defined shape typical of ionization cones.
They also discovered a radio-jet aligned with this ENLR and they suggested that the two structures are interacting.

The existence of the ionization cones was finally confirmed by \citet{Cracco11} through integral field spectroscopic observations.
This structure extends from the nucleus to about $3-4\,\si{kpc}$ in both directions and it is almost aligned with the optical minor axis of the galaxy ($\rm PA=\ang{150}$).
\citet{Cracco11} discovered clear asymmetries in the emission lines of the ENLR, a sign of the presence of radial motions of the gas. 
This property, together with a possible sub-solar metallicity, induced these authors to suggest that the ENLR in NGC 7212 formed through the interaction of the active galaxy with the other members of the system. 
In addition \citet{Cracco11} found quite high [\ion{N}{II}]/$\Ha$ and [\ion{S}{II}]/$\Ha$ ratios, especially in the region of interaction between NGC 7212 and one of the other galaxies of the triplet. 
This suggested that the ionization mechanism of the gas is a combination of photo-ionization from a non-thermal continuum and shocks \citep{Contini12}.

\section{Data Reduction}
\label{sec:datared}

We performed the data reduction using IRAF 2.15\footnote{http://iraf.noao.edu} (Image Reduction and Analysis Facility).
After subtracting the bias from all the images, we extracted the spectrum of each dispersion order using the \verb!strip! option of the \verb!apall! task and we processed each extracted spectrum (from now on also called apertures) following a standard long-slit reduction.
We used a Th--Ar lamp for wavelength calibration and the standard star HR5501 for flux calibration and telluric correction.

The final goal of the data reduction was to obtain one-dimensional (1D) spectra of different regions of the galaxies.
In our analysis it is essential to have straight and aligned spectra, to extract the emission of the same region at all wavelengths. 
Therefore, we first rectified the two dimensional spectra using the tasks \verb!identify! and \verb!reidentify! to trace the central peak of each spectrum along the spatial direction, secondly we applied the correction to the images with \verb!fitcoords! and \verb!transform!. 
Then we used \verb!xregister! to align all the apertures.

After that, we divided the 2D spectra of the galaxies in regions, looking for the best trade-off between a good SNR in all spectra and a good spatial sampling of the extended emission.
We used, as a reference, the $\Ha$ emission line (Fig.\,\ref{fig:IC5063} and Fig.\,\ref{fig:NGC7212}, right panels).
Then, we summed up the flux of each region to obtain a 1D spectrum for each one of them.
The names and the positions of the extracted regions are shown in Table\,\ref{tab:reg}.

\begin{table*}
\caption[]{Projected distances in arcseconds and kiloparsec of the regions from the nucleus of the galaxy, recessional velocities measured from the spectra and velocity corrections from STARLIGHT.}
\label{tab:reg}
\centering
\begin{tabular}[width=\columnwidth]{lcccccccc}
\hline
Region& \multicolumn{4}{c}{IC 5063}&\multicolumn{4}{c}{NGC 7212}\\
& arcsec& pc&$V\,(\si{\kms})$&$\Delta V\,(\si{\kms})$&arcsec&pc&$V\,(\si{\kms})$&$\Delta V\,(\si{\kms})$\\
\hline
N2&$4.50$&$1044$&$3328$&$-38.51$&$4.50$&$2408$&$8062$&$-44.02$\\
N1&$3.00$&$696$&$3337$&$-14.58$&$3.00$&$1605$&$8016$&$4.11$\\
N0&$1.80$&$417$&$3357$&$-28.78$&$1.80$&$963$&$8074$&$-55.91$\\
CN&$0.60$&$139$&$3364$&$-19.01$&$0.60$&$321$&$7963$&$-38.77$\\
CS&$0.60$&$139$&$3380$&$-15.52$&$0.60$&$321$&$7916$&$-59.07$\\
S0&$1.65$&$382$&$3417$&$-28.51$&$1.80$&$963$&$7799$&$54.31$\\
S1&$2.55$&$591$&$3438$&$-40.09$&$3.00$&$1605$&$7703$&$137.68$\\
S2&$3.90$&$904$&$3448$&$-23.02$&-&-&-&-\\
\hline
\end{tabular}
\end{table*}


\subsection{Subtraction of the stellar continuum}
\label{sec:starcont}

The galaxy continuum is a combination of several components: the stellar continuum, the AGN continuum and the nebular continuum, which is mainly due to hydrogen free-free and free-bound emission.
AGN emission is negligible in Seyfert 2 galaxies because the AGN is obscured by the dusty torus \citep[e.g.][]{Beckmann12} and the nebular continuum is estimated to be at least one order of magnitude fainter than the observed stellar emission \citep{OsterbrockAGN89}.
Therefore, in our cases the main component of the galaxy continuum is the stellar continuum.
To obtain reliable line flux measurements it is important to subtract this component from the spectra, mainly because of the typical stellar absorption features (e.g. hydrogen Balmer lines) which can fall at the same wavelength of emission lines, modifying their observed flux.

One of the best method to subtract this component is to use STARLIGHT \citep{Fernandes05, Mateus06, Fernandes07}.
STARLIGHT is a software developed to fit the stellar component of galaxy spectra using a linear combination of simple stellar populations spectra, taking into account both extinction and velocity dispersion.

To work properly, STARLIGHT needs spectra with fixed characteristics: they have to be corrected for Galactic extinction, they must be shifted to rest frame and they must also have a $1\,\si{\angstrom\,px^{-1}}$ dispersion.
We corrected for Galactic extinction with A(V) $= 0.165$ and A(V) $=0.195$ (Table\,\ref{tab:obs}) for IC\,5063 and NGC\,7212 respectively. 
We then shifted the spectra to rest-frame wavelengths.
It is not easy to do accurate redshift measurements on these objects because the emission lines are broad and they sometimes show multiple peaks. 
For this reason we averaged the values obtained from some of the deepest stellar absorptions: the \ion{Mg}{I} triplet ($\lambda5167,\,\lambda5177,\,\lambda5184$\,\AA) and the \ion{Na}{I} doublet ($\lambda5889,\,\lambda5895$\,\AA).
In the most external regions of the galaxies, where the SNR of the continuum was too low to detect those lines, we  used the strongest peak of some emission lines such as $\Hb$ or [\ion{O}{III}]$\lambda5007$.
The obtained recessional velocities are shown in Table\,\ref{tab:reg}.
To estimate the errors on the velocity measurements, we calculated the accuracy of the wavelength calibration comparing the position of the sky lines with respect to the expected position.
We obtained $\sigma = 0.2\,\si{\angstrom}$, which correspond to $\sigma_V=10\,\si{\kms}$.
However, STARLIGHT can independently measure small velocity shifts from the rest frame spectra ($< 500\,\si{\kms}$) when it fits the galaxy continuum with the template spectra.
This allows to fine tune the velocity correction, especially where emission lines are used.
The fine-tuning is very important because we needed to remove every single velocity component due to galaxy rotation to study the kinematics of the gas.

Once applied the listed corrections, we ran STARLIGHT using $150$ spectra of stellar populations with 25 different ages (from $10^6$ to $1.8\times10^8\,\si{yr}$) and 6 different metallicities (from $\rm Z=10^{-4}$ to $\rm Z=5\times 10^{-2}$) to obtain the spectrum of the stellar component and the velocity correction.
In the regions where the SNR was too low to obtain a good fit of the continuum, we used the results of the software to correct the velocity measurements but we subtracted the stellar continuum fitting it with a simple function, using IRAF \verb!splot! task.

\subsection{Deblending}
\label{sec:deblending}

\begin{figure}
\centering
\includegraphics[width=0.48\textwidth]{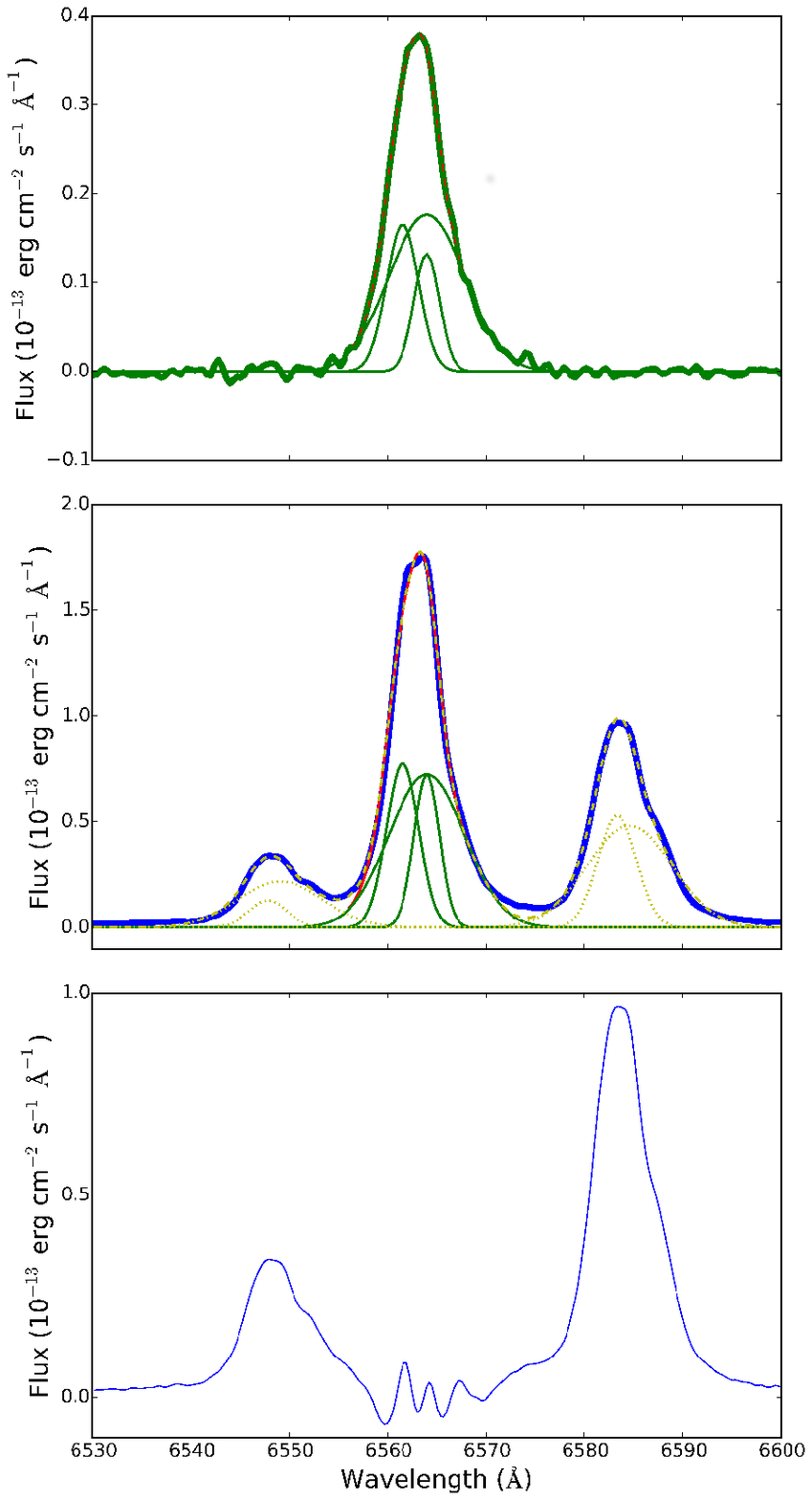} \\
\caption[]{The deblending process of $\Ha + [\ion{N}{II}]$ in the CS region of IC\,5063. The top panel shows $\Hb$ shifted to $\Ha$ wavelength and fitted with $3$ Gaussians (the thin curves). The middle panel shows the best fit of $\Ha+[\ion{N}{II}]$. The thin solid curves are the Gaussians fitting $\Ha$ and the dotted ones are the function fitting the [\ion{N}{II}] doublet. In the bottom panel there is the result of the subtraction of the $\Ha$ fit from the spectrum. }
\label{fig:hadeb}
\end{figure}

\begin{figure}
\centering
\includegraphics[width=0.45\textwidth]{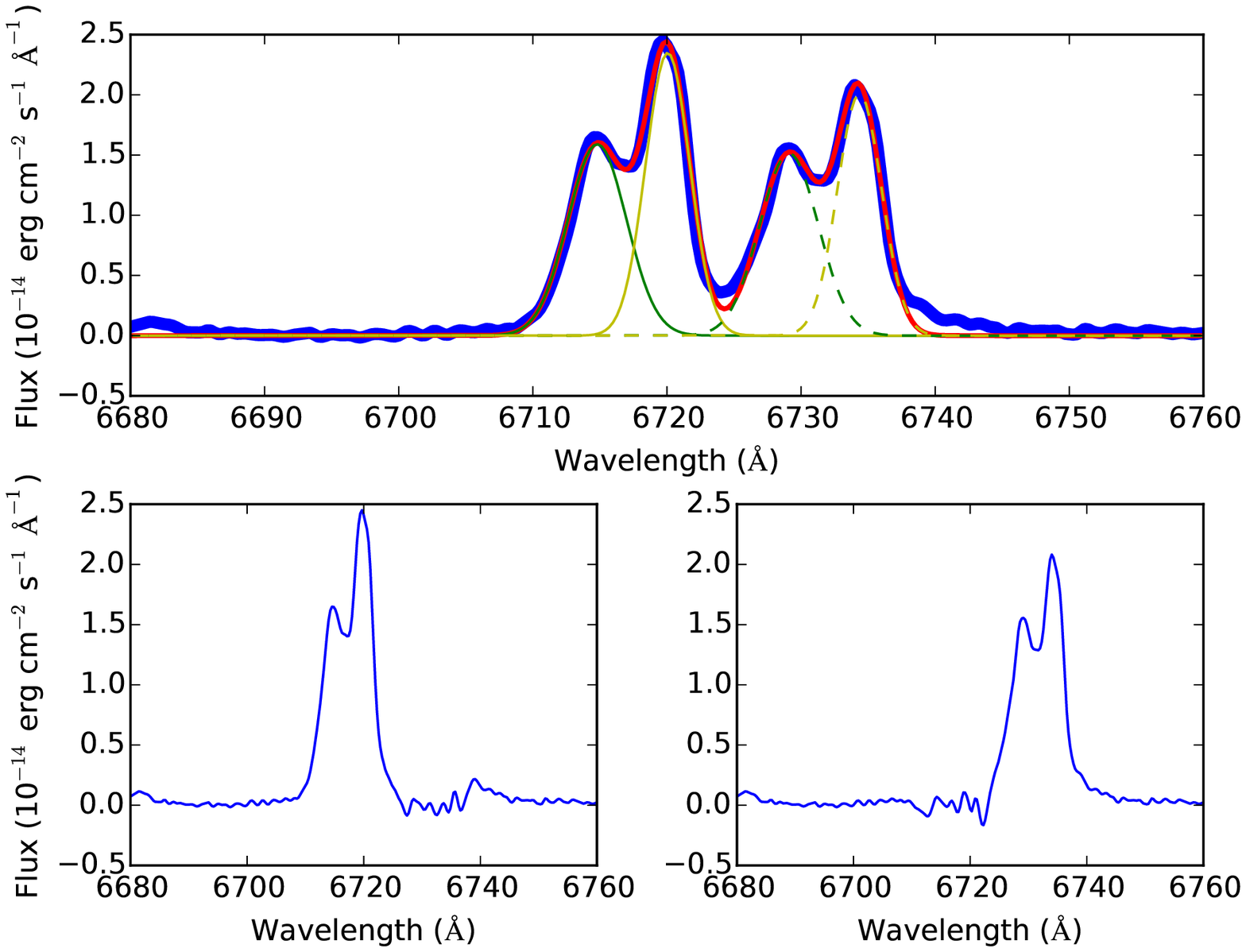} \\
\vskip0.2cm
\includegraphics[width=0.45\textwidth]{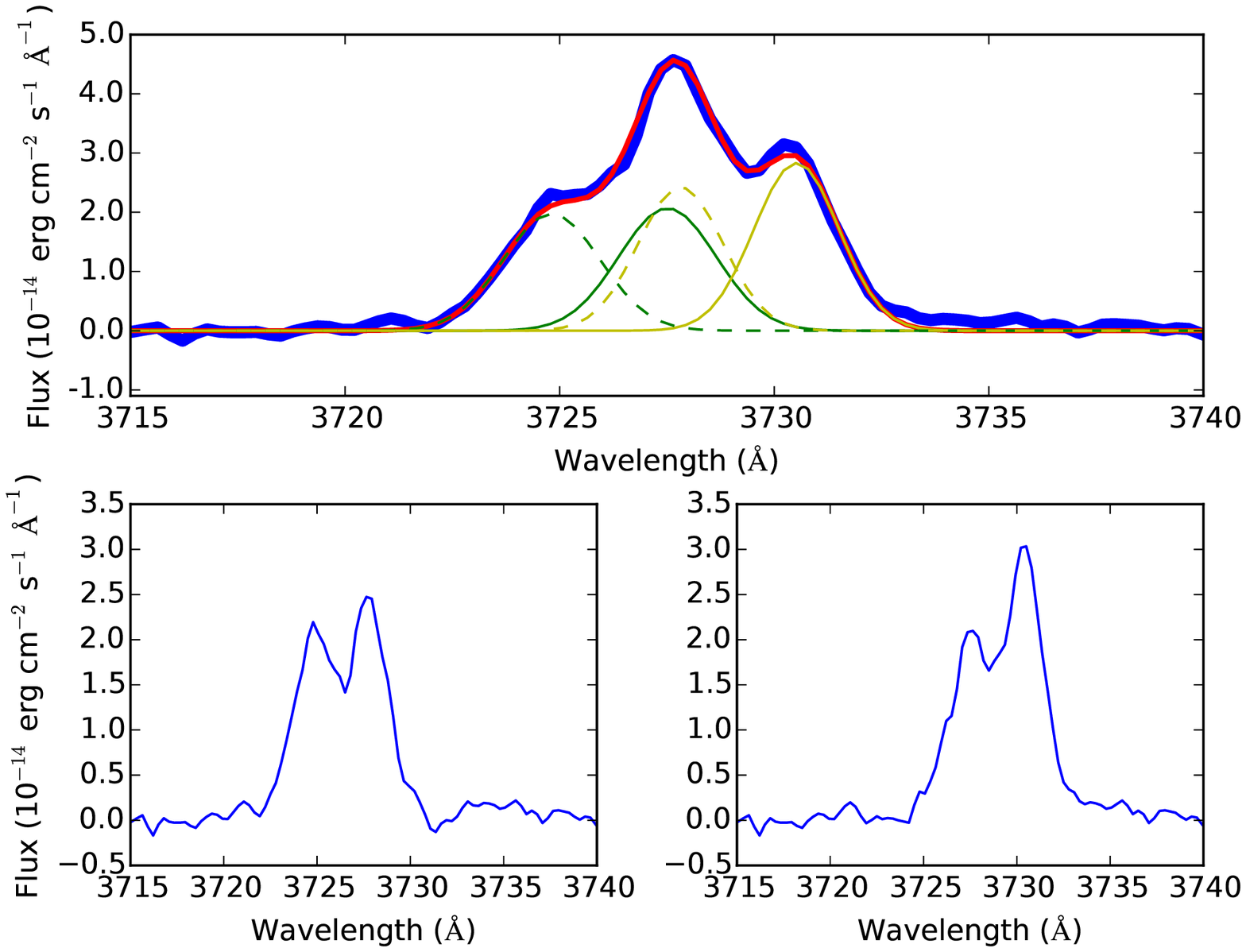}\\
\caption[]{The final results of the deblending process for the [\ion{S}{II}] doublet (top) and [\ion{O}{II}] doublet (bottom) of the CS region of IC\,5063. For each doublet is shown the original spectrum, with the fitting functions, and the two lines after the deblending. The continuous lines represent the functions with the free parameters ([\ion{S}{II}]$\lambda6716$ (top), [\ion{O}{II}]$\lambda3729$ (bottom)), the dashed lines represent the functions fitted with the constraints ([\ion{S}{II}]$\lambda6731$ (top),[\ion{O}{II}]$\lambda3726$ (bottom)). The original spectrum is shown in blue, the fitted spectrum is shown in red, the green Gaussians are the main components of each line, the yellow ones are their secondary components. }
\label{fig:deb}
\end{figure}

Because of their intrinsic width, some prominent spectral lines are partially overlapped (e.g. $\Ha$ and [\ion{N}{II}]$\lambda\lambda6548,6584$; [\ion{S}{II}]$\lambda\lambda6716,6731$).
These lines are essential to study the gas properties: $\Ha$ is used in the diagnostic diagrams \citep{Baldwin81,Veilleux87}  and to measure the extinction coefficient, the ratio between the [\ion{S}{II}]$\lambda\lambda6716,6731$ doublet is used to estimate the electron density.
Therefore, to measure their fluxes, especially on the line wings, we had to deblend them.
To perform the deblending we adapted to our case the method developed by \citet{Schirmer13}.
They assumed that if $\Ha$ and $\Hb$ are produced by the very same gas, they must have the same profile, except for extinction effects.
Therefore, they multiplied $\Hb$ for a constant factor in order to match the $\Ha$ peak intensity, they subtracted the multiplied $\Hb$ from the $\Ha +$ [\ion{N}{II}] group, obtaining only the [\ion{N}{II}] doublet emission.
To recover the final $\Ha$ line they subtracted the doublet from the original spectrum, after removing the residuals of the previous subtraction, due to intrinsic noise and to the assumption of constant extinction in the whole line.

To deblend $\Ha$ from [\ion{N}{II}] we wrote a \textit{python 2.7} script, modifying this process in the following way. 
We first shifted $\Hb$ to the wavelength of $\Ha$ and we fitted it with $2$ or $3$ Gaussians, trying to reproduce the line profile and selecting the result showing the minimum residual.
Then, we fitted $\Ha$ with the same number of functions, keeping the same FWHM and the same relative positions but varying their intensities.
In this way, we considered that the extinction could change in each kinematic component.
To get a better result we also fitted the [\ion{N}{II}] doublet using $2$ Gaussians for each line.
Then, we subtracted the $\Ha$ fit from the spectrum, obtaining the [\ion{N}{II}] doublet and we removed the residuals (Fig.\,\ref{fig:hadeb}).
Finally, we subtracted again the [\ion{N}{II}] doublet from the original spectrum, obtaining the final $\Ha$ line.
We used the same process to remove [\ion{S}{III}]$\lambda6310$ from [\ion{O}{I}]$\lambda6300$ in the regions where the [\ion{S}{III}] line was strong enough to be detected.

To deblend the [\ion{S}{II}] doublet and the [\ion{O}{II}]$\lambda\lambda3726,3729$ doublet we further modified the method.
We started deblending the [\ion{S}{II}] doublet, because their separation in wavelength is $\sim 14\,\si{\angstrom}$ and they usually are partially resolved.
To fit these lines we assumed that they have the same profile, except for the intensity, because they are emitted by the same gas but their ratio is influenced by the electron density.
We reproduced each line with $2$ or $3$ Gaussian functions, depending on the quality of the spectrum.
To reduce the number of free parameters, we fitted the [\ion{S}{II}]$\lambda6716$ line letting them free to vary and we fixed the position and the width of the Gaussians for the [\ion{S}{II}]$\lambda6731$ line with respect to the corresponding quantities for the [\ion{S}{II}]$\lambda6716$ line, according to the theoretical properties of the doublet ($\Delta \lambda = 14.3$ and same FWHM).
Fig.\,\ref{fig:deb} (top) shows the result of the deblending of these lines for the CS region of IC\,5063.

The [\ion{O}{II}] lines are much closer to each other ($\sim 2.5$\,\AA) and it is not possible to deblend them without using any reference line.
We adopted the [\ion{S}{II}] doublet because, considering only the ions showing strong emission lines in our spectra, it has the closest ionization potential to that of the \ion{O}{II} (Table\,\ref{tab:ion}).
Therefore, we can suppose that their emission comes from spatially close regions and we can assume for them a similar profile.
The two doublets also depend on the electron density in a similar way \citep[see][]{OsterbrockAGN}.
Therefore, we measured the ratio between the central intensity of each component in the two lines of the [\ion{S}{II}] fit and we used those ratios to put constraints on the Gaussians used to fit the [\ion{O}{II}] doublet.
After that, we proceeded as before, fitting one line and constraining the parameters of the second one.
However, since the [\ion{O}{II}] lines are almost totally overlapped, in several spectra it was not possible to use the same number of Gaussians used to fit the [\ion{S}{II}] doublet.
In these cases we applied the process using only $2$ Gaussians and refitting the [\ion{S}{II}] lines.
Fig.\,\ref{fig:deb} (bottom) shows the final result after the deblending of the [\ion{O}{II}] doublet in the CS region of IC\,5063.

\section{Data Analysis and Results}
\label{sec:data_an}

\begin{figure}
\centering
\includegraphics[width=0.40\textwidth]{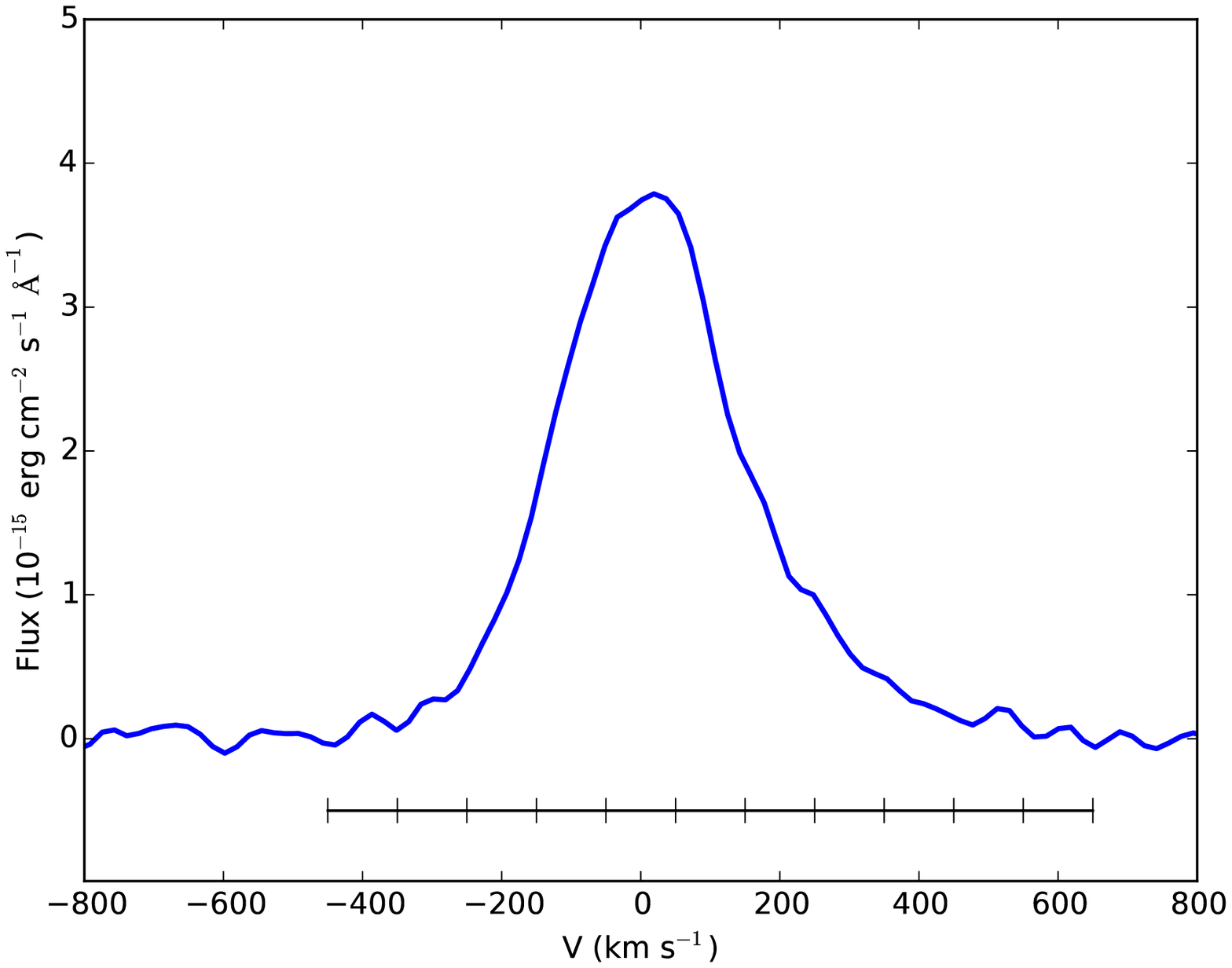} \\
\includegraphics[width=0.40\textwidth]{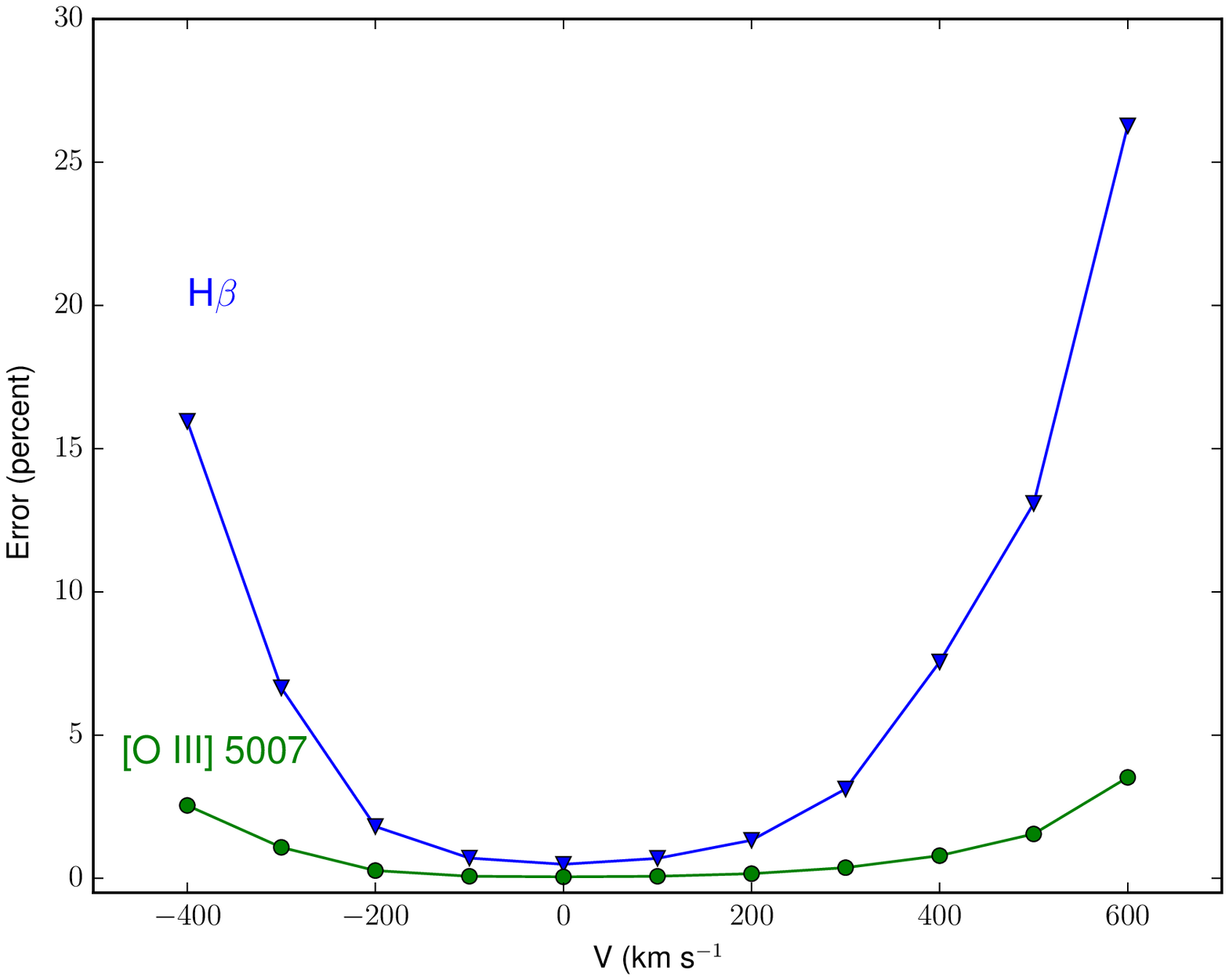}
\caption[]{\textbf{Top:} Plot of $\Hb$ taken from the spectrum of the CS region of IC\,5063. In the lower part of the figure the adopted velocity bins are shown. \textbf{Bottom:} Profile of the relative error in $\Hb$ and [\ion{O}{III}]$\lambda5007$ in the same region.}
\label{fig:bins}
\end{figure}

We divided all the line profiles in velocity bins (Fig.\,\ref{fig:bins}) and we measured the fluxes inside them, as done by \citet{Ozaki09}.
While this author took into account the morphology of the profile of the brightest emission line ($[\ion{O}{III}]\lambda5007$) for each region, we preferred to adopt a less arbitrary method and to use a fixed width of $100\,\si{\kms}$ which is about three times the theoretical instrumental velocity resolution ($\rm R\sim 37.5\,\si{\kms}$).
We performed the measurement using an IRAF script. 
To take into account the different line widths in different regions, we assumed the velocity bins of $\Hb$ as reference for the other emission lines. 
This choice was based on the need of calculating flux ratios, deriving physical parameters and plotting diagnostic diagrams. 
To this aim it is essential to have good measurements of $\Hb$ flux in each bin. 
Being $\Hb$ one of the weakest lines we are interested in, we are obviously losing high velocity bins detectable only in the strongest lines. 
We also measured the total flux of each line, to compare the results obtained dividing the lines in velocity bins to the average properties of the gas.

Fig.\,\ref{fig:bins} (bottom) shows the behaviour of the relative error on the flux as a function of velocity in two lines ($\Hb$ and [\ion{O}{III}]$\lambda5007$) of the CS region of IC\,5063.
To estimate the error we used the following equation:
\begin{equation}
\frac{\Delta I}{I}=\frac{rms}{I_0}
\end{equation}
where $I$ is the bin flux and $\Delta I$ its error, $rms$ is the root mean square of the continuum measured near the line and $I_0$ is the central peak of the bin.
The error increases, as expected, in the line wings both in a weak line ($\Hb$) and in a strong line ([\ion{O}{III}]$\lambda5007$).
In $\Hb$ we have a maximum error of $\sim25$ per cent of the flux in the line wings, while in [\ion{O}{III}]$\lambda5007$ the error is much lower ($\sim4$ per cent).
A similar behaviour is expected for all the measured lines.

\begin{table}
\caption[]{Measured emission lines and energy existence interval of ions.}
\label{tab:ion}
\centering
\begin{tabular}{lcc}
\hline
Ion&Wavelength (\AA)&$\Delta E$ (eV)\\
\hline
$[\ion{O}{II}]$& $3727$&$13.6$ -- $35.1$\\
$[\ion{O}{III}]$& $4363$&$35.1$ -- $54.9$\\
$[\ion{Ar}{IV}]$& $4711$&$40.9$ -- $59.8$\\
$[\ion{Ar}{IV}]$&$4740$&$40.9$ -- $59.8$\\
H& $4861$&-\\
$[\ion{O}{III}]$& $4959$&$35.1$ -- $54.9$\\
$[\ion{O}{III}]$& $5007$&$35.1$ -- $54.9$\\
$[\ion{O}{I}]$& $6300$&-- $13.6$\\
H&$6563$&-\\
$[\ion{N}{II}]$& $6584$&$14.5$ -- $29.6$\\
$[\ion{S}{II}]$& $6717$&$10.4$ -- $23.3$\\
$[\ion{S}{II}]$&$6731$&$10.4$ -- $23.3$\\
$[\ion{O}{II}]$& $7319$&$13.6$ -- $35.1$\\
$[\ion{O}{II}]$&  $7330$&$13.6$ -- $35.1$\\
\hline
\end{tabular}
\end{table}

We first estimated the local extinction by using the Balmer decrement and assuming a theoretical $\Ha/\Hb$ ratio of $2.86$.
To recover the visual extinction A(V), we applied the CCM (Cardelli--Clayton--Mathis) extinction law \citep{Cardelli89}.
We measured A(V) both for each velocity bin and for the total line flux and we corrected  the measured fluxes of all the lines.

Then, we studied the mechanisms responsible for heating and ionizing the gas in the different regions of both galaxies  directly from diagnostic diagrams \citep{Baldwin81,Veilleux87}. 
\citet{Penston90} empirical relation was adopted to determine the ionization parameter, while the physical parameters such as the temperature and the density of the emitting gas were obtained from the characteristic line ratios.
Finally, to have a more complete picture of the physical parameters and abundances of the elements throughout the regions, we carried out a detailed modelling of the spectra accounting for both the photo-ionization from the active centre and for shocks.

\subsection{Diagnostic Diagrams}
\label{sec:diag}

\begin{figure*} 
\centering
\includegraphics[width=0.8\textwidth]{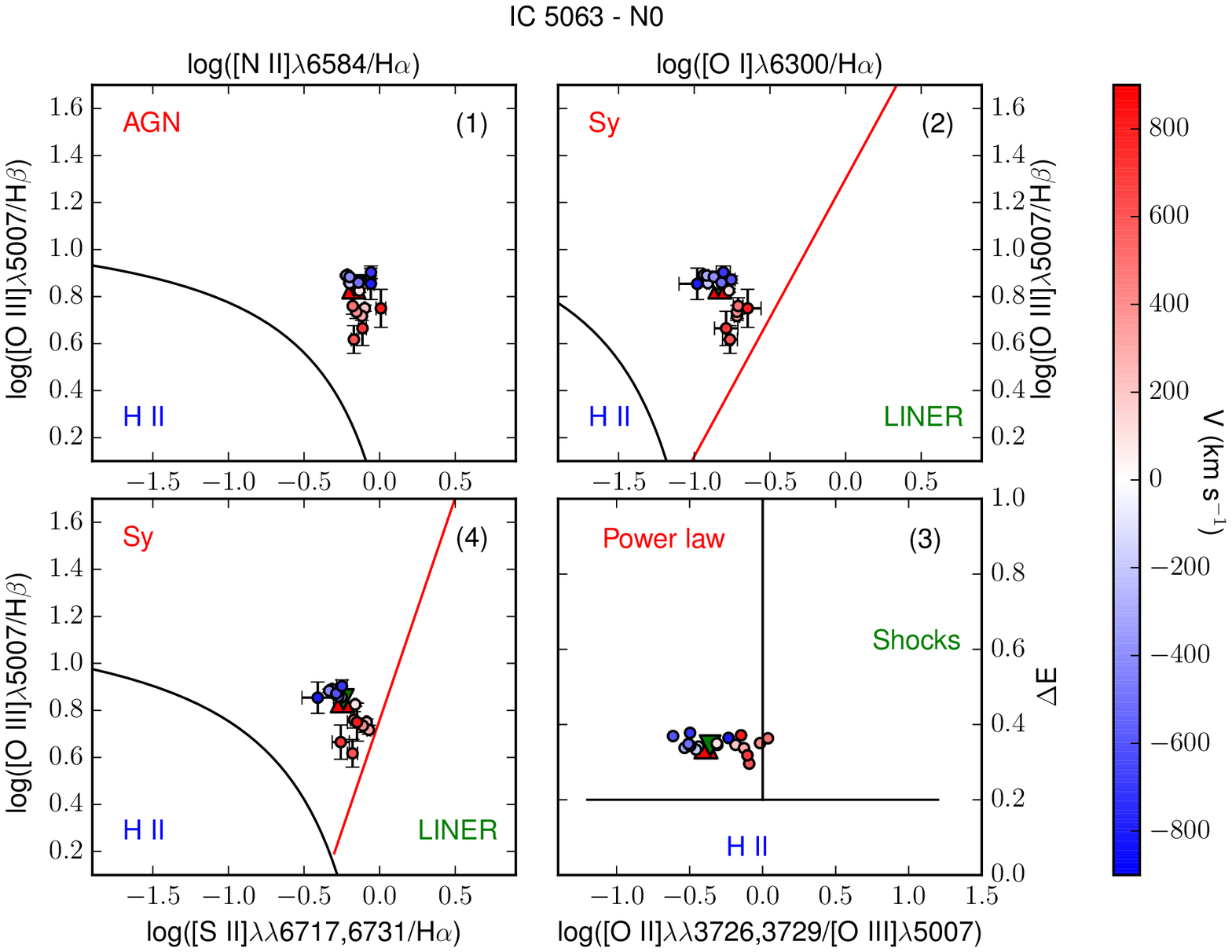}\\
\includegraphics[width=0.8\textwidth]{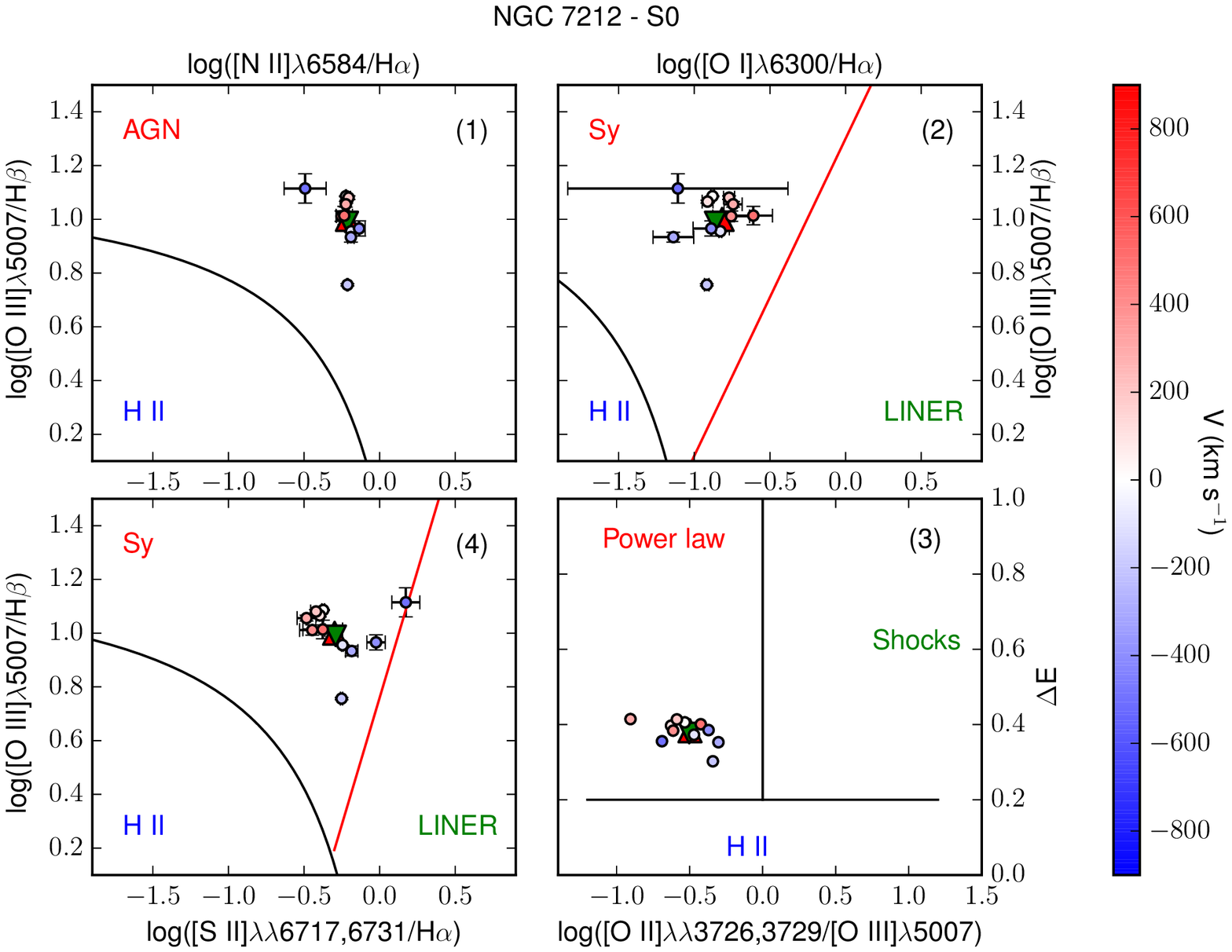}\\
\caption[]{Diagnostic diagrams of the N0 region of IC\,5063 (top) and of the S0 region of NGC\,7212 (bottom).  In each plot we show, from the top left panel clockwise: (1) $\log([\ion{O}{III}] \Hb)$ vs $\log([\ion{N}{II}]/ \Ha)$, (2) $\log([\ion{O}{III}]/ \Hb)$ vs $\log([\ion{O}{I}]/ \Ha)$, (3) $\Delta E$ vs $\log([\ion{O}{II}]/[\ion{O}{III}])$, (4) $\log([\ion{O}{III}]/ \Hb)$ vs $\log([\ion{S}{II}]/ \Ha)$ \citep{Baldwin81, Veilleux87}. The colorbar shows the velocity of each bin. The black curves in (1), (2), (4) divide power-law ionized regions (top) and HII regions (bottom). The red lines divide Seyfert-like regions (left) and LINER-like regions (right) \citep{Kewley06}. The black lines in (3) divide HII regions (bottom), power-law ionized region (left) and shock ionized regions (right) \citep{Baldwin81}.  }
\label{fig:diag_es}
\end{figure*}

\begin{figure*} 
\centering
\includegraphics[width=0.8\textwidth]{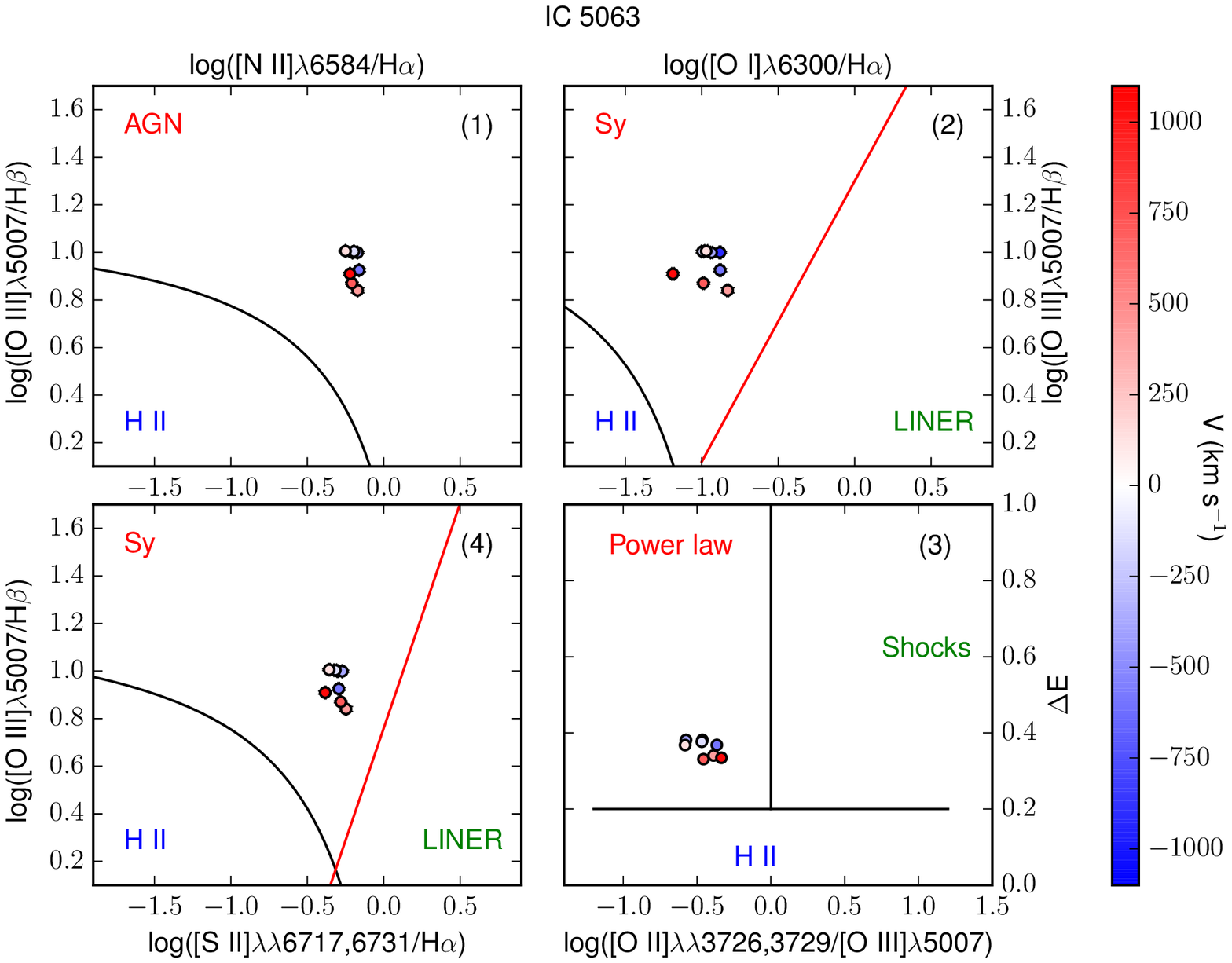}\\
\includegraphics[width=0.8\textwidth]{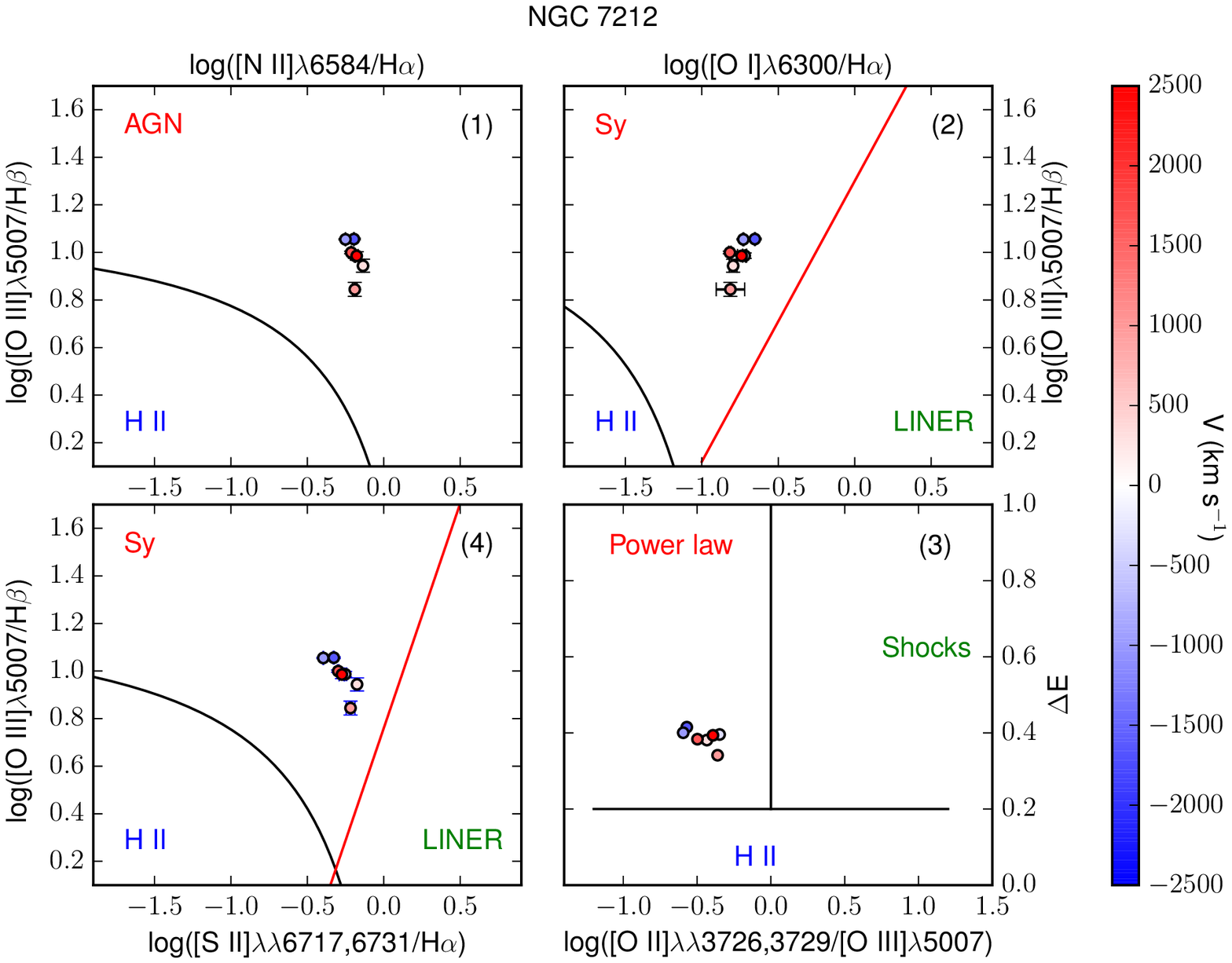}\\
\caption[]{Diagnostic diagrams of IC\,5063 (top) and NGC\,7212 (bottom) calculated with the total flux of the lines.  In each plot we show, from the top left panel clockwise: (1) $\log([\ion{O}{III}] \Hb)$ vs $\log([\ion{N}{II}]/ \Ha)$, (2) $\log([\ion{O}{III}]/ \Hb)$ vs $\log([\ion{O}{I}]/ \Ha)$, (3) $\Delta E$ vs $\log([\ion{O}{II}]/[\ion{O}{III}])$, (4) $\log([\ion{O}{III}]/ \Hb)$ vs $\log([\ion{S}{II}]/ \Ha)$ \citep{Baldwin81, Veilleux87}. The colorbar shows the distance from the nucleus of each region. The black curves in (1), (2), (4) divide power-law ionized regions (top) and HII regions (bottom). The red lines divide Seyfert-like regions (left) and LINER-like regions (right) \citep{Kewley06}. The black lines in (3) divide HII regions (bottom), power-law ionized region (left) and shock ionized regions (right) \citep{Baldwin81}.  }
\label{fig:diag_estot}
\end{figure*}

With the extinction corrected fluxes we proceeded plotting the diagnostic diagrams.
We used four different diagrams: three are the traditional BPT diagrams from \citet{Baldwin81} and \citet{Veilleux87} which compare the [\ion{O}{III}]$\lambda5007/\Hb$ ratio to [\ion{O}{I}]$\lambda6300/\Ha$, [\ion{N}{II}]$\lambda6584/\Ha$ and [\ion{S}{II}]$\lambda\lambda6716,6731/\Ha$.
The fourth one is a less used diagram developed by \citet{Baldwin81} which involves the [\ion{O}{II}]$\lambda\lambda3726,3729/$[\ion{O}{III}]$\lambda5007$ ratio to evaluate the ionization degree of the gas and a $\Delta E$ parameter which combines the previous ratios to discern between different ionization mechanisms (Eq.\,\ref{eq:deltae1}, \ref{eq:deltae2}, \ref{eq:deltae3}, \ref{eq:deltae}, where $x=([\ion{O}{II}]\lambda3726+[\ion{O}{II}]\lambda3729)/[\ion{O}{III}]\lambda5007$):

\begin{equation}
\label{eq:deltae1}
\Delta E_1=\log\left(\frac{\rm[\ion{O}{III}]\lambda5007}{\rm \Hb}\right)+\log\left(0.32+x\right)-0.44;
\end{equation}
\begin{equation}
\label{eq:deltae2}
\Delta E_2= \frac{1}{2}\left[\log\left(\frac{\rm[\ion{N}{II}]\lambda6584}{\rm \Ha}\right)-\log\left(\frac{x}{x+1.93}\right) + 0.37\right];
\end{equation}
\begin{equation}
\label{eq:deltae3}
\Delta E_3=\frac{1}{5}\left[\log\left(\frac{\rm  [\ion{O}{I}]\lambda6300}{\rm \Ha}\right)+2.23\right];
\end{equation}
\begin{equation}
\label{eq:deltae}
\Delta E=\frac{1}{3}\left[\Delta E_1+\Delta E_2+\Delta E_3 \right].
\end{equation}

In the first three diagrams, we used the functions defined by \citet{Kewley06} to discern between \ion{H}{II} regions and power-law ionized regions and to divide the latter in Seyfert-like emission regions and LINER-like emission regions.
We used the $\Delta E$ vs. [\ion{O}{II}]$\lambda\lambda3726,3729/$[\ion{O}{III}]$\lambda5007$ diagrams to investigate the influence of shock-waves in ionizing the NLR/ENLR gas.
Fig.\,\ref{fig:diag_es} shows an example of diagnostic diagrams for IC\,5063 and NGC\,7212.
The diagrams of the other regions are shown in Appendix\,\ref{sec:Adiag}.
Fig.\,\ref{fig:diag_estot} also shows the same diagrams calculated using the total flux of the lines for each region.
From these plots it is possible to see that all the points lie in the AGN region.
Moreover, the majority of the points in the $\log([\ion{O}{III}]\lambda5007/\Hb)$ vs $\log([\ion{O}{I}]\lambda6300/\Ha)$ and in the $\log([\ion{O}{III}]\lambda5007/\Hb)$ vs $\log([\ion{S}{II}]\lambda\lambda6716,6731/\Ha)$ diagrams are located to the left of the diagram, usually occupied by Seyfert galaxies \citep{Kewley06}.
In the $\Delta E$ vs $\log([\ion{O}{II}]\lambda\lambda3726,3729/[\ion{O}{III}]\lambda5007)$ the points are located in the side of the diagram occupied by clouds ionized by a power-law continuum.
This means that the main ionization mechanism in the NLR/ENLR of both galaxies is the photo-ionization by the active nucleus.
However, it must be noticed that in most diagrams the points are not randomly distributed but they follow a well defined trend.
The points corresponding to high values of $\lvert V \lvert$ tend to have a lower [\ion{O}{III}]$\lambda5007/\Hb$ ratio but a higher ratio between low ionization lines and $\Ha$.  
For this reason some points are located close to the LINER region (e.g. Fig.\,\ref{fig:diag_es}).
This behaviour might be explained as a consequence of shocks, which affect the spectrum lowering the ionization degree of the gas (low [\ion{O}{III}]$\lambda5007/\Hb$ ratio) and increasing the amount of ions in a low ionization state (high [\ion{O}{I}]$\lambda6300/\Ha$, [\ion{N}{II}]$\lambda6584/\Ha$ and [\ion{S}{II}]$\lambda\lambda6716,6731/\Ha$).
This is in agreement with the $\Delta E$ vs. [\ion{O}{II}]$\lambda\lambda3726,3729/$[\ion{O}{III}]$\lambda5007$ where the points corresponding to those bins are close to the shocks side of the plot.
Moreover, gas moving at relatively low speed does not show any sign of shocks.
Most of the line fluxes is contained within these bins, therefore, when the diagnostic ratios are calculated using the whole flux of the lines almost no sign of shock-ionized gas is observed.
In fact, analysing the diagnostic diagrams in Fig.\,\ref{fig:diag_estot}, it is possible to see that both in IC\,5063 and NGC\,7212 all the points are clustered in the Seyfert region of the plots, without any peculiar behaviour.
This does not mean that there is no contribution by shocks at low $V$, but that it is negligible with respect to photo-ionization.

\begin{figure*}
\centering
\includegraphics[width=0.43\textwidth]{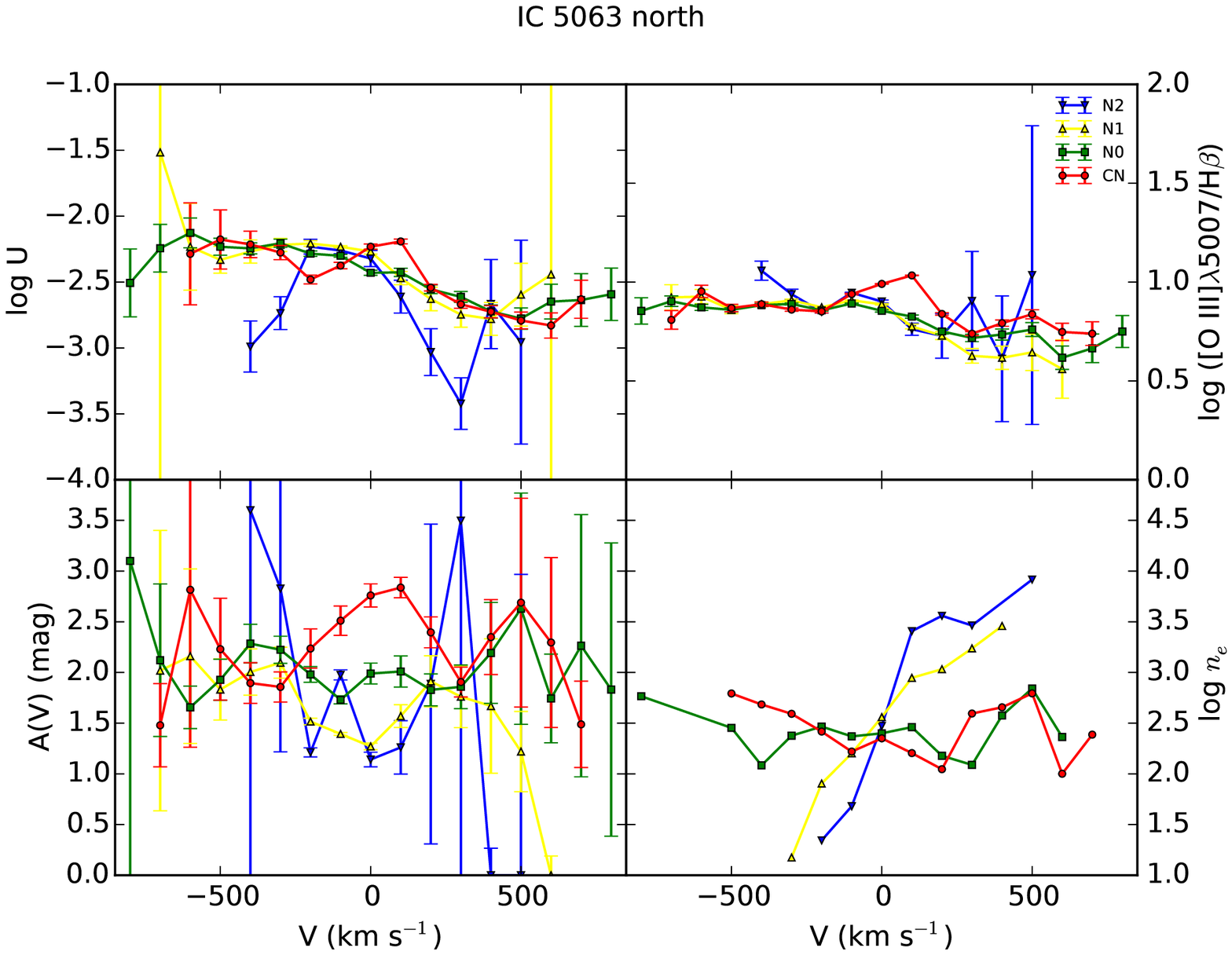} \quad
\includegraphics[width=0.43\textwidth]{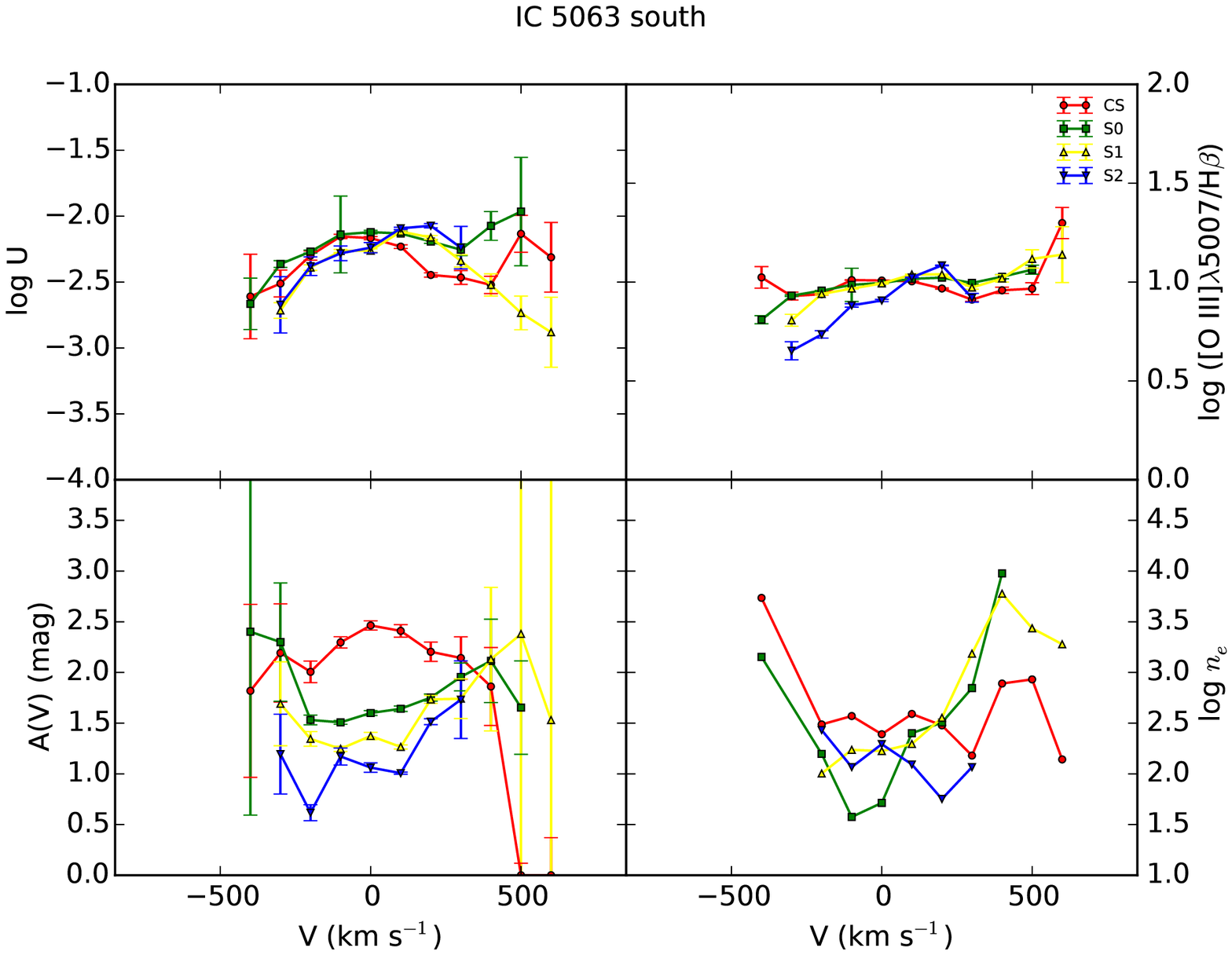}\\
\caption[]{\textbf{Left:} From the top left panel, clockwise: ionization parameter, $\log([\ion{O}{III}]\lambda5007/ \Hb)$, electron density ($\si{cm^{-3}}$) and extinction coefficient (mag) as a function of velocity for the northern regions of IC\,5063. \textbf{Right:} same quantities for the southern regions of IC\,5063.}
\label{fig:all_ns}
\end{figure*}

\begin{figure*}
\centering
\includegraphics[width=0.43\textwidth]{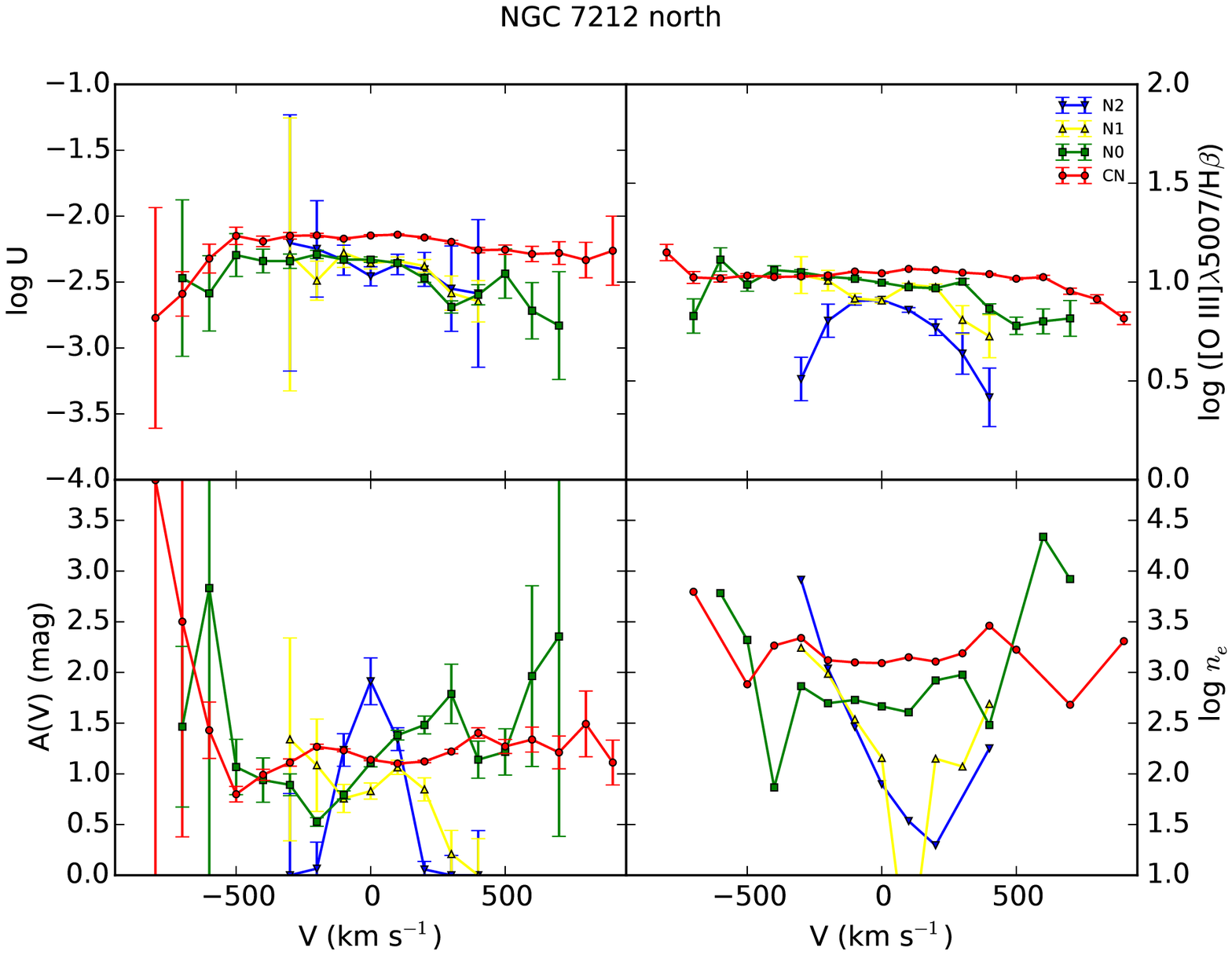} \quad
\includegraphics[width=0.43\textwidth]{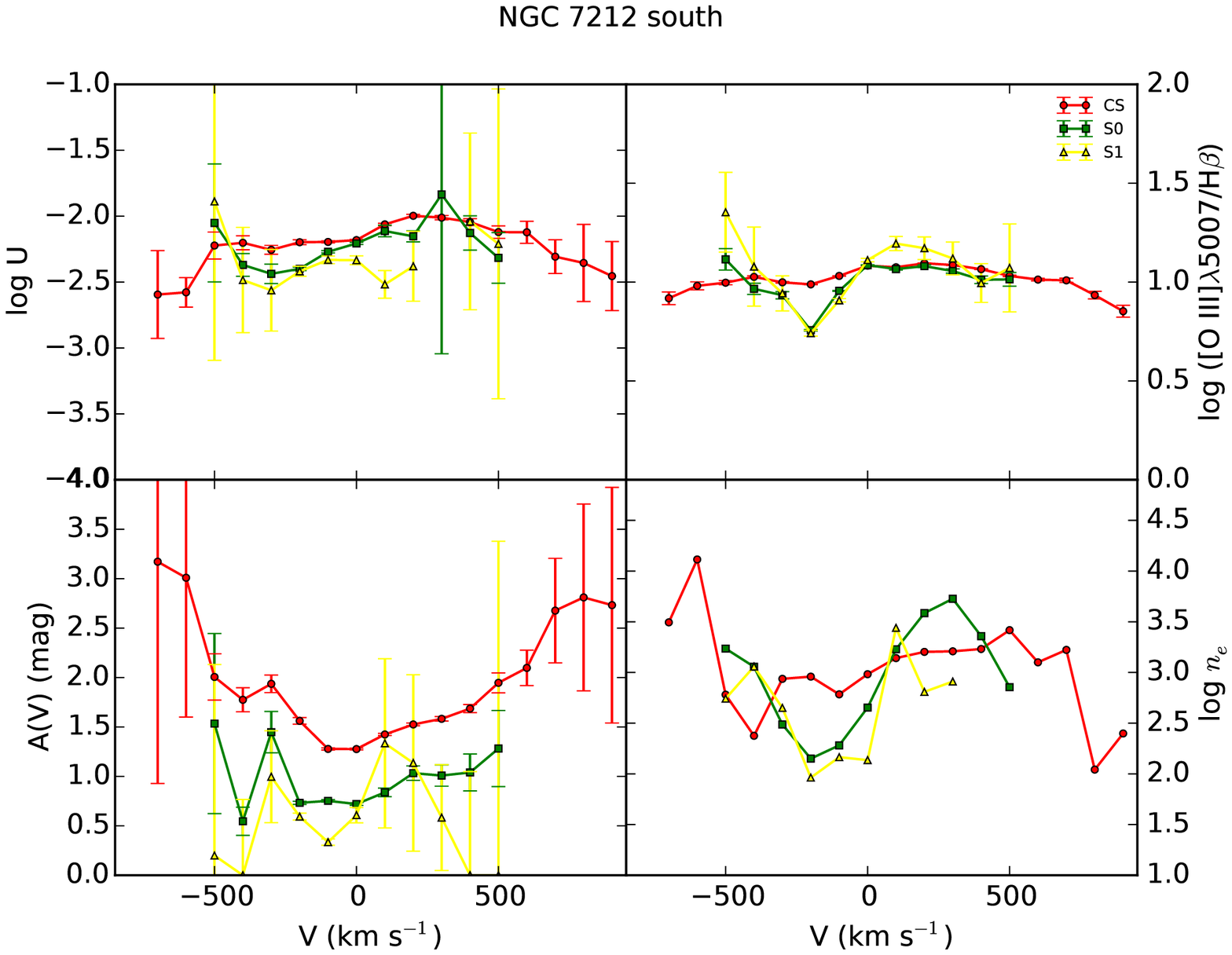}\\
\caption[]{\textbf{Left:} From the top left panel, clockwise: ionization parameter, $\log([\ion{O}{III}]\lambda5007/ \Hb)$, electron density ($\si{cm^{-3}}$) and extinction coefficient (mag) as a function of velocity for the northern regions of NGC\,7212. \textbf{Right:} same quantities for the southern regions of NGC\,7212.}
\label{fig:all_ns_N}
\end{figure*}

\begin{figure*}
\centering
\includegraphics[width=0.43\textwidth]{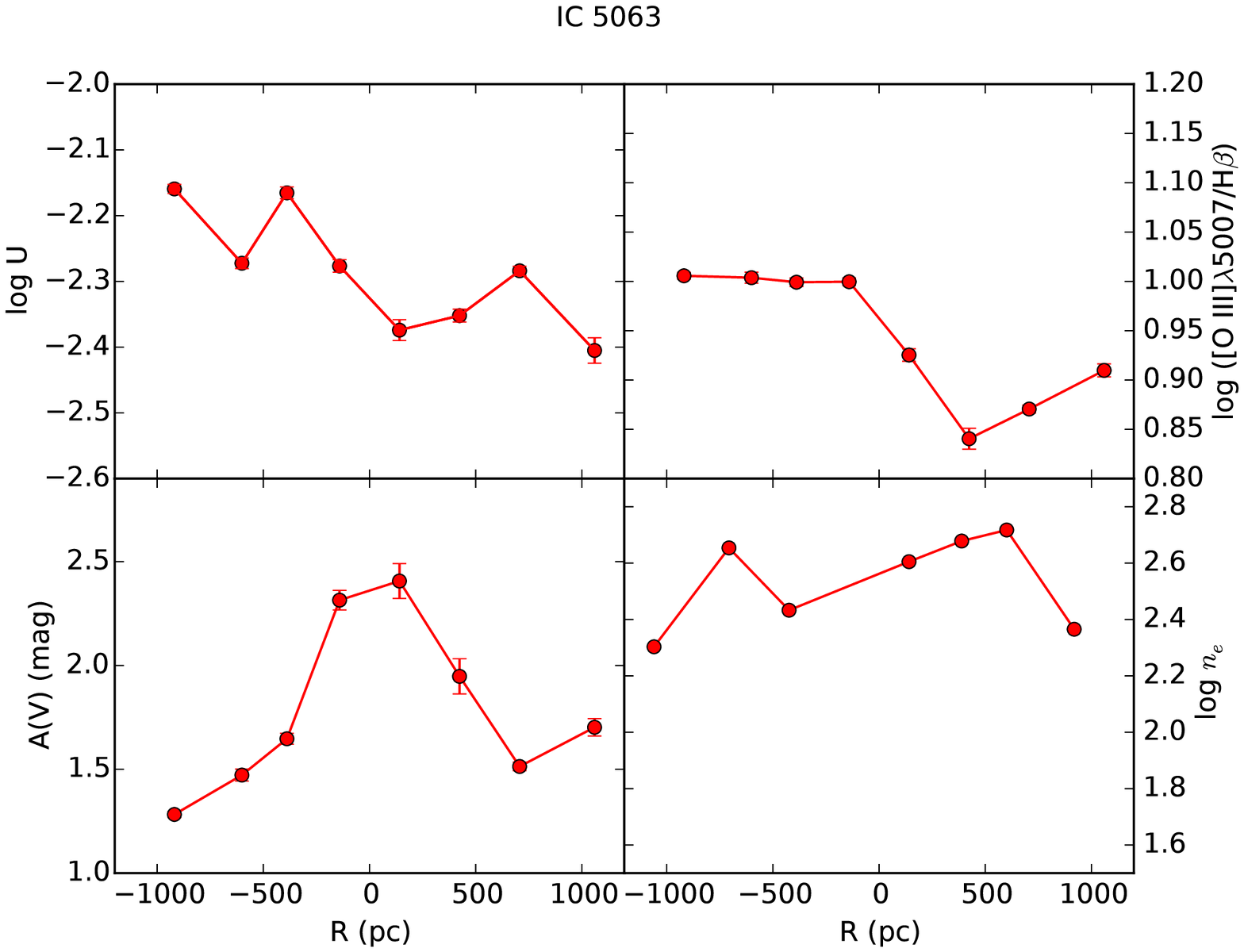} \quad
\includegraphics[width=0.43\textwidth]{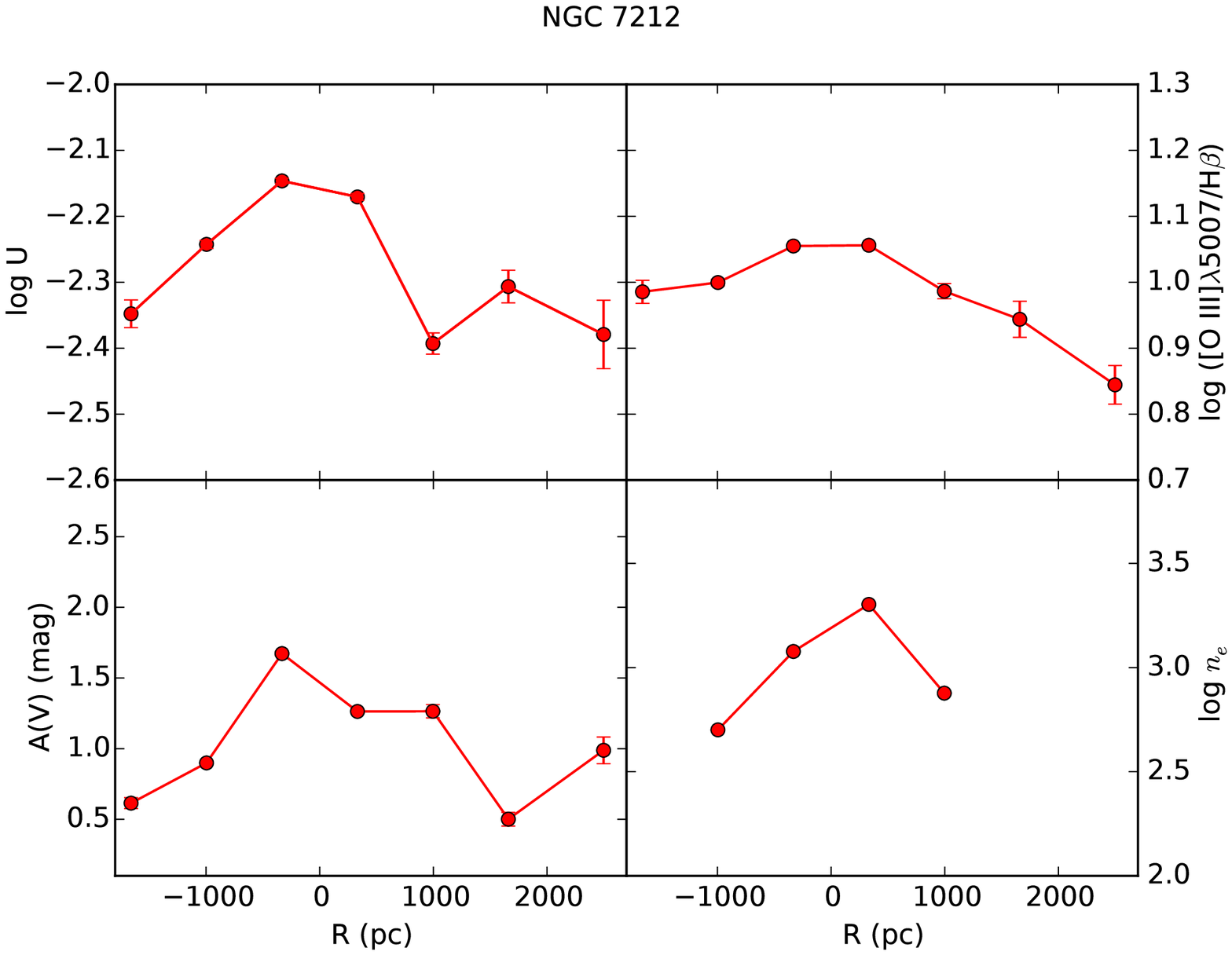}\\
\caption[]{From the top left panel, clockwise: ionization parameter, $\log([\ion{O}{III}]\lambda5007/\Hb)$, electron density ($\si{cm^{-3}}$) and extinction coefficient (mag) as a function of the distance from the galactic nucleus for IC\,5063 (left) and NGC\,7212 (right).}
\label{fig:all_N}
\end{figure*}

\subsection{Ionization parameter and extinction}
\label{sec:ion_par}

The ionization parameter, defined as the number of ionizing photons reaching a cloud per number of electrons in the gas, can be used to estimate the ionization degree of the gas.
For an isotropic source it is defined as :
\begin{equation}
U=\frac{Q}{4\pi r^2 c n_e}\;,
\end{equation}
where $Q$ is the total number of ionizing photons emitted by the source, $r$ is the distance of the cloud from the ionizing source, $c$ is the speed of light and $n_e$ is the electron density.
To estimate this parameter we used the empirical relation from \citet{Penston90}:
\begin{equation}
\label{eq:U}
\log U= -2.74 - \log\left(\frac{[\ion{O}{II}]\lambda\lambda3726,3729}{[\ion{O}{III}]\lambda5007}\right)\;.
\end{equation}
The main problem of this relation is that the $[\ion{O}{II}]\lambda\lambda3726,3729/ [\ion{O}{III}]\lambda5007$ ratio strongly depends on the extinction correction.
To check the reliability of the obtained values we compared the $\log U$ vs velocity ($V$) and $\log U$ vs distance from the nucleus ($R$) plots with similar plots involving the $[\ion{O}{III}]\lambda5007/\Hb$ ratio.
Even if this ratio is not a real ionization parameter, it is sensitive to the ionization degree of the gas and it has the advantage to be independent from the extinction correction, because the two lines are close in wavelength.
When the two diagrams are compatible, the extinction correction applied to the fluxes is reliable.

Fig.\,\ref{fig:all_ns} and Fig.\,\ref{fig:all_ns_N} show the plots of  $\log U$, $\log([\ion{O}{III}]\lambda5007/\Hb)$, A(V) and $n_e$ as a function of velocity.
Similar plots obtained with the total flux of the lines are shown in Fig.\,\ref{fig:all_N}.

The behaviours of $\log U$ and $\log([\ion{O}{III}]/\Hb)$ are often similar.
This confirms the overall reliability of the extinction correction, even though the errorbars on A(V) increase rapidly with $\lvert V \rvert$.
Almost all the differences between the two plots can be explained as an effect of errors in measuring A(V) at high $\lvert V \rvert$ due to a low SNR in $\Hb$ wings.
An example is the region N2 of IC\,5063 (Fig.\,\ref{fig:all_ns} left, blue solid line):  the behaviour of $\log U$ reflects perfectly that of the extinction.

In IC\,5063, A(V) usually shows a peak at low velocities, then it decreases and the errorbars are compatible with a flat trend.
The A(V) vs distance ($R$) plot is similar, we found a maximum extinction of $2.4\,\si{mag}$ in the CN region and then A(V) progressively decreases at increasing $R$.
In NGC\,7212 we observed the same behaviour of the A(V) vs $R$ plot, but the extinction is generally $1\,\si{mag}$ lower.
The high A(V) values of IC\,5063 are expected because the galaxy shows a large dust lane which crosses the whole galaxy that is aligned with the ENLR.
The A(V) vs. $V$ plots show that the extinction is almost constant at all velocities.
However, the errorbars are compatible with a constant behaviour also in those velocity bins.

In principle, $U$ (eq.\,\ref{eq:U}) is not expected to be dependent on velocity because there is not a direct link between these two quantities.
However, if the clouds are accelerated by radiation pressure and ionized by a continuum which is attenuated by an absorbing medium with varying column density, $\log U$ is expected to depend on the gas velocity \citep{Ozaki09}.
This kind of behaviour is observed in some regions of our galaxies.
In Fig.\,\ref{fig:all_ns} it is possible to see that in the CN, N0 and N1 regions of IC\,5063, the ionization parameter decreases of about one order of magnitude between $V=-600\,\si{\kms}$ and $V=500\,\si{\kms}$.
On the basis of the results from \citet{Kraemer00} and \citet{Ozaki09}, this might mean that the blueshifted gas is irradiated directly by the AGN, while the redshifted gas is ionized by an attenuated continuum.
In the S0, S1 and S2 regions $\log U$ increases in the opposite direction, from $V=-400$ to $V=200\,\si{\kms}$.
The velocity range is quite narrow because at higher $\lvert V \rvert$ the effect of the extinction correction dominates with respect to the real behaviour of the ionization parameter.
However, the fact that in these regions, the ionization increases with $V$ is clearly confirmed by the $\log([\ion{O}{III}]/\Hb)$ vs $V$ plot.

\citet{Ozaki09} linked  the ionization parameter dependence on velocity to the geometry of the ionization cones.
If the ionization cones have a hollow bi-conical shape and one of their edges is aligned to the galaxy disk, such as in NGC 1068 \citep{Cecil02, Das06}, this edge should be irradiated by a continuum attenuated by the dust present in the disk while the rest of the cone should be excited by the non-absorbed continuum.
Therefore, if the line of sight intercepts the ionization cone at the proper angle, it is possible to see the velocity dependence of the ionization parameter.

This explanation seems to be in contrast with what said in the previous section.
However, it is possible for the two phenomena to coexist. 
The attenuation of the continuum contributes in decreasing the ionization degree of the gas, also seen in the diagnostic diagrams, while the presence of shocks causes the points in the $\Delta E$ diagram to get close, and often to cross, the shock-power law threshold (Fig.\,\ref{fig:diag_css0}).

The geometrical shape of the ionization cones of IC 5063 is not known but, combining our results with Ozaki's conclusions, the behaviour of the ionization parameter is likely the result of a hollow bi-conical shape. 
This shape could be explained as the effect of the radio-jet expanding in a clumpy medium and forming a cocoon around it, which drives away the gas from the jet axis \citep{Morganti15,Dasyra15} . 

In NGC\,7212, $U$ can be considered independent from the velocity in every region.
All the small deviations from a constant value are related to variations of the extinction.
This might mean that the three-dimensional shape of the ionization cones is different from that of IC\,5063, or that our line of sight intercepts the object at a different angle.
The only region with a peculiar behaviour is N2 (Fig.\,\ref{fig:all_ns_N}).
In this region $\log U$ and $\log([\ion{O}{III}]/\Hb)$ are quite different.
$U$ slightly decreases from $V=-300$ to $V= 400\,\si{\kms}$ while $\log([\ion{O}{III}]/\Hb)$ has a parabolic shape.
A possible reason might be the low SNR of $\Hb$ wings which causes an overestimation of its flux at high velocities and an underestimation of the $\log([\ion{O}{III}]/\Hb)$ ratio.

Our results are well in agreement with \citet{Cracco11}, who found $\log([\ion{O}{III}]/\Hb)\sim1$ in the whole ENLR.
The observed scatter of the ionization parameter is compatible with an almost constant trend.

\subsection{Temperature and densities}

We were able to obtain, in most regions, the average temperature and density of the gas in two different ionization regimes.
In particular, we used [\ion{O}{III}]$\lambda\lambda4959,5007$, [\ion{O}{III}]$\lambda4363$ and  [\ion{Ar}{IV}]$\lambda\lambda4711,4740$ for the gas with medium ionization degree, and [\ion{O}{II}]$\lambda\lambda3726,3729$, [\ion{O}{II}]$\lambda7320$ and  [\ion{S}{II}]$\lambda\lambda6716,6731$ for the gas in low ionization degree.

Unfortunately, some of the lines necessary to estimate temperatures and densities were so faint that they could not be divided in velocity bins (e.g. [\ion{O}{III}]$\lambda4363$, [\ion{Ar}{IV}]$\lambda\lambda4711,4740$).
Therefore, we were forced to use the total flux of these lines and calculate the averaged temperature and density.
Only with the [\ion{S}{II}]$\lambda\lambda6716,6731$ lines it was possible to estimate the density as a function of velocity. 

We proceeded using iteratively the \verb!temden! IRAF task, until we obtained stable values for both temperature and density.
Once reached the final results we used the average temperature of the low ionization gas to attempt to calculate $n_e$ from the [\ion{S}{II}] lines in the velocity bins.
All the obtained results are shown in Fig.\,\ref{fig:all_ns}, \ref{fig:all_ns_N} and \ref{fig:all_N}.

Table\,\ref{tab:ted_ic}  and Table\,\ref{tab:ted_n} report the values obtained with the total flux of the lines.
Due to the errors on the line ratios, what we really are interested in is the order of magnitude of these quantities.
For this reason we rounded off the measured quantities to the nearest hundreds K for the temperature and to the nearest multiple of $50$ for the density.
The observed temperature is of the order of $10000\,\si{K}$, a typical temperature in photo-ionized regions \citep{OsterbrockAGN}.
In IC\,5063 both temperature and density of the medium ionization gas are slightly higher than those of the low ionization gas, while for NGC 7212 the situation is less clear.
NGC\,7212 spectra had a lower SNR than IC 5063 ones, especially in the external regions, therefore it was not possible to measure the weak lines needed for the calculations.
In this galaxy it is worth noticing that the temperature of the low ionization gas measured in the central region of the galaxy is close to $15000\,\si{K}$. 
Following \citet{Roche16}, high temperatures of the low ionization gas might be a hint of jet-ISM interaction.

To calculate the density in the regions where we could not directly measure the temperature, we used the typical value $\rm{T} = 10000\,\si{K}$ for photo-ionized gas \citep{OsterbrockAGN}. 
The density measurement depends weakly on the temperature \citep{OsterbrockAGN} and this value is close enough to the mean $T_{[\ion{O}{II}]}$ measured from from Table\,\ref{tab:ted_ic} and \ref{tab:ted_n} ($\sim 11800\,\si{K}$) that the difference is negligible for our purposes.

After that, we used the temperatures obtained with the [\ion{O}{II}] lines (Table\,\ref{tab:ted_ic} and Table\,\ref{tab:ted_n}) to measure the electron density with the [\ion{S}{II}] lines as a function of velocity.
The results are shown in Fig.\,\ref{fig:all_ns} and Fig\,\ref{fig:all_ns_N}.
Typical values are $\log n_e([\ion{S}{II}]) = 2.0$--$3.0$, but in some cases we measured larger and smaller values, especially at high $\lvert V \rvert$ where the SNR is low and the effects of the deblending can affect the results.
In some regions the situation is not clear (e.g. N1 and N2 regions in IC 5063, Fig.\,\ref{fig:all_ns}) and these properties will be further investigated with SUMA simulations.

\begin{table}
\caption[]{Temperature and electron density of the medium and low ionization gas of IC 5063. 
The temperature are rounded off to the nearest hundred K, the electron density to the nearest multiple of $50$. 
We used the [\ion{O}{II}] doublet when it was not possible to measure $n_e$ with the [\ion{S}{II}] doublet. }
\label{tab:ted_ic}
\centering
\begin{tabular}{lcccc}
\hline
Region&T$_{[\ion{O}{III}]}\,(\si{K})$&n$_{[\ion{Ar}{IV}]}\,(\si{cm^{-3}})$&T$_{[\ion{O}{II}]}\,(\si{K})$&n$_{[\ion{S}{II}]}\,(\si{cm^{-3}})$\\
\hline
N2&-&-&$9100$&$2000$\\
N1&$12400$&$8500$&$7800$&$4500$\\
N0&$13600$&$1400$&$15500$&$250$\\
CN&$13700$&$12750$&$11100\,^a$&$250\,^b$\\
CS&$15200$&$3400$&$10700$&$400$\\
S0&$13800$&$6200$&$11700$&$500$\\
S1&$13300$&$3150$&$10900$&$500$\\
S2&$15300$&$4250$&$12400$&$250$\\
\hline
\multicolumn{5}{l}{$^a$ measured with n$_{[\ion{O}{II}]}$ }\\
\multicolumn{5}{l}{$^b$ measured from the [\ion{O}{II}] doublet}\\
\end{tabular}
\end{table}

\begin{table}
\caption[]{Temperature and electron density of the medium and low ionization gas of NGC 7212.
The temperature are rounded off to the nearest hundred K, the electron density to the nearest multiple of $50$. 
We measured $n_e$ of the low ionization gas assuming $\rm{T}=10000\,\si{K}$ in the regions where it was not possible to measure the temperature.}
\label{tab:ted_n}
\centering
\begin{tabular}{lcccc}
\hline
Region&T$_{[\ion{O}{III}]\,(\si{K})}$&$n_{[\ion{Ar}{IV}]}\,(\si{cm^{-3}})$&T$_{[\ion{O}{II}]\,(\si{K})}$&$n_{[\ion{S}{II}]}\,(\si{cm^{-3}})$\\
\hline
N2&-&-&-&$200\,^a$\\
N1&-&-&-&$150\,^a$\\
N0&-&-&$10800$&$750$\\
CN&$14800$&$5600$&$13000$&$2000$\\
CS&$15200$&$3100$&$14500$&$1200$\\
S0&$14000$&$2450$&$14200$&$500$\\
S1&-&-&-&$350\,^a$\\
\hline
\multicolumn{5}{l}{$^a$ measured with T$_{[\ion{O}{II}]}= 10000\,\si{K}$ }\\
\end{tabular}
\end{table}

\subsection{Line profiles}
\label{sec:lineprof}

The analysis of the line profiles, on the basis of the emitting gas conditions estimated in previous sections, can provide detailed information on the kinematics and the distribution of the clouds.
For each region we compared the profile of some of the brightest lines normalized to their peak emission, to study how they change.
Fig.\,\ref{fig:n2l1_I}--\ref{fig:s2l1_I} and in Fig.\,\ref{fig:n2l1_N}--\ref{fig:s1l1_N} show, from left to right, the comparison between: $[\ion{O}{III}]\lambda5007$ and $\Hb$; $[\ion{O}{I}]\lambda6300$, $[\ion{O}{II}]\lambda3726$ and $[\ion{O}{III}]\lambda5007$; $[\ion{O}{II}]\lambda3726$, $[\ion{O}{II}]\lambda3729$ and $[\ion{O}{III}]\lambda5007$; $[\ion{S}{II}]\lambda6716$, $[\ion{S}{II}]\lambda6731$ and $[\ion{O}{III}]\lambda5007$, for IC\,5063 and NGC\,7212 respectively.
The A(V) profile measured in each region is reported under the corresponding panel, to evaluate if possible differences between the line profiles could be caused by extinction.
We also used these plots to check the results of the deblending process, comparing the lines of the deblended doublets to the $[\ion{O}{III}]\lambda5007$ to highlight potential differences.

The line profile can significantly change from region to region.
In the external regions they are quite narrow and relatively smooth, but they often show a prominent blue or red wing (Fig.\,\ref{fig:n2l1_I}, \ref{fig:s2l1_I}).
NGC\,7212 shows broader and more disturbed profiles than IC\,5063 in the external regions: the FWHM of the N2 region of NGC\,7212 and IC\,5063 are $300\,\si{\kms}$ and $170\,\si{\kms}$ respectively.

The spectra were all shifted to rest frame using stellar kinematics, but, outside the nucleus, the main peak of each line is often shifted toward longer or shorter wavelengths of $\sim100\,\si{\kms}$, probably caused by bulk motions of gas.

In general, the complexity of the line profiles might be related to the interaction of jets with ISM, as better explained at the end of Sec.\,\ref{sec:parameters}, considering the result of the composite modelling.

\subsubsection{IC\,5063}
IC\,5063 shows an emission outside the nucleus (regions S0 and S1, Fig.\,\ref{fig:IC5063}, bottom panel) characterized by two well resolved peaks at $\Delta V\sim 200\,\si{\kms}$ which can be observed in all the emission lines (Fig.\,\ref{fig:s0l1_I} and \ref{fig:s1l1_I}).
A further analysis of the emission lines of the southern regions (CS, S0, S1, S2) shows a connection between their profiles.
In the CS region, which represents the southern part of the nucleus (Fig.\,\ref{fig:csl1_I}), the emission lines are quite narrow and smooth, but there is a small asymmetry at $V\sim 180\,\si{\kms}$.
In the next region (S0, Fig.\,\ref{fig:s0l1_I}) there is a secondary peak at $V\sim 140\,\si{\kms}$, weaker than the main peak.  
In the S1 region (Fig.\,\ref{fig:s1l1_I}) the red peak becomes the strongest one and the blue peak starts to weaken and it becomes a blue wing in the S2 region (Fig.\,\ref{fig:s2l1_I}).
This evolution indicates the presence of two distinct kinematic components.
The blue one could be associated with an outflow which is the dominant component of the nuclear emission and then it starts to become weaker at higher distances from the nucleus.
These two peaks correspond to the two narrow components found by \citet{Morganti07} nearby the South-east hotspot.
In each region the line profiles are very similar for all the considered lines except in S1.
Here, all the lines show a secondary peak which is $10$ to $20$ per cent stronger with respect to the same feature visible in $[\ion{O}{III}]\lambda5007$  (Fig.\,\ref{fig:s1l1_I}), with the exception of $[\ion{O}{I}]\lambda6300$ which is very similar to the $[\ion{O}{III}]$ line.

On the other side of the galaxy the line profiles do not show any secondary peak, but in the CN and N0 regions we observe a very broad component (FWHM$\sim 900\,\si{\kms}$) which can be identified with the well known gas outflow of the galaxy West hotspot  \citep[e.g.][]{Morganti07,Morganti15,Dasyra15}.

In the CN region each line has a different shape (Fig.\,\ref{fig:cnl1_I}).
All the lines show clearly a narrow component and the broad asymmetric component of the outflow.
The relative peak intensity of the two changes as a function of the line.
The $[\ion{O}{III}]\lambda5007$ is characterized by the weakest broad component, while the $[\ion{S}{II}]\lambda6716$ has the strongest one.
Finally the N0 region (Fig.\,\ref{fig:n0l1_I}) shows very similar profiles of $\Hb$ and $[\ion{O}{III}]\lambda5007$, but broader profiles of the low ionization oxygen and sulphur lines.

\subsubsection{NGC\,7212}
NGC\,7212 is characterized by a more complex gas kinematics.
The emission lines are more asymmetric and disturbed.
In the northern regions there is a good agreement between the shape of all the studied lines, except for the N0 region (Fig.\,\ref{fig:n0l1_N}) where the [\ion{O}{I}]$\lambda6300$ is significantly narrower than the other oxygen lines.
All the lines show a blue component which becomes weaker towards the external region of the galaxy.
The southern regions deserve a more detailed analysis.
The lines have a complex profile characterized by multiple kinematic components.
The CS region (Fig.\,\ref{fig:csl1_N}) has a double peak, separated by $\Delta V\sim 150\,\si{\kms}$, which is visible in all lines except in the [\ion{O}{II}] doublet where it becomes an asymmetry.
It is not clear whether this asymmetry is real or an effect of the deblending process which is not able to recover the secondary peak starting from the blended lines.
Such a feature could be caused by a strong extinction of the weaker peak, but this is not confirmed by the A(V) profile.
In the S0 region (Fig.\,\ref{fig:s0l1_N}) the lines show very different profiles.
$\Hb$ has a double peaked shape ($\Delta V\sim 120$) not visible in $[\ion{O}{III}]\lambda5007$ which shows only a wing.
The secondary peak is observed in the [\ion{S}{II}] doublet but not in the oxygen lines.
They have the blue bump, but its relative intensity changes: the $[\ion{O}{II}]\lambda3726$ has the strongest bump ($\sim80\%$ of the peak intensity), followed by the $[\ion{O}{I}]\lambda6300$ ($\sim60\%$) and by the $[\ion{O}{III}]\lambda5007$ ($\sim40\%$).
Finally, in the S1 region (Fig.\,\ref{fig:s1l1_N}) $\Hb$ and all the other low ionization lines (except $[\ion{O}{II}]\lambda3729$) show an asymmetric profile with a blue-shifted peak ($V\sim-150\,\si{\kms}$) and two bumps in the red part of the lines.
On the other hand, $[\ion{O}{III}]\lambda5007$ shows a peak at $V\sim -50\,\si{\kms}$ and a relatively strong bump ($>80$ per cent of the peak intensity) in the velocity where the other lines have the peak.

\subsection{Detailed modelling of the observed spectra}
\label{sec:models}

Modelling the spectra by pure photo-ionization models gives satisfying results for intermediate ionization level lines.
However, collisional phenomena can be critical in the calculation of the spectra emitted by high velocity gas.
The physical properties of the galaxies, such as the complex structure of the emission lines, the presence of merging and possible interaction between jets and ISM among others, suggested us to use a code which takes into account the effects of both photo-ionization and shocks.

The SUMA code \citep[][and references therein]{Contini15} simulates the physical conditions in an emitting gaseous cloud under the coupled effect of photo-ionization from the radiation source and shocks. The line and continuum emission from the gas are calculated consistently with dust-reprocessed radiation in a plane-parallel geometry.
The main physical properties of the emitting gas and the element abundances are accounted for.

\subsubsection{Input parameters}

\begin{figure*}
\includegraphics[width=8.0cm]{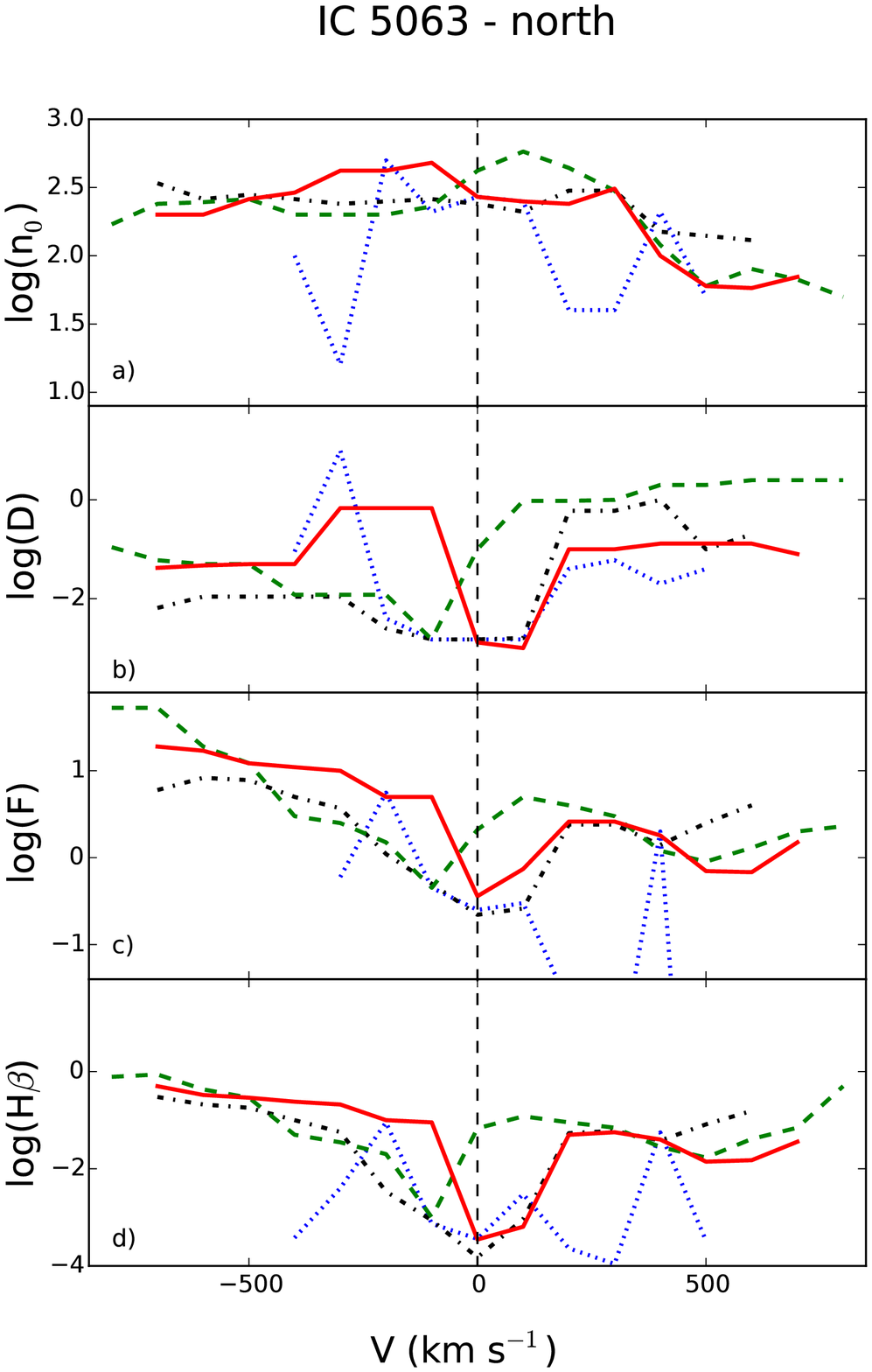}
\includegraphics[width=8.0cm]{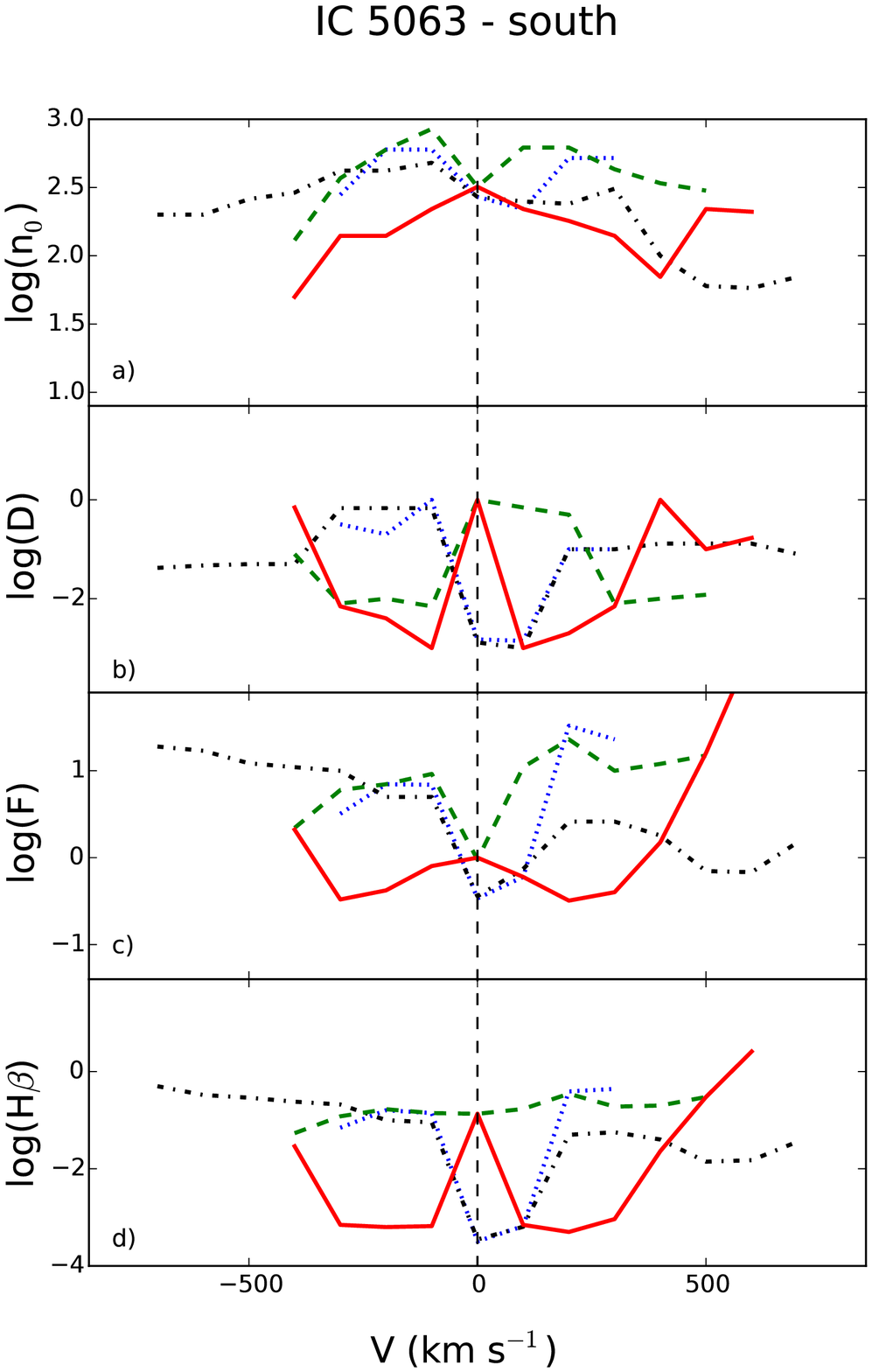}
\caption{The physical quantities in the IC\,5063 regions. From top to bottom: $n_0$ is in \cm3, $D$ is in $10^{18}\,\si{cm}$, $F$ is in $10^{10}\,\si{photons\,cm^{-2} s^{-1} eV^{-1}}$ at the Lyman limit, $\Hb$ is in \erg. Left : north; right: south. Red solid line: CN (CS), green dashed: N0 (S0), black dash-dotted: N1 (S1), blue dotted line: N2 (S2).}
\label{fig:IC_m}
\end{figure*}

\begin{figure*}
\includegraphics[width=8.0cm]{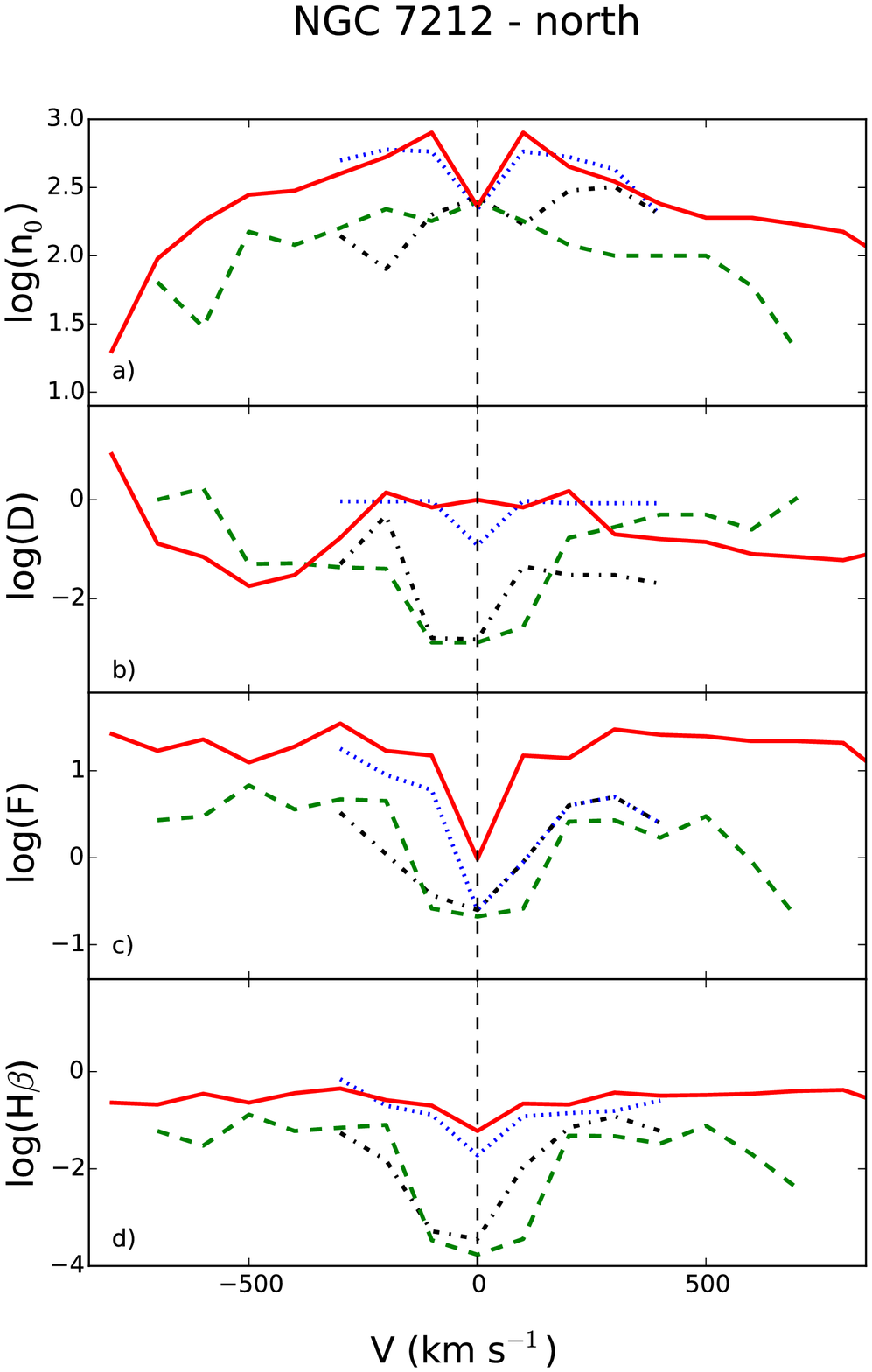}
\includegraphics[width=8.0cm]{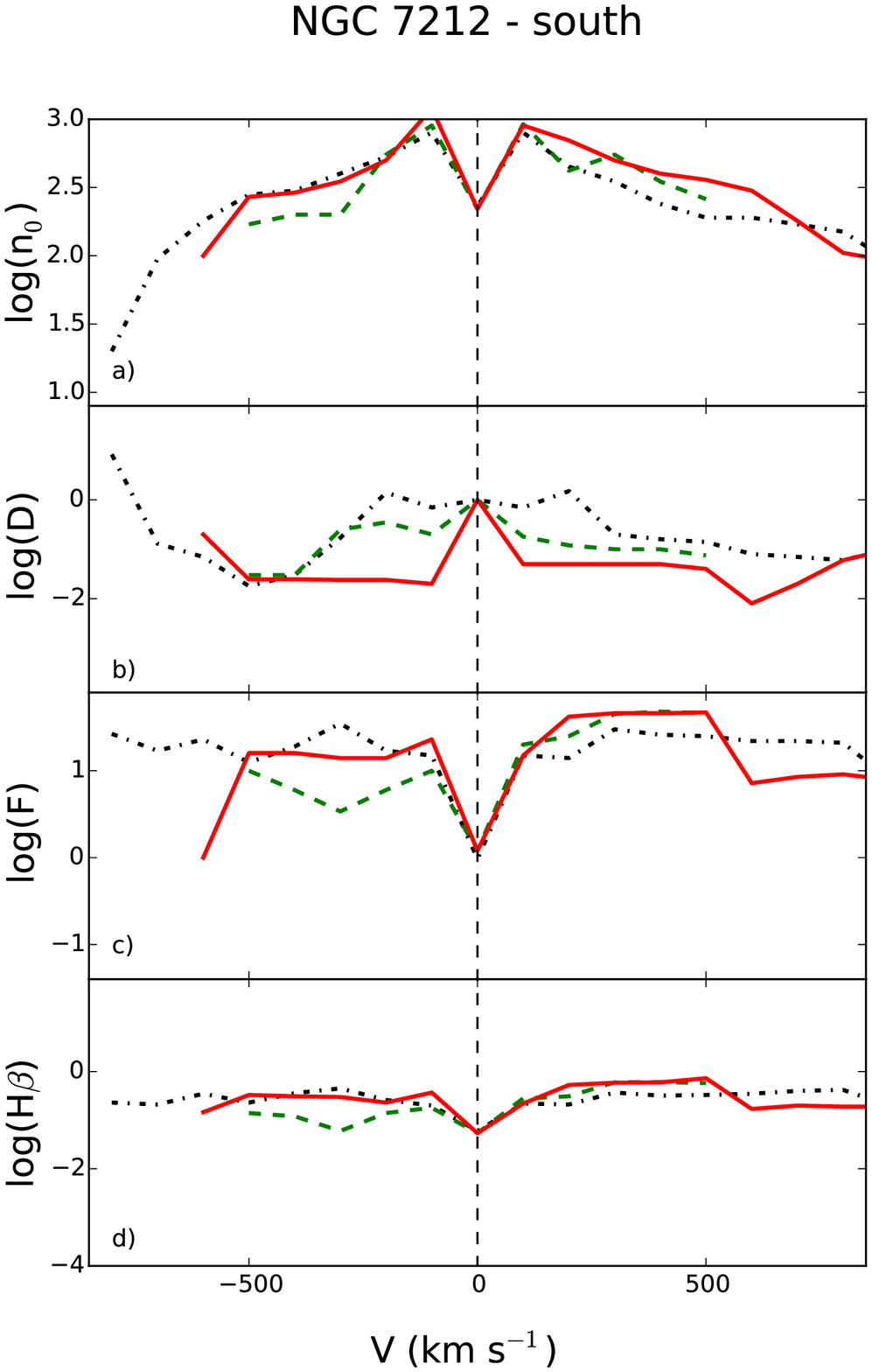}
\caption{The physical quantities in the NGC\,7212 regions. Left: north; right: south. Red solid line: CN (CS), green dashed: N0 (S0), black dash-dotted: N1 (S1), blue dotted line: N2.}
\label{fig:NGC_m}
\end{figure*}

The parameters which characterise the shock are roughly suggested by the data, e.g. the  shock velocity $V_s$ by the velocity bin and the pre-shock density $n_0$ by the characteristic line ratios and by the pre-shock magnetic field $B_0$.
We adopt $B_0  = 10^{-4}$ G, which is suitable to the NLR of AGN \citep{Beck12}.
Changes in $B_0$  are compensated by opposite changes in $n_0$.

The ionizing radiation from a source, external to the emitting cloud, is characterized by its spectrum and by the flux intensity. 
The flux is calculated at $440$ energy bands ranging between a few eV to keV.
If the photo-ionization source is an active nucleus, the input parameter that refers to the radiation field is the power-law flux  from the active center $F$ in number of photons $\si{cm^{-2} s^{-1} eV^{-1}}$ at the Lyman limit with spectral indices  $\alpha_{UV}=-1.5$ and $\alpha_X=-0.7$.
It was found by modelling the spectra of many different AGN that these indices were the most suitable, in general \citep[see, e.g.][ and references therein]{Contini83,Aldrovandi84, Rodriguez05}.
For UV line ratios of a sample of galaxies at $z> 1.7$, \citet{Villar97} found that $\alpha_{UV}=-1$ provided a very good fit.
The power-law in the X-ray domain was found flatter by the observations of local galaxies \citep[$\alpha_X=-1$, e.g.][]{Crenshaw02,Turner01}.
Nevertheless, for all the models presented in the following, we will adopt $\alpha_{UV}=-1.5$ and $\alpha_{X} = -0.7$, recalling that the shocked zone also contributes to the emission-line intensities.  
Therefore our results are less dependent on the shape of the ionizing radiation.
$F$ is combined with the ionization parameter $U$ by: 
\begin{equation}
\label{eq:Umarc}
U= \left(\frac{F}{n c (\alpha_{UV} -1)}\right) \left(E_H^{-\alpha_{UV} +1} - E_C^{-\alpha_{UV} +1}\right),
\end{equation}
where $E_H$ is H ionization potential  and $E_C$ is the high energy cutoff, $n$ the density, $\alpha_{UV}$ the spectral index, and $c$ the speed of light.

In addition to the radiation from the primary source, the diffuse secondary radiation created by the hot gas is also calculated, using $240$ energy bands for the spectrum.
The primary radiation source is independent but it affects the surrounding gas.
In contrast, the secondary diffuse radiation is emitted from the slabs of gas heated by the radiation flux reaching the gas and, in particular, by the shock.
Primary and secondary radiations are calculated by radiation transfer downstream.

In our model the gas region surrounding the radiation source is not considered as a unique cloud, but as an ensemble of filaments. 
The geometrical thickness of these filaments is an input parameter of the code ($D$) which is calculated consistently with the physical conditions and element abundances of the emitting gas.
$D$ determines whether the model is radiation-bounded or matter-bounded.
The abundances of He, C, N, O, Ne, Mg, Si, S, Ar, Fe, relative to H, are accounted for as input parameters. 
Previous results lead to O/H close to solar for most AGNs and for most of the \ion{H}{II} regions \citep[e.g.][]{Contini17}.
We will conventionally define ``solar''  relative abundances (O/H)$_{\odot }=6.6 - 6.7 \times10^{-4}$ and (N/H)$_{\odot }= 9.\times10^{-5}$ \citep{Allen76,Grevesse98} that were found suitable to local galaxy nebulae. 
Moreover, these values are included between those of \citet{Anders89} ($8.5\times10^{-4}$ and $1.12\times10^{-4}$, respectively) and \citet{Asplund09} ($4.9\times10^{-4}$ and $6.76\times10^{-5}$, respectively ).

The fractional abundances  of all the ions in all ionization levels are calculated  in each slab resolving the ionization equations.
The dust-to-gas ratio ($d/g$)  is an input parameter. 
Mutual heating and cooling between dust and gas affect the temperatures of gas in the emitting region of the cloud.
Fig.\,\ref{fig:IC_m} and Fig.\,\ref{fig:NGC_m} show the input parameter as a function of $V$ used to model the observed spectra.

\subsubsection{Calculation process}

The code accounts for the direction of the cloud motion relative to the external photo-ionizing source.  
A parameter switches between inflow (the radiation flux from the source reaches the shock front edge of the cloud) and outflow (the flux reaches the edge opposite to the shock front).
The calculations start at the shock front where the gas is compressed and thermalized adiabatically, reaching the maximum temperature in the immediate post-shock region (eq.\,\ref{eq:T}).
\begin{equation}
\label{eq:T}
T\sim 1.5 \times 10^5 \left(\frac{V_s}{100\,\si{\kms}}\right)^2. 
\end{equation}
$T$ decreases downstream by the cooling rate and the gas recombines. 
The downstream region is cut into a maximum of 300 plane-parallel slabs with different geometrical widths calculated automatically, in order to account for the temperature gradient.
In each slab, compression ($n/n_0$) is calculated by combining the Rankine--Hugoniot equations for conservation of mass, momentum and energy throughout the shock front \citep{Cox72}.
Compression ranges between $4$ (the adiabatic jump) and $> 100$, depending on $V_s$ and $B_0$. 
The stronger the magnetic field, the lower the compression downstream, while a higher shock velocity corresponds to a higher compression.

In pure photo-ionization models, the density $n$ is constant throughout the nebula.
In models accounting for the shocks, both the electron temperature  $T_{\rm e}$  and  density $n_e$ show a characteristic profile throughout each cloud (see for instance Fig.\,\ref{fig:sumamod} left and right panels). 
After the shock, the temperature reaches its upper limit at a certain distance from the shock-front and remains nearly constant, while $n_e$ decreases following recombination.
The cooling rate is calculated in each slab by free-free (bremsstrahlung), free-bound and line emission. 
Therefore, the most significant lines must be calculated in each slab even if only a few ones are observed because they contribute to the temperature slope downstream.

The primary and secondary radiation spectra change throughout the downstream slabs, each of them contributing to the optical depth. 
In each slab of gas the fractional abundance of the ions of each chemical element is obtained by solving the ionization equations which account for photo-ionization (by the primary and diffuse secondary radiations and collisional ionization) and for  recombination (radiative, dielectronic), as well as for charge transfer effects, etc. 
The ionization equations are coupled to the energy equation when collision processes dominate \citep{Cox72} and to the thermal balance if radiative processes dominate. 
The latter balances the heating of the gas due to the primary and diffuse radiations reaching the slab with the cooling due to  
line emission, dust collisional ionization and thermal bremsstrahlung. 
The line intensity contributions from all the slabs are integrated throughout the cloud.
In particular, the absolute line fluxes referring to the ionization level i of element K are calculated by the term $n_K$(i) which represents the density of the ion X(i).
We consider that $n_K$(i)=X(i)[K/H]$n_H$, where X(i) is the fractional abundance of the ion i calculated by the ionization equations, [K/H] is the relative abundance of the element K to H and $n_H$ is the density of H (by number \cm3). 
In models including shock, $n_H$ is calculated by the compression equation in each slab downstream. 
So the element abundances relative to H appear as input parameters.
To obtain the N/H relative abundance for each galaxy, we consider the charge exchange reaction  N$^+$+H $\rightleftharpoons$ N+H$^+$. 
Charge exchange reactions occur between ions with similar ionization potential ($\rm I(H^+)=13.54\,\si{ eV}$, $\rm I(N^+)=14.49\, \si{eV}$ and $\rm I(O^+)=13.56\,\si{ eV}$). 
It was found that N ionization equilibrium in the ISM is strongly affected by charge exchange.  
This process as well as O$^+$+H $\rightleftharpoons$ O+H$^+$  are included in the SUMA code.
The N$^+$/N ion fractional abundance follows the behaviour of O$^+$/O so, comparing the [\ion{N}{II}]/$\Hb$   and the [\ion{O}{II}]/$\Hb$ line ratios with the data, the N/H relative abundances can be easily determined \citep[see][]{Contini12}.

Dust grains are coupled to the gas across the shock front by the magnetic field. 
They are heated radiatively by photo-ionization and collisionally by the gas up to the evaporation temperature ($\rm T_{\rm dust} \geq 1500\,\si{K}$). 
The distribution of the grain radii in each of the downstream slabs is determined by sputtering, which depends on the shock velocity and on the gas density. 
Throughout shock fronts and downstream, the grains might be completely destroyed by sputtering.

The calculations proceed until the gas cools down to a temperature below $10^3\,\si{K}$ (the model is radiation bounded) or the calculations are interrupted when all the lines reproduce the observed line ratios (the model is matter bounded).
In case that photo-ionization and shocks act on opposite edges, i.e. when the cloud propagates outwards from the 
radiation source, the calculations require some iterations, until the results converge.
In this case  the cloud geometrical thickness plays an important role.  
Actually, if the cloud is very thin, the cool gas region may disappear leading to low or negligible low ionization level lines. 

Summarizing, the calculations start in the first slab downstream adopting the input parameters  given by the model.
Then, it calculates the density, the fractional abundances of the ions from each level for each element, free-free,  free-bound  and line emission fluxes. 
It calculates $T_e$ by thermal balancing or the enthalpy equation, and the optical depth of the slab  in order to obtain the primary and secondary fluxes by radiation transfer for the next slab. 
Finally,  the parameters  calculated in slab i are adopted as initial conditions for slab i+1. 
Integrating the line intensities from each slab, the absolute fluxes of the lines and of bremsstrahlung are obtained at the nebula. 
The line ratios to a certain line (generally $\Hb$ for the optical-UV spectrum) are then calculated and compared with the observed data, in order to avoid problems of distances, absorption, etc.
The number of the lines calculated by the code (over 300) does not depend on the number of
the observed lines nor does it depend on the number of  input parameters,
but rather on the elements composing the gas.

\subsubsection{Grids of models}

The physical parameters are combined throughout the calculation of forbidden and permitted lines emitted from a shocked nebula.
The ranges of the physical conditions in the gas are deduced, as a first guess, from the observed line ratios because they are more constraining than the continuum SED.
Grids of models are calculated for each spectrum, modifying the input parameters gradually, in order to reproduce as close as possible all the observed line ratios.
At each stage of the modelling process, if a satisfactory fit is not found for all the lines, a new iteration is initiated with a different set of input parameters. 
When one of the line ratios is not reproduced, we check how it depends on the physical parameters and decide accordingly how to change them, considering that each ratio has a different weight.
The input parameters are therefore refined by the detailed modelling of the spectra.
The spectra of NGC\,7212 and IC\,5063 are rich in number of lines, therefore the calculated spectra are strongly constrained by the observed lines. 
They are different in each of the observed spectra, revealing different physical conditions from region to region. 
The models selected  by the fit of the line spectrum  are cross-checked by fitting the continuum SED.
In the UV range the bremsstrahlung from the  nebula is blended with black body emission from the star population background. 
The maximum frequency of the bremsstrahlung peak in the UV -- X-ray domain depends on the shock velocity. 
In the IR range dust reprocessed radiation is generally seen.  
In the radio range synchrotron radiation by the Fermi mechanism at the shock front is easily recognized by the slope of the SED.

We generally consider that the observed spectrum is satisfactorily fitted by a model when the strongest lines are
reproduced by the calculation within 20 per cent and the weak ones within 50 per cent.
The final gap between observed and calculated line ratios is due to observational errors both random and systematic, as well as to the uncertainties of the atomic parameters  adopted by the code, such as recombination coefficients, collision strengths etc., which are continuously updated, and to the choice of the model itself.
The set of the input parameters which leads to the best fit of the observed line ratios and continuum SED determines the physical and chemical properties of the emitting gas. 
They are considered as the "results" of modelling (Table\,\ref{tab:sim_s0-}). 

\begin{table*}
\centering
\caption{Simulation for $V < 0$ of the S0 region of IC\,5063. For each bin there are the observed quantities and the results of the models. The first nine rows show the comparison between the observed and the synthetic spectra, the remaining rows show the input parameters of each model.}
\label{tab:sim_s0-}
\begin{tabular}{lcccccccc} 
\hline
\ &\multicolumn{2}{c}{bin 1}&\multicolumn{2}{c}{bin 2}&\multicolumn{2}{c}{bin 3}&\multicolumn{2}{c}{bin 4}\\
\   line              & obs.  &mod. &obs.&mod.  &obs.  &mod. &obs.  &mod.  \\ 
\hline
\ [\ion{O}{II}]$\lambda3727+ $         &$2.43   $&$ 2.68 $&$3.06 $&$2.8   $&$3.58  $&$3.5  $&$5.43  $&$5.47    $\\            
\ [\ion{Ne}{III}]$\lambda3869+ $       &$1.27   $&$ 1.4  $&$1.35 $&$1.6   $&$2.2   $&$1.6   $&$0.12  $&$1.0    $\\            
\ [\ion{He}{II}]$\lambda4686$           &$0.15   $&$ 0.37 $&$0.12 $&$0.3   $&$0.02  $&$0.28  $&$0.0   $&$0.2    $\\            
\ [\ion{O}{III}]$\lambda5007+ $        &$12.97  $&$ 13.0 $&$12.0 $&$12.0  $&$11.4  $&$11.1  $&$ 8.98 $&$ 8.88  $\\            
\ [\ion{O}{I}]$\lambda6300$,[\ion{S}{III}]$\lambda6310$  &$0.32   $&$0.46  $&$0.3  $&$0.3   $&$0.41  $&$0.46  $&$0.53  $&$0.4$\\            
\ $\Ha$ &$2.86   $&$ 2.9  $&$2.86 $&$2.9   $&$2.86  $&$2.9   $&$2.86  $&$2.96   $\\            
\ [\ion{N}{II}]$\lambda6584$ &$1.9    $&$ 1.9  $&$2.07 $&$2.2   $&$2.18  $&$2.    $&$3.22  $&$3.86    $\\            
\ [\ion{S}{II}]$\lambda6716$  &$0.9    $&$ 0.53 $&$1.09 $&$0.54  $&$0.92  $&$0.5   $&$0.98  $&$0.82   $\\            
\ [\ion{S}{II}]$\lambda6731$  &$0.8    $&$ 1.1  $&$0.9  $&$1.1   $&$1.0   $&$1.0   $&$0.72  $&$1.3     $\\            
\ V ($\si{\kms}$) &$-100   $&$-     $&$-200 $&$-     $&$-300  $&$-     $&$-400  $&$-        $\\            
\ \Vs($\si{\kms}$)  &$-      $&$ 100  $&$-    $&$170   $&$-     $&$300   $&$-     $&$400     $\\            
\ \n0(\cm3)  &$-      $&$ 850  $&$-    $&$600   $&$-     $&$370   $&$-     $&$130     $\\            
\ D($10^{18}\,\si{cm}$)      &$-      $&$ 0.007$&$-    $&$0.01  $&$-     $&$0.008 $&$-     $&$0.08  $\\            
\ F (units$^1$)       &$-      $&$ 9.2  $&$-    $&$7.    $&$-     $&$6.    $&$-    $&$2.2     $\\            
\ O/H ($10^{-4}$)     &$-      $&$ 5.   $&$-    $&$6.6   $&$-     $&$6.6   $&$-     $&$6.1     $\\            
\ N/H ($10^{-4}$)     &$-      $&$ 0.5  $&$-    $&$0.6   $&$-     $&$0.5   $&$-     $&$0.8     $\\            
\ Ne/H($10^{-4}$)     &$-      $&$ 0.7  $&$-    $&$1.    $&$-     $&$1.    $&$-     $&$0.7    $\\            
\ S/H ($10^{-4}$)     &$-      $&$ 0.2  $&$-    $&$0.2   $&$-     $&$0.18  $&$-     $&$0.12    $\\            
\ $\Hb$ ($\si{erg}$)          &$-      $&$0.14  $&$-    $&$0.168 $&$-     $&$0.12  $&$-     $&$0.054   $\\            
\hline
\end{tabular}
\\
$^1$ in $10^{10}\,\si{phot\,cm^{-2} s^{-1} eV^{-1}}$ at the Lyman limit
(${\alpha}_{UV}=-1.5$, ${\alpha}_X=-0.7$);

\end{table*}

\subsubsection{Choice of  NGC\,7212 and IC\,5063 parameters}
\label{sec:parameters}

We start modelling by trying to reproduce the observed [\ion{O}{III}]$\lambda\lambda5007,4959/\Hb$ line ratio ($\lambda\lambda5007,4959$, hereafter $5007$+; the + indicates that the doublet 5007, 4959 is summed up), which is in general  the highest  ratio, by readjusting $F$ and $V_s$ ($V_s$, however, is constrained through a small range by the observed $V$). 
The higher $F$, the higher the [\ion{O}{III}]/$\Hb$ and the [\ion{O}{III}]/[\ion{O}{II}] line ratios, as well as \ion{He}{II}/$\Hb$. 
Moreover, a high $F$ maintains the gas ionized far from the source, yielding enhanced [\ion{O}{I}] and [\ion{S}{II}] lines. 
These lines behave similarly because the first ionization potential of S ($10.36\,\si{ eV}$) is lower than that of O ($13.61\,\si{ eV}$).
Then, we consider the [\ion{O}{II}]$\lambda\lambda3726,3729$ doublet (hereafter $3726$+). 
If the flux from the active centre is low ($F \leq 10^9\,\si{ph.cm^{-2} s^{-1} eV^{-1}}$), a shock-dominated regime is found, which is characterised by relatively high [\ion{O}{II}]/[\ion{O}{III}] ($\geq 1$). 
[\ion{O}{II}] can be drastically reduced by collisional de-excitation at high electron densities ($n_e >  3000\,\si{cm^{-3}}$).

The gas density is a crucial parameter. 
In each cloud, it reaches its upper limit downstream and remains nearly constant, while the electron density decreases following recombination. 
A high density, increasing the cooling rate, speeds up the recombination process of the gas, enhancing the low-ionization lines.
Indeed, each line is produced in a region of gas at a different $n_e$ and $T_e$, depending on the ionization level and the atomic parameters characteristic of the ion. 
The density n, which can be roughly inferred from the [\ion{S}{II}]$6716/6731$ doublet ratio, is related with $n_0$ by compression downstream ($n/n_0$), which ranges between $4$ and $\sim100$, depending on $V_s$ and $B_0$. 
The [\ion{S}{II}] lines are also characterised by a relatively low critical density for collisional de-excitation.  
In some cases the [\ion{S}{II}]6716/6731 line ratio varies from $> 1$ to $< 1$ throughout a relatively small region, since the [\ion{S}{II}] line ratios depend on both the temperature and electron density of the emitting gas \citep{OsterbrockAGN}, which in models accounting for the shock are far from constant throughout the clouds.
Thus, even sophisticated calculations which reproduce approximately the high inhomogeneous conditions of the gas
lead to some discrepancies between the calculated and observed line ratios.
Unfortunately, there are no data for S lines from higher ionization levels which could indicate whether the choice of the model is misleading or different relative abundances should be adopted. 
We recall that sulphur can be easily depleted from the gaseous phase and trapped into dust grains and molecules.

\begin{figure*}
\centering
\includegraphics[width=8.8cm]{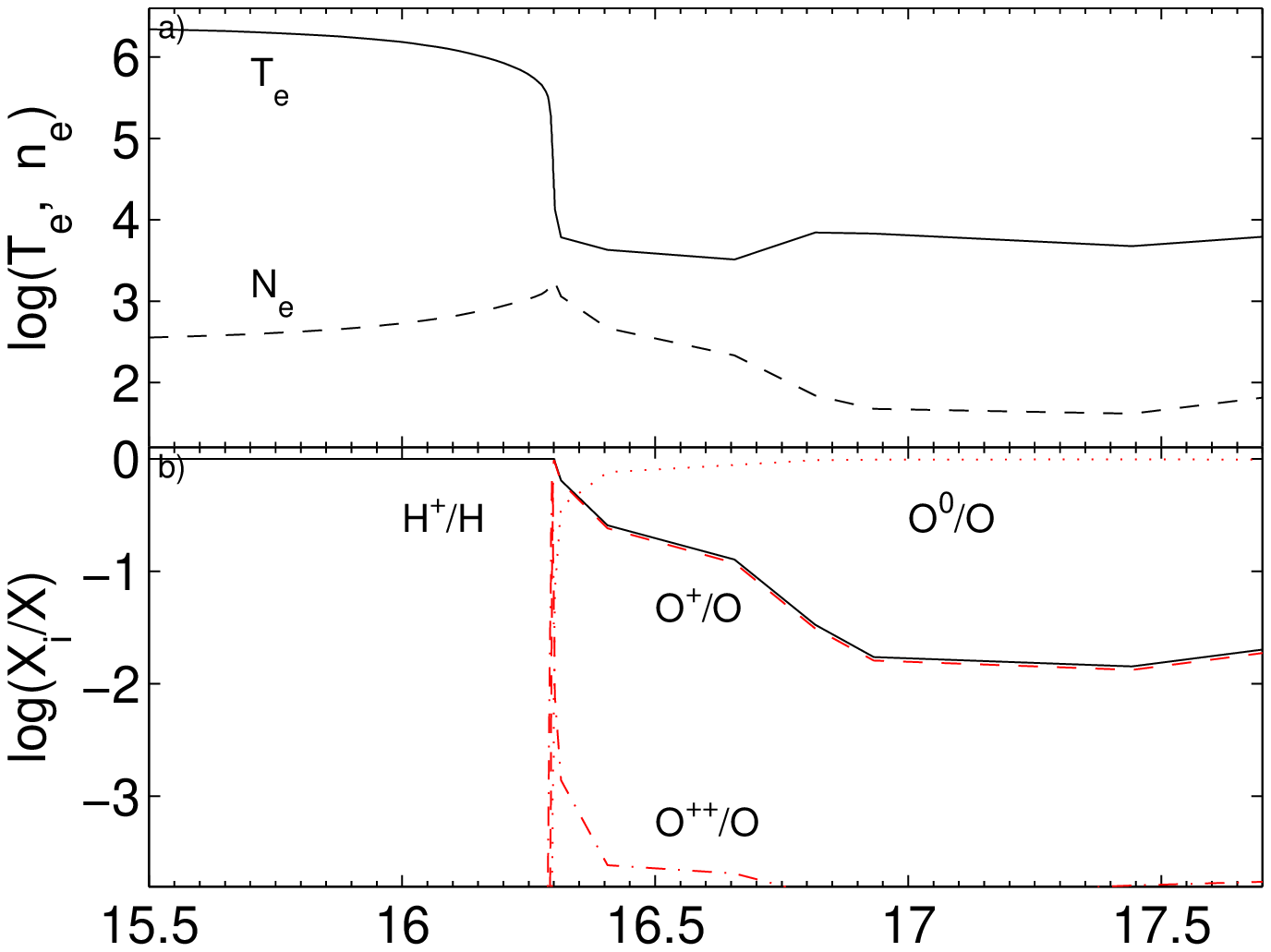}
\includegraphics[width=8.8cm]{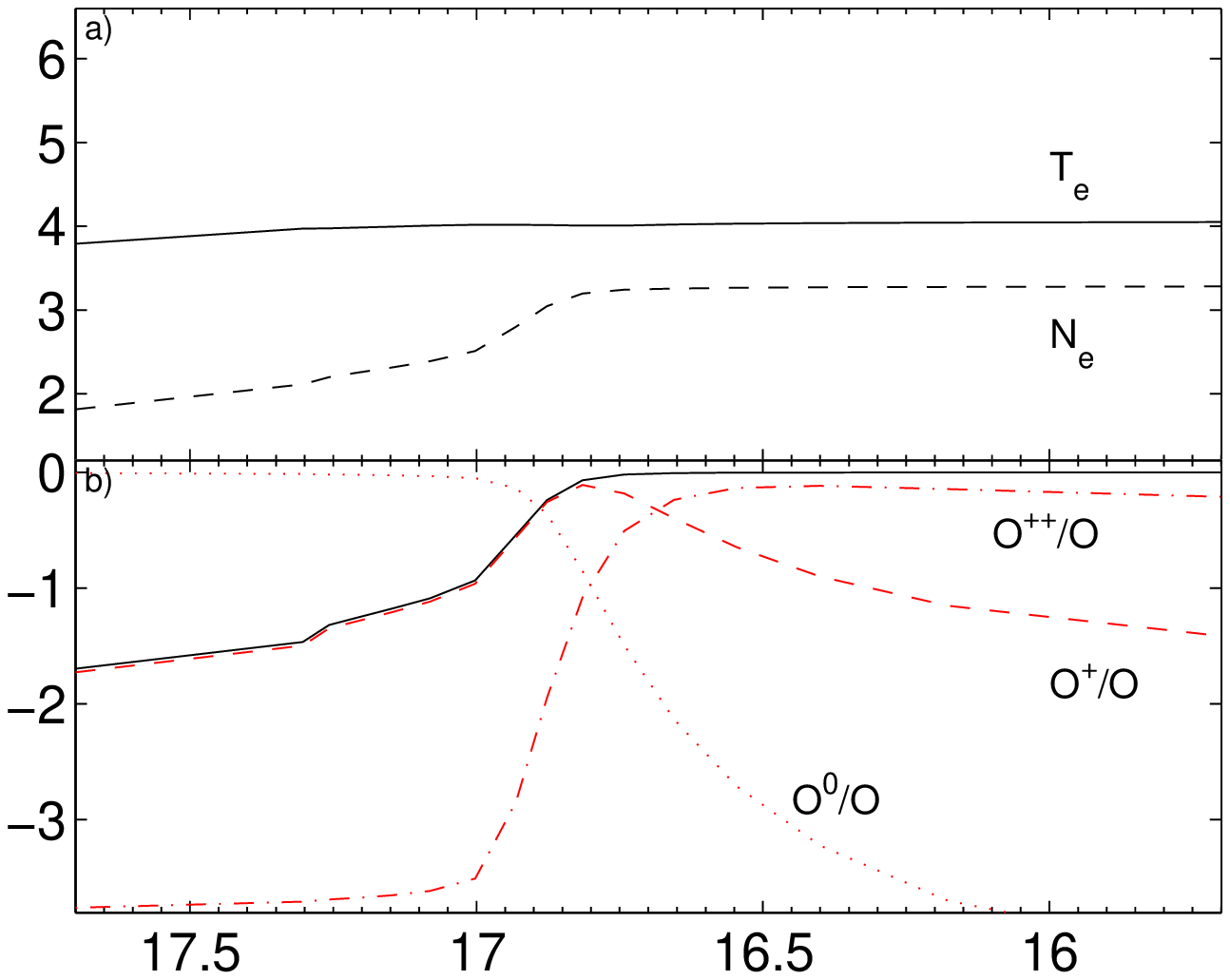}
\caption{The emitting cloud is divided into two halves represented by the left and right  diagrams.  
The left diagram shows the region close to the shock front (left edge) and the distance from the shock front on the X-axis scale is logarithmic.
The right diagram shows the conditions downstream far from the shock front, close to the (right) edge reached by the photo-ionization flux which is opposite to the shock front.  
The distance from the illuminated edge is given by a reverse logarithmic X-axis scale.
Top panels: the electron temperature and  the electron density throughout the emitting cloud.
Bottom panels: red lines: O$^{++}$/O (dot-dashed),  O$^+$/O (dashed) and O$^0$/O (dotted); black solid lines: H$^+$/H. 
}
\label{fig:sumamod}
\end{figure*}

Finally, the results of the modelling are shown in Table C1-C32 of the on-line material.
We show here Table\,\ref{tab:sim_s0-}, as an example of a typical table. 
We can conclude that composite models are able to reproduce the observed spectra accurately, in particular they well reproduce the high [\ion{O}{I}]/$\Hb$ line ratios.
For instance, the spectrum in Table C9 of the on-line material, corresponding to the emitting cloud in the CS region of IC5063 (bin 4), shows $V_s$ of $\sim 400\,\si{\kms}$. 
The cloud moves outwards from the active centre, therefore the photo-ionizing radiation reaches the edge opposite to the shock front.
The profiles of $T_e$, $n_e$ and of the O$^{++}$/O, O$^+$/O and O$^0$/O fractional abundances throughout the cloud are shown in Fig.\,\ref{fig:sumamod}.
Reducing the shock velocity to $V_s =50\,\si{\kms}$ and even lower we can simulate the case of pure photo-ionization.
With these models we can still obtain a good fit of [\ion{O}{III}]/$\Hb$ and [\ion{O}{II}]/$\Hb$ by increasing the pre shock density and the photo-ionization flux.
However,  [\ion{O}{I}]/$\Hb$  will be lower than observed by a factor of $\sim 10$, indicating that the shock velocity constrains the spectra, particularly at high $V_s$.

Another interesting result of the modelling is the high fragmentation of matter which is revealed by the large range of geometrical thickness ($D$) used to model the spectra.
This might be explained by the interaction between jets and ISM.
The interaction causes shocks and it creates turbulence at the shock-front producing fragmentation of matter.
These clouds move in a turbulent regime which can cause the complex line profile described in Sec.\,\ref{sec:lineprof}. 

\subsection{The spectral energy distribution of the continuum}

\begin{figure}
\includegraphics[width=7.5cm]{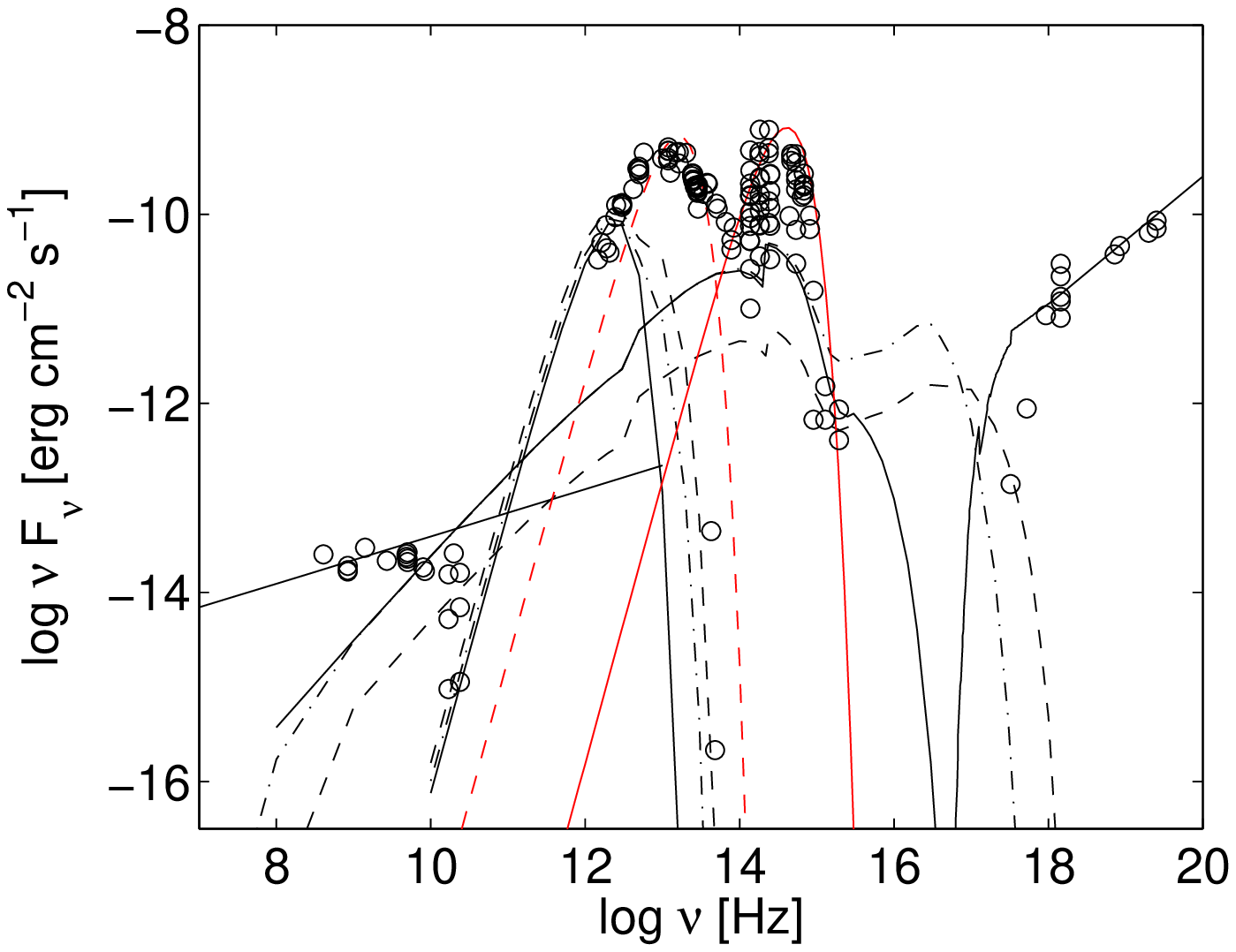}
\includegraphics[width=7.5cm]{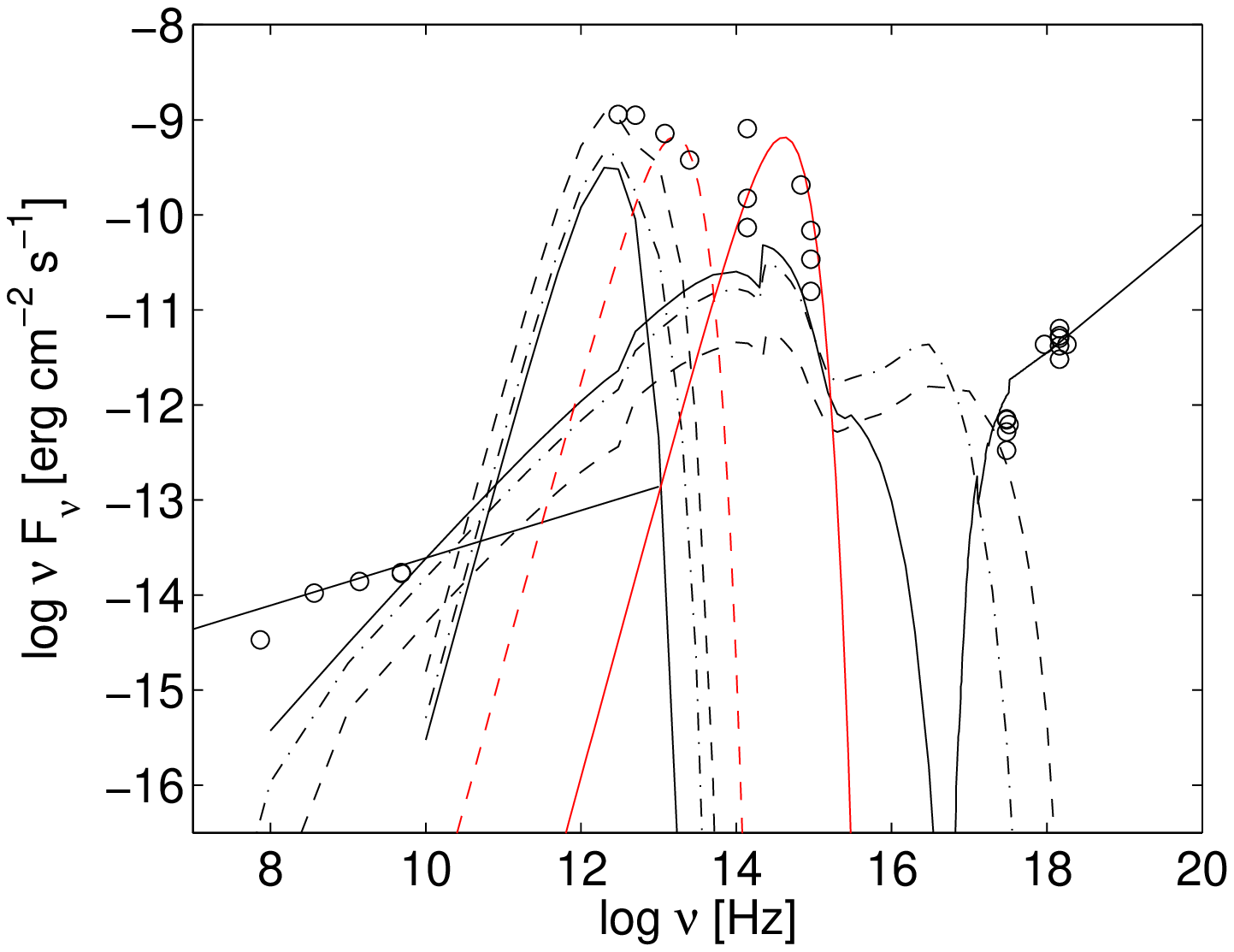}
\caption{The continuum SED of IC\,5063 (top) and NGC\,7212 (bottom).
Black circles: observed data described in the NED.
Black solid line in the X-ray domain represents the flux from the AGN.
Black lines at lower frequencies represent bremsstrahlung and dust emission from clouds at $V_s=100\,\si{\kms}$, $n_0=200\,\si{cm^{-3}}$, $F=6 \cdot 10^9$ photons $\si{cm^{-2} s^{-1} eV^{-1}}$ at the Lyman limit, $ D=10^{17}\,\si{cm}$ (solid), at $V_s=300\,\si{\kms}$, $n_0=300 \,\si{cm^{-3}}$, $F= 10^{10}$ units (dash-dotted) , $ D=3\cdot10^{17}\,\si{ cm}$ and at $V_s= 600\,\si{\kms}$,  $n_0=200\,\si{cm^{-3}}$, $F=2\cdot10^{10}$  units (dashed), $ D=2\cdot10^{16}\,\si{ cm}$. 
For all models O/H $=6.6\cdot10^{-4}$, N/H $= 5\cdot10^{-5}$ and Ne/H $= 10^{-4}$.
The power-law black solid line in the radio represents synchrotron radiation.
Solid red line corresponds to a black body for $T=5000\,\si{K}$, red dashed for $T=200\,\si{K}$.}
\label{fig:SED}
\end{figure}

To cross-check the results obtained by the detailed modelling of the line spectra, we have gathered from the NED the data corresponding to the continuum spectral energy distribution (SED) of IC\,5063 and NGC\,7212. 
Fig.\,\ref{fig:SED} shows the SED obtained from these data. 
Error bars are not shown for sake of clarity at frequencies $< 10^{17}\,\si{Hz}$.
We have selected some models which, on average, best reproduce the line ratios at different   $V_s$ ($100$, $300$ and $600 \,\si{\kms}$) and we have compared them with the data in Fig.\,\ref{fig:SED}.
At $\nu>10^{17}\,\si{Hz}$ the data are fitted by the power-law flux from the AGN.
The flux is reprocessed by gas and dust within the clouds and it is emitted as free-free and free-bound (and line) at lower frequencies. 
The reprocessed radiation by dust grains appears in the IR.
In the specific case of IC\,5063 the model corresponding to low $V_s$ and low $F$ reproduces the data lower limit in the UV-optical range.
Most of the data are nested inside the black body radiation flux corresponding to a temperature of $\rm T=5000\,\si{K}$, which represents the background contribution of relatively old stars as, e.g. red giants and Mira. 
The near IR side of the IR bump is fitted by the black body re-radiation flux corresponding to $ T=200\,\si{K}$. 
It represents emission from a large amount of warm dust produced in the stellar wind of red giant stars. 
Dust is heated collisionally and radiatively in the clouds.
The grains are destroyed by sputtering at high   $V_s$ ($> 200 \,\si{\kms}$) throughout the shock front.
Dust is heated to a maximum $T=66\,\si{K}$ in the $V_s =300 \,\si{\kms}$ cloud and to a maximum of $90\,\si{K}$ in the $V_s=600 \,\si{\kms}$ cloud, so the reprocessed radiation peaks at lower $\nu$. 
In the radio ranges, the data at $\nu\geq 10^{10}\,\si{Hz}$ are fitted by the bremsstrahlung and reradiation by dust, while the data at lower $\nu$ follow a power-law flux with spectral index $=0.75$. 
It represents the synchrotron radiation flux created by the Fermi mechanism at the shock front.

Interestingly, the same models are used to reproduce the continuum SED of  both galaxies.
In the far radio range of NGC\,7212 self absorption of the flux is evident. 
Unfortunately the data for IC\,5063 are lacking.

\section{Concluding remarks}
\label{sec:conclusions}

We studied the NLR/ENLR gas of two nearby Seyfert 2 galaxies: IC\,5063 and NGC\,7212.
We analysed high resolution spectra to highlight the different kinematic components of the emission lines and to study the properties of the gas as a function of velocity.
We produced diagnostic diagrams, we studied the ionization parameter and the physical conditions
of the gas resulting from the line ratios and we compared the observations to detailed models of the spectra, obtaining the following results: 

\begin{enumerate}
\item The diagnostic diagrams show that the main ionization mechanism of the gas is photo-ionization from the power-law continuum produced by the AGN.
However, high velocity gas seems to lie closer to the LINER/shock region of the diagrams than low velocity gas. 
This could suggest that there might be some contribution of shocks in the ionization of high velocity gas.

\item In the CN, N0 and N1 regions of IC\,5063 the ionization parameter decreases of about one order of magnitude between $V=-600\,\si{\kms}$ and $V=500\,\si{kms}$, which means that the blueshifted gas is irradiated directly by the AGN, while the redshifted gas is ionized by an attenuated continuum.
In the S0, S1 and S2 regions $U$ increases in the opposite direction, from $V=-400$ to $V=200\,\si{\kms}$.
The velocity range is quite narrow because at higher $\lvert V \rvert$ the effects of the extinction correction dominates with respect to the real behaviour of the ionization parameter.
The ionization increase with $V$ is clearly confirmed by the $\log([\ion{O}{III}]/\Hb)$ vs $V$ plot.
This behaviour of the ionization parameter might be explained assuming a hollow bi-conical shape of the ENLR with one of the edges aligned with the galaxy disk.

\item NGC\,7212 shows an ionization parameter which does not depend on velocity.
Therefore, it is not possible to say anything about the real geometrical shape of the ENLR.

\item The electron temperature and density are obtained from measured line ratios where possible (Table\,\ref{tab:ted_ic} and \ref{tab:ted_n}).
A value of the density was also calculated for each bin by the detailed modelling of the spectra, which accounts also for the shocks (Table\,\ref{tab:sim_s0-} and on-line material).
The two results are typically in agreement and they are consistent with properties of photo-ionized gas.

\item The SUMA composite models results show that the O/H relative abundances are close to solar \citep{Allen76}.

\item Analizing the SED we noticed that, although the multiwavelength dataset is not complete, the power-law flux in the radio range created by the Fermi mechanism at the shock-front, the bremsstrahlung emitted by the gas downstream and the black-body fluxes corresponding to dust reprocessed radiation and to the old star population.
Therefore, the Fermi mechanism and the bremsstrahlung radiation confirm the presence of shocks in both galaxies.

\item The analysis of the line profiles shows that the kinematics of both galaxies is quite complex.
The profiles change significantly from region to region. 
In the nucleus of the galaxies they are often relatively broad and characterized by multiple peaks and bumps. 
In the external regions they become narrower but they usually shows a red or blue wing.
NGC\,7212 lines are broader and more disturbed than IC\,5063 lines, but they are characterized by less prominent wings.
Those are all signs of gas in turbulent regime.

\item The high fragmentation of the clouds derived by SUMA is an index of interaction between jets and ISM and it might also explain the complex gas kinematics.

\item The high temperatures that seem to characterize the low ionization gas of NGC\,7212 might be explained by the jet-ISM interaction \citep{Roche16}.

\item The main peak of the lines outside the nucleus is shifted towards longer or shorter wavelengths, depending on the observed region.
It is a sign of consistent bulk motions of gas with respect to the galaxy stellar component.

\item The profile of the lines within each region does not change, with some exception due to variations in the ionization degree of the gas or difficulties in recovering the original shape of the lines during the deblending process.
\end{enumerate}

Finally, we confirmed that this kind of analysis of the line profiles can be a powerful tool to investigate the properties of gas in such complex conditions.
It can show  gas properties that a standard analysis would miss, for example the peculiar behaviour of the ionization profile in IC\,5063 and the shift of the points corresponding to high $\lvert V\rvert$ in the diagnostic diagrams, which can be associated to the presence of shocks.
In particular, the latter cannot be easily observed, when the diagnostic diagrams are produced with the whole line flux, because the contribution of the high $\lvert V\rvert$ gas is negligible with respect to the whole line flux.

\section*{Acknowledgements}

The authors would like to thank prof. Raffaella Morganti for her useful comments, and the referee, Dr. Andrew Humphrey, which helped to increase the quality of the paper with his review.
This research has made use of the NASA/IPAC Extragalactic Database (NED) which is operated by the Jet Propulsion Laboratory, California Institute of Technology, under contract with the National Aeronautics and Space Administration. 
This paper includes data gathered with the $6.5$-m Magellan Telescopes located at Las Campanas Observatory, Chile.
The STARLIGHT project is supported by the Brazilian agencies CNPq, CAPES and FAPESP and by the France--Brazi CAPES/Cofecub program. Based on observations made with the NASA/ESA Hubble Space Telescope, and obtained from the Hubble Legacy Archive, which is a collaboration between the Space Telescope Science Institute (STScI/NASA), the Space Telescope European Coordinating Facility (ST-ECF/ESA) and the Canadian Astronomy Data Centre (CADC/NRC/CSA).


\bibliographystyle{mnras}
\bibliography{bibliografia_def} 

\begin{thebibliography}{}
\makeatletter
\relax
\def\mn@urlcharsother{\let\do\@makeother \do\$\do\&\do\#\do\^\do\_\do\%\do\~}
\def\mn@doi{\begingroup\mn@urlcharsother \@ifnextchar [ {\mn@doi@}
  {\mn@doi@[]}}
\def\mn@doi@[#1]#2{\def\@tempa{#1}\ifx\@tempa\@empty \href
  {http://dx.doi.org/#2} {doi:#2}\else \href {http://dx.doi.org/#2} {#1}\fi
  \endgroup}
\def\mn@eprint#1#2{\mn@eprint@#1:#2::\@nil}
\def\mn@eprint@arXiv#1{\href {http://arxiv.org/abs/#1} {{\tt arXiv:#1}}}
\def\mn@eprint@dblp#1{\href {http://dblp.uni-trier.de/rec/bibtex/#1.xml}
  {dblp:#1}}
\def\mn@eprint@#1:#2:#3:#4\@nil{\def\@tempa {#1}\def\@tempb {#2}\def\@tempc
  {#3}\ifx \@tempc \@empty \let \@tempc \@tempb \let \@tempb \@tempa \fi \ifx
  \@tempb \@empty \def\@tempb {arXiv}\fi \@ifundefined
  {mn@eprint@\@tempb}{\@tempb:\@tempc}{\expandafter \expandafter \csname
  mn@eprint@\@tempb\endcsname \expandafter{\@tempc}}}

\bibitem[\protect\citeauthoryear{{Aldrovandi} \& {Contini}}{{Aldrovandi} \&
  {Contini}}{1984}]{Aldrovandi84}
{Aldrovandi} S.~M.~V.,  {Contini} M.,  1984, \aap, \href
  {http://adsabs.harvard.edu/abs/1984A%26A...140..368A} {140, 368}

\bibitem[\protect\citeauthoryear{{Allen}}{{Allen}}{1976}]{Allen76}
{Allen} C.~W.,  1976, {Astrophysical Quantities}

\bibitem[\protect\citeauthoryear{{Anders} \& {Grevesse}}{{Anders} \&
  {Grevesse}}{1989}]{Anders89}
{Anders} E.,  {Grevesse} N.,  1989, \mn@doi [\gca]
  {10.1016/0016-7037(89)90286-X}, \href
  {http://adsabs.harvard.edu/abs/1989GeCoA..53..197A} {53, 197}

\bibitem[\protect\citeauthoryear{{Antonucci}}{{Antonucci}}{1993}]{Antonucci93}
{Antonucci} R.,  1993, \mn@doi [ARA\&A] {10.1146/annurev.aa.31.090193.002353},
  \href {http://adsabs.harvard.edu/abs/1993ARA%26A..31..473A} {31, 473}

\bibitem[\protect\citeauthoryear{{Antonucci} \& {Miller}}{{Antonucci} \&
  {Miller}}{1985}]{Antonucci85}
{Antonucci} R.~R.~J.,  {Miller} J.~S.,  1985, \mn@doi [\apj] {10.1086/163559},
  \href {http://adsabs.harvard.edu/abs/1985ApJ...297..621A} {297, 621}

\bibitem[\protect\citeauthoryear{{Asplund}, {Grevesse}, {Sauval}  \&
  {Scott}}{{Asplund} et~al.}{2009}]{Asplund09}
{Asplund} M.,  {Grevesse} N.,  {Sauval} A.~J.,   {Scott} P.,  2009, \mn@doi
  [\araa] {10.1146/annurev.astro.46.060407.145222}, \href
  {http://adsabs.harvard.edu/abs/2009ARA%26A..47..481A} {47, 481}

\bibitem[\protect\citeauthoryear{{Baldwin}, {Phillips}  \&
  {Terlevich}}{{Baldwin} et~al.}{1981}]{Baldwin81}
{Baldwin} J.~A.,  {Phillips} M.~M.,   {Terlevich} R.,  1981, \mn@doi [\pasp]
  {10.1086/130766}, \href {http://adsabs.harvard.edu/abs/1981PASP...93....5B}
  {93, 5}

\bibitem[\protect\citeauthoryear{{Baldwin}, {Wilson}  \& {Whittle}}{{Baldwin}
  et~al.}{1987}]{Baldwin87}
{Baldwin} J.~A.,  {Wilson} A.~S.,   {Whittle} M.,  1987, \mn@doi [\apj]
  {10.1086/165435}, \href {http://adsabs.harvard.edu/abs/1987ApJ...319...84B}
  {319, 84}

\bibitem[\protect\citeauthoryear{{Beck}}{{Beck}}{2012}]{Beck12}
{Beck} R.,  2012, in Journal of Physics Conference Series. p. 012051
  (\mn@eprint {arXiv} {1112.1823}), \mn@doi{10.1088/1742-6596/372/1/012051}

\bibitem[\protect\citeauthoryear{{Beckmann} \& {Shrader}}{{Beckmann} \&
  {Shrader}}{2012}]{Beckmann12}
{Beckmann} V.,  {Shrader} C.~R.,  2012, {Active Galactic Nuclei}

\bibitem[\protect\citeauthoryear{{Cardelli}, {Clayton}  \& {Mathis}}{{Cardelli}
  et~al.}{1989}]{Cardelli89}
{Cardelli} J.~A.,  {Clayton} G.~C.,   {Mathis} J.~S.,  1989, \mn@doi [\apj]
  {10.1086/167900}, \href {http://adsabs.harvard.edu/abs/1989ApJ...345..245C}
  {345, 245}

\bibitem[\protect\citeauthoryear{{Cecil}, {Dopita}, {Groves}, {Wilson},
  {Ferruit}, {P{\'e}contal}  \& {Binette}}{{Cecil} et~al.}{2002}]{Cecil02}
{Cecil} G.,  {Dopita} M.~A.,  {Groves} B.,  {Wilson} A.~S.,  {Ferruit} P.,
  {P{\'e}contal} E.,   {Binette} L.,  2002, \mn@doi [\apj] {10.1086/338950},
  \href {http://adsabs.harvard.edu/abs/2002ApJ...568..627C} {568, 627}

\bibitem[\protect\citeauthoryear{{Cid Fernandes}, {Mateus}, {Sodr{\'e}},
  {Stasi{\'n}ska}  \& {Gomes}}{{Cid Fernandes} et~al.}{2005}]{Fernandes05}
{Cid Fernandes} R.,  {Mateus} A.,  {Sodr{\'e}} L.,  {Stasi{\'n}ska} G.,
  {Gomes} J.~M.,  2005, \mn@doi [\mnras] {10.1111/j.1365-2966.2005.08752.x},
  \href {http://adsabs.harvard.edu/abs/2005MNRAS.358..363C} {358, 363}

\bibitem[\protect\citeauthoryear{{Cid Fernandes}, {Asari}, {Sodr{\'e}},
  {Stasi{\'n}ska}, {Mateus}, {Torres-Papaqui}  \& {Schoenell}}{{Cid Fernandes}
  et~al.}{2007}]{Fernandes07}
{Cid Fernandes} R.,  {Asari} N.~V.,  {Sodr{\'e}} L.,  {Stasi{\'n}ska} G.,
  {Mateus} A.,  {Torres-Papaqui} J.~P.,   {Schoenell} W.,  2007, \mn@doi
  [\mnras] {10.1111/j.1745-3933.2006.00265.x}, \href
  {http://adsabs.harvard.edu/abs/2007MNRAS.375L..16C} {375, L16}

\bibitem[\protect\citeauthoryear{{Ciroi}, {Afanasiev}, {Moiseev}, {Botte}, {Di
  Mille}, {Dodonov}, {Rafanelli}  \& {Smirnova}}{{Ciroi}
  et~al.}{2005}]{Ciroi05}
{Ciroi} S.,  {Afanasiev} V.~L.,  {Moiseev} A.~V.,  {Botte} V.,  {Di Mille} F.,
  {Dodonov} S.~N.,  {Rafanelli} P.,   {Smirnova} A.~A.,  2005, \mn@doi [\mnras]
  {10.1111/j.1365-2966.2005.09031.x}, \href
  {http://adsabs.harvard.edu/abs/2005MNRAS.360..253C} {360, 253}

\bibitem[\protect\citeauthoryear{{Cisternas} et~al.,}{{Cisternas}
  et~al.}{2011}]{Cisternas11}
{Cisternas} M.,  et~al., 2011, \mn@doi [\apj] {10.1088/0004-637X/726/2/57},
  \href {http://adsabs.harvard.edu/abs/2011ApJ...726...57C} {726, 57}

\bibitem[\protect\citeauthoryear{{Colina}, {Sparks}  \& {Macchetto}}{{Colina}
  et~al.}{1991}]{Colina91}
{Colina} L.,  {Sparks} W.~B.,   {Macchetto} F.,  1991, \mn@doi [\apj]
  {10.1086/169795}, \href {http://adsabs.harvard.edu/abs/1991ApJ...370..102C}
  {370, 102}

\bibitem[\protect\citeauthoryear{{Contini}}{{Contini}}{2015}]{Contini15}
{Contini} M.,  2015, \mn@doi [\mnras] {10.1093/mnras/stv1451}, \href
  {http://adsabs.harvard.edu/abs/2015MNRAS.452.3795C} {452, 3795}

\bibitem[\protect\citeauthoryear{{Contini}}{{Contini}}{2017}]{Contini17}
{Contini} M.,  2017, preprint, \href
  {http://adsabs.harvard.edu/abs/2017arXiv170407604C} {} (\mn@eprint {arXiv}
  {1704.07604})

\bibitem[\protect\citeauthoryear{{Contini} \& {Aldrovandi}}{{Contini} \&
  {Aldrovandi}}{1983}]{Contini83}
{Contini} M.,  {Aldrovandi} S.~M.~V.,  1983, \aap, \href
  {http://adsabs.harvard.edu/abs/1983A%26A...127...15C} {127, 15}

\bibitem[\protect\citeauthoryear{{Contini}, {Cracco}, {Ciroi}  \& {La
  Mura}}{{Contini} et~al.}{2012}]{Contini12}
{Contini} M.,  {Cracco} V.,  {Ciroi} S.,   {La Mura} G.,  2012, \mn@doi [\aap]
  {10.1051/0004-6361/201219184}, \href
  {http://adsabs.harvard.edu/abs/2012A%26A...545A..72C} {545, A72}

\bibitem[\protect\citeauthoryear{{Cox}}{{Cox}}{1972}]{Cox72}
{Cox} D.~P.,  1972, \mn@doi [\apj] {10.1086/151774}, \href
  {http://adsabs.harvard.edu/abs/1972ApJ...178..143C} {178, 143}

\bibitem[\protect\citeauthoryear{{Cracco} et~al.,}{{Cracco}
  et~al.}{2011}]{Cracco11}
{Cracco} V.,  et~al., 2011, \mn@doi [\mnras]
  {10.1111/j.1365-2966.2011.19654.x}, \href
  {http://adsabs.harvard.edu/abs/2011MNRAS.418.2630C} {418, 2630}

\bibitem[\protect\citeauthoryear{{Crenshaw} et~al.,}{{Crenshaw}
  et~al.}{2002}]{Crenshaw02}
{Crenshaw} D.~M.,  et~al., 2002, \mn@doi [\apj] {10.1086/338058}, \href
  {http://adsabs.harvard.edu/abs/2002ApJ...566..187C} {566, 187}

\bibitem[\protect\citeauthoryear{{Das}, {Crenshaw}, {Kraemer}  \& {Deo}}{{Das}
  et~al.}{2006}]{Das06}
{Das} V.,  {Crenshaw} D.~M.,  {Kraemer} S.~B.,   {Deo} R.~P.,  2006, \mn@doi
  [\aj] {10.1086/504899}, \href
  {http://adsabs.harvard.edu/abs/2006AJ....132..620D} {132, 620}

\bibitem[\protect\citeauthoryear{{Dasyra}, {Bostrom}, {Combes}  \&
  {Vlahakis}}{{Dasyra} et~al.}{2015}]{Dasyra15}
{Dasyra} K.~M.,  {Bostrom} A.~C.,  {Combes} F.,   {Vlahakis} N.,  2015, \mn@doi
  [\apj] {10.1088/0004-637X/815/1/34}, \href
  {http://adsabs.harvard.edu/abs/2015ApJ...815...34D} {815, 34}

\bibitem[\protect\citeauthoryear{{Di Mille}}{{Di Mille}}{2007}]{DiMille07}
{Di Mille} F.,  2007, PhD thesis, {Univ. Padova}

\bibitem[\protect\citeauthoryear{{Dietrich} \& {Wagner}}{{Dietrich} \&
  {Wagner}}{1998}]{Dietrich98}
{Dietrich} M.,  {Wagner} S.~J.,  1998, \aap, \href
  {http://adsabs.harvard.edu/abs/1998A%26A...338..405D} {338, 405}

\bibitem[\protect\citeauthoryear{{Evans}, {Tsvetanov}, {Kriss}, {Ford},
  {Caganoff}  \& {Koratkar}}{{Evans} et~al.}{1993}]{Evans93}
{Evans} I.~N.,  {Tsvetanov} Z.,  {Kriss} G.~A.,  {Ford} H.~C.,  {Caganoff} S.,
   {Koratkar} A.~P.,  1993, \mn@doi [\apj] {10.1086/173292}, \href
  {http://adsabs.harvard.edu/abs/1993ApJ...417...82E} {417, 82}

\bibitem[\protect\citeauthoryear{{Falcke}, {Wilson}  \& {Simpson}}{{Falcke}
  et~al.}{1998}]{Falcke98}
{Falcke} H.,  {Wilson} A.~S.,   {Simpson} C.,  1998, \mn@doi [\apj]
  {10.1086/305886}, \href {http://adsabs.harvard.edu/abs/1998ApJ...502..199F}
  {502, 199}

\bibitem[\protect\citeauthoryear{{Gallimore}, {Axon}, {O'Dea}, {Baum}  \&
  {Pedlar}}{{Gallimore} et~al.}{2006}]{Gallimore06}
{Gallimore} J.~F.,  {Axon} D.~J.,  {O'Dea} C.~P.,  {Baum} S.~A.,   {Pedlar} A.,
   2006, \mn@doi [\aj] {10.1086/504593}, \href
  {http://adsabs.harvard.edu/abs/2006AJ....132..546G} {132, 546}

\bibitem[\protect\citeauthoryear{{Grevesse} \& {Sauval}}{{Grevesse} \&
  {Sauval}}{1998}]{Grevesse98}
{Grevesse} N.,  {Sauval} A.~J.,  1998, \mn@doi [\ssr]
  {10.1023/A:1005161325181}, \href
  {http://adsabs.harvard.edu/abs/1998SSRv...85..161G} {85, 161}

\bibitem[\protect\citeauthoryear{{Hough}, {Brindle}, {Axon}, {Bailey}  \&
  {Sparks}}{{Hough} et~al.}{1987}]{Hough87}
{Hough} J.~H.,  {Brindle} C.,  {Axon} D.~J.,  {Bailey} J.,   {Sparks} W.~B.,
  1987, \mnras, \href {http://adsabs.harvard.edu/abs/1987MNRAS.224.1013H} {224,
  1013}

\bibitem[\protect\citeauthoryear{{Inglis}, {Brindle}, {Hough}, {Young}, {Axon},
  {Bailey}  \& {Ward}}{{Inglis} et~al.}{1993}]{Inglis93}
{Inglis} M.~D.,  {Brindle} C.,  {Hough} J.~H.,  {Young} S.,  {Axon} D.~J.,
  {Bailey} J.~A.,   {Ward} M.~J.,  1993, \mnras, \href
  {http://adsabs.harvard.edu/abs/1993MNRAS.263..895I} {263, 895}

\bibitem[\protect\citeauthoryear{{Kewley}, {Groves}, {Kauffmann}  \&
  {Heckman}}{{Kewley} et~al.}{2006}]{Kewley06}
{Kewley} L.~J.,  {Groves} B.,  {Kauffmann} G.,   {Heckman} T.,  2006, \mn@doi
  [\mnras] {10.1111/j.1365-2966.2006.10859.x}, \href
  {http://adsabs.harvard.edu/abs/2006MNRAS.372..961K} {372, 961}

\bibitem[\protect\citeauthoryear{{Komatsu} et~al.,}{{Komatsu}
  et~al.}{2011}]{Komatsu11}
{Komatsu} E.,  et~al., 2011, \mn@doi [\apjs] {10.1088/0067-0049/192/2/18},
  \href {http://adsabs.harvard.edu/abs/2011ApJS..192...18K} {192, 18}

\bibitem[\protect\citeauthoryear{{Kraemer} \& {Crenshaw}}{{Kraemer} \&
  {Crenshaw}}{2000}]{Kraemer00}
{Kraemer} S.~B.,  {Crenshaw} D.~M.,  2000, \mn@doi [\apj] {10.1086/317246},
  \href {http://adsabs.harvard.edu/abs/2000ApJ...544..763K} {544, 763}

\bibitem[\protect\citeauthoryear{{Marshall} et~al.,}{{Marshall}
  et~al.}{2008}]{Marshall08}
{Marshall} J.~L.,  et~al., 2008, in Society of Photo-Optical Instrumentation
  Engineers (SPIE) Conference Series. p.~54 (\mn@eprint {arXiv} {0807.3774}),
  \mn@doi{10.1117/12.789972}

\bibitem[\protect\citeauthoryear{{Mateus}, {Sodr{\'e}}, {Cid Fernandes},
  {Stasi{\'n}ska}, {Schoenell}  \& {Gomes}}{{Mateus} et~al.}{2006}]{Mateus06}
{Mateus} A.,  {Sodr{\'e}} L.,  {Cid Fernandes} R.,  {Stasi{\'n}ska} G.,
  {Schoenell} W.,   {Gomes} J.~M.,  2006, \mn@doi [\mnras]
  {10.1111/j.1365-2966.2006.10565.x}, \href
  {http://adsabs.harvard.edu/abs/2006MNRAS.370..721M} {370, 721}

\bibitem[\protect\citeauthoryear{{Morganti}, {Oosterloo}  \&
  {Tsvetanov}}{{Morganti} et~al.}{1998}]{Morganti98}
{Morganti} R.,  {Oosterloo} T.,   {Tsvetanov} Z.,  1998, \mn@doi [\aj]
  {10.1086/300236}, \href {http://adsabs.harvard.edu/abs/1998AJ....115..915M}
  {115, 915}

\bibitem[\protect\citeauthoryear{{Morganti}, {Holt}, {Saripalli}, {Oosterloo}
  \& {Tadhunter}}{{Morganti} et~al.}{2007}]{Morganti07}
{Morganti} R.,  {Holt} J.,  {Saripalli} L.,  {Oosterloo} T.~A.,   {Tadhunter}
  C.~N.,  2007, \mn@doi [\aap] {10.1051/0004-6361:20077888}, \href
  {http://adsabs.harvard.edu/abs/2007A%26A...476..735M} {476, 735}

\bibitem[\protect\citeauthoryear{{Morganti}, {Oosterloo}, {Oonk}, {Frieswijk}
  \& {Tadhunter}}{{Morganti} et~al.}{2015}]{Morganti15}
{Morganti} R.,  {Oosterloo} T.,  {Oonk} J.~B.~R.,  {Frieswijk} W.,
  {Tadhunter} C.,  2015, \mn@doi [\aap] {10.1051/0004-6361/201525860}, \href
  {http://adsabs.harvard.edu/abs/2015A%26A...580A...1M} {580, A1}

\bibitem[\protect\citeauthoryear{{Mu{\~n}oz Mar{\'{\i}}n}, {Gonz{\'a}lez
  Delgado}, {Schmitt}, {Cid Fernandes}, {P{\'e}rez}, {Storchi-Bergmann},
  {Heckman}  \& {Leitherer}}{{Mu{\~n}oz Mar{\'{\i}}n} et~al.}{2007}]{Munoz07}
{Mu{\~n}oz Mar{\'{\i}}n} V.~M.,  {Gonz{\'a}lez Delgado} R.~M.,  {Schmitt}
  H.~R.,  {Cid Fernandes} R.,  {P{\'e}rez} E.,  {Storchi-Bergmann} T.,
  {Heckman} T.,   {Leitherer} C.,  2007, \mn@doi [\aj] {10.1086/519448}, \href
  {http://adsabs.harvard.edu/abs/2007AJ....134..648M} {134, 648}

\bibitem[\protect\citeauthoryear{{Mulchaey}, {Wilson}  \&
  {Tsvetanov}}{{Mulchaey} et~al.}{1996}]{Mulchaey96}
{Mulchaey} J.~S.,  {Wilson} A.~S.,   {Tsvetanov} Z.,  1996, \mn@doi [\apj]
  {10.1086/177595}, \href {http://adsabs.harvard.edu/abs/1996ApJ...467..197M}
  {467, 197}

\bibitem[\protect\citeauthoryear{{Nagar}, {Wilson}, {Mulchaey}  \&
  {Gallimore}}{{Nagar} et~al.}{1999}]{Nagar99}
{Nagar} N.~M.,  {Wilson} A.~S.,  {Mulchaey} J.~S.,   {Gallimore} J.~F.,  1999,
  \mn@doi [\apjs] {10.1086/313183}, \href
  {http://adsabs.harvard.edu/abs/1999ApJS..120..209N} {120, 209}

\bibitem[\protect\citeauthoryear{{Netzer}}{{Netzer}}{2015}]{Netzer15}
{Netzer} H.,  2015, \mn@doi [\araa] {10.1146/annurev-astro-082214-122302},
  \href {http://adsabs.harvard.edu/abs/2015ARA%26A..53..365N} {53, 365}

\bibitem[\protect\citeauthoryear{{Osterbrock}}{{Osterbrock}}{1989}]{OsterbrockAGN89}
{Osterbrock} D.~E.,  1989, {Astrophysics of gaseous nebulae and active galactic
  nuclei}

\bibitem[\protect\citeauthoryear{{Osterbrock} \& {Ferland}}{{Osterbrock} \&
  {Ferland}}{2006}]{OsterbrockAGN}
{Osterbrock} D.~E.,  {Ferland} G.~J.,  2006, {Astrophysics of gaseous nebulae
  and active galactic nuclei}

\bibitem[\protect\citeauthoryear{{Ozaki}}{{Ozaki}}{2009}]{Ozaki09}
{Ozaki} S.,  2009, \mn@doi [\pasj] {10.1093/pasj/61.2.259}, \href
  {http://adsabs.harvard.edu/abs/2009PASJ...61..259O} {61, 259}

\bibitem[\protect\citeauthoryear{{Penston} et~al.,}{{Penston}
  et~al.}{1990}]{Penston90}
{Penston} M.~V.,  et~al., 1990, \aap, \href
  {http://adsabs.harvard.edu/abs/1990A\%26A...236...53P} {236, 53}

\bibitem[\protect\citeauthoryear{{Roche}, {Humphrey}, {Lagos}, {Papaderos},
  {Silva}, {Cardoso}  \& {Gomes}}{{Roche} et~al.}{2016}]{Roche16}
{Roche} N.,  {Humphrey} A.,  {Lagos} P.,  {Papaderos} P.,  {Silva} M.,
  {Cardoso} L.~S.~M.,   {Gomes} J.~M.,  2016, \mn@doi [\mnras]
  {10.1093/mnras/stw765}, \href
  {http://adsabs.harvard.edu/abs/2016MNRAS.459.4259R} {459, 4259}

\bibitem[\protect\citeauthoryear{{Rodr{\'{\i}}guez-Ardila}, {Contini}  \&
  {Viegas}}{{Rodr{\'{\i}}guez-Ardila} et~al.}{2005}]{Rodriguez05}
{Rodr{\'{\i}}guez-Ardila} A.,  {Contini} M.,   {Viegas} S.~M.,  2005, \mn@doi
  [\mnras] {10.1111/j.1365-2966.2005.08628.x}, \href
  {http://adsabs.harvard.edu/abs/2005MNRAS.357..220R} {357, 220}

\bibitem[\protect\citeauthoryear{{Schirmer}, {Diaz}, {Holhjem}, {Levenson}  \&
  {Winge}}{{Schirmer} et~al.}{2013}]{Schirmer13}
{Schirmer} M.,  {Diaz} R.,  {Holhjem} K.,  {Levenson} N.~A.,   {Winge} C.,
  2013, \mn@doi [\apj] {10.1088/0004-637X/763/1/60}, \href
  {http://adsabs.harvard.edu/abs/2013ApJ...763...60S} {763, 60}

\bibitem[\protect\citeauthoryear{{Schmitt}, {Donley}, {Antonucci}, {Hutchings},
  {Kinney}  \& {Pringle}}{{Schmitt} et~al.}{2003}]{Schmitt03}
{Schmitt} H.~R.,  {Donley} J.~L.,  {Antonucci} R.~R.~J.,  {Hutchings} J.~B.,
  {Kinney} A.~L.,   {Pringle} J.~E.,  2003, \mn@doi [\apj] {10.1086/381224},
  \href {http://adsabs.harvard.edu/abs/2003ApJ...597..768S} {597, 768}

\bibitem[\protect\citeauthoryear{{Singh}, {Ishwara-Chandra}, {Wadadekar},
  {Beelen}  \& {Kharb}}{{Singh} et~al.}{2015}]{Singh15}
{Singh} V.,  {Ishwara-Chandra} C.~H.,  {Wadadekar} Y.,  {Beelen} A.,   {Kharb}
  P.,  2015, \mn@doi [\mnras] {10.1093/mnras/stu2124}, \href
  {http://adsabs.harvard.edu/abs/2015MNRAS.446..599S} {446, 599}

\bibitem[\protect\citeauthoryear{{Tadhunter}, {Morganti}, {Rose}, {Oonk}  \&
  {Oosterloo}}{{Tadhunter} et~al.}{2014}]{Tadhunter14}
{Tadhunter} C.,  {Morganti} R.,  {Rose} M.,  {Oonk} J.~B.~R.,   {Oosterloo} T.,
   2014, \mn@doi [\nat] {10.1038/nature13520}, \href
  {http://adsabs.harvard.edu/abs/2014Natur.511..440T} {511, 440}

\bibitem[\protect\citeauthoryear{{Tran}}{{Tran}}{1995}]{Tran95}
{Tran} H.~D.,  1995, \mn@doi [\apj] {10.1086/175297}, \href
  {http://adsabs.harvard.edu/abs/1995ApJ...440..578T} {440, 578}

\bibitem[\protect\citeauthoryear{{Turner}, {Romano}, {George}, {Edelson},
  {Collier}, {Mathur}  \& {Peterson}}{{Turner} et~al.}{2001}]{Turner01}
{Turner} T.~J.,  {Romano} P.,  {George} I.~M.,  {Edelson} R.,  {Collier} S.~J.,
   {Mathur} S.,   {Peterson} B.~M.,  2001, \mn@doi [\apj] {10.1086/323232},
  \href {http://adsabs.harvard.edu/abs/2001ApJ...561..131T} {561, 131}

\bibitem[\protect\citeauthoryear{{Unger}, {Pedlar}, {Axon}, {Whittle}, {Meurs}
  \& {Ward}}{{Unger} et~al.}{1987}]{Unger87}
{Unger} S.~W.,  {Pedlar} A.,  {Axon} D.~J.,  {Whittle} M.,  {Meurs} E.~J.~A.,
  {Ward} M.~J.,  1987, \mnras, \href
  {http://adsabs.harvard.edu/abs/1987MNRAS.228..671U} {228, 671}

\bibitem[\protect\citeauthoryear{{Veilleux} \& {Osterbrock}}{{Veilleux} \&
  {Osterbrock}}{1987}]{Veilleux87}
{Veilleux} S.,  {Osterbrock} D.~E.,  1987, \mn@doi [\apjs] {10.1086/191166},
  \href {http://adsabs.harvard.edu/abs/1987ApJS...63..295V} {63, 295}

\bibitem[\protect\citeauthoryear{{Veilleux}, {Bland-Hawthorn}  \&
  {Cecil}}{{Veilleux} et~al.}{1999}]{Veilleux99}
{Veilleux} S.,  {Bland-Hawthorn} J.,   {Cecil} G.,  1999, \mn@doi [\aj]
  {10.1086/301095}, \href {http://adsabs.harvard.edu/abs/1999AJ....118.2108V}
  {118, 2108}

\bibitem[\protect\citeauthoryear{{Villar-Martin}, {Tadhunter}  \&
  {Clark}}{{Villar-Martin} et~al.}{1997}]{Villar97}
{Villar-Martin} M.,  {Tadhunter} C.,   {Clark} N.,  1997, \aap, \href
  {http://adsabs.harvard.edu/abs/1997A%26A...323...21V} {323, 21}

\bibitem[\protect\citeauthoryear{{Wasilewski}}{{Wasilewski}}{1981}]{Wasilewski81}
{Wasilewski} A.~J.,  1981, \mn@doi [\pasp] {10.1086/130887}, \href
  {http://adsabs.harvard.edu/abs/1981PASP...93..560W} {93, 560}

\bibitem[\protect\citeauthoryear{{Wilson} \& {Tsvetanov}}{{Wilson} \&
  {Tsvetanov}}{1994}]{Wilson94}
{Wilson} A.~S.,  {Tsvetanov} Z.~I.,  1994, \mn@doi [\aj] {10.1086/116935},
  \href {http://adsabs.harvard.edu/abs/1994AJ....107.1227W} {107, 1227}

\makeatother
\end{thebibliography}


\newpage
\appendix

\section{Lines profiles}
\label{sec:figures}

\subsection{IC 5063}

\begin{figure*}
\centering
\includegraphics[width=0.45\textwidth]{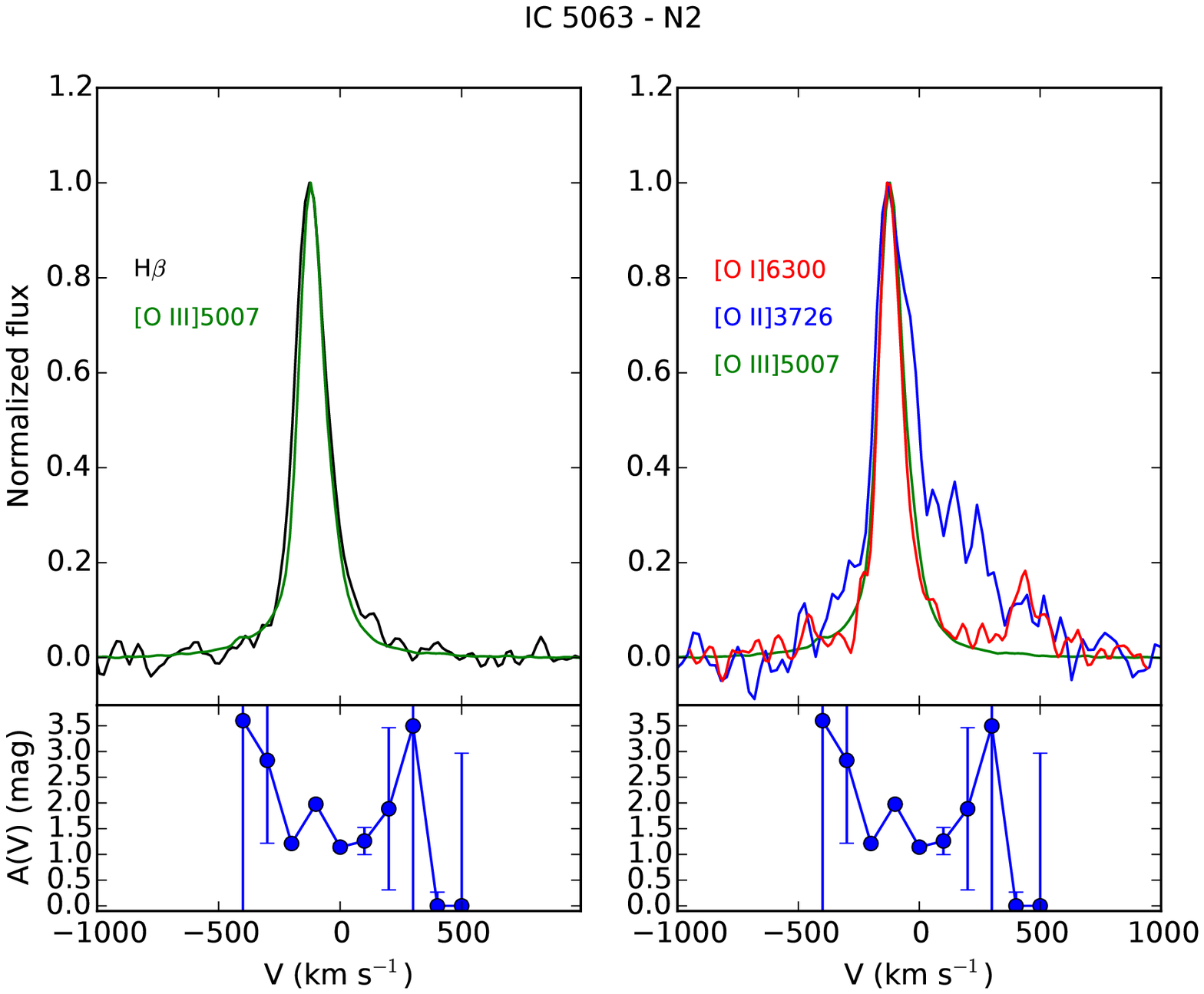} \quad
\includegraphics[width=0.45\textwidth]{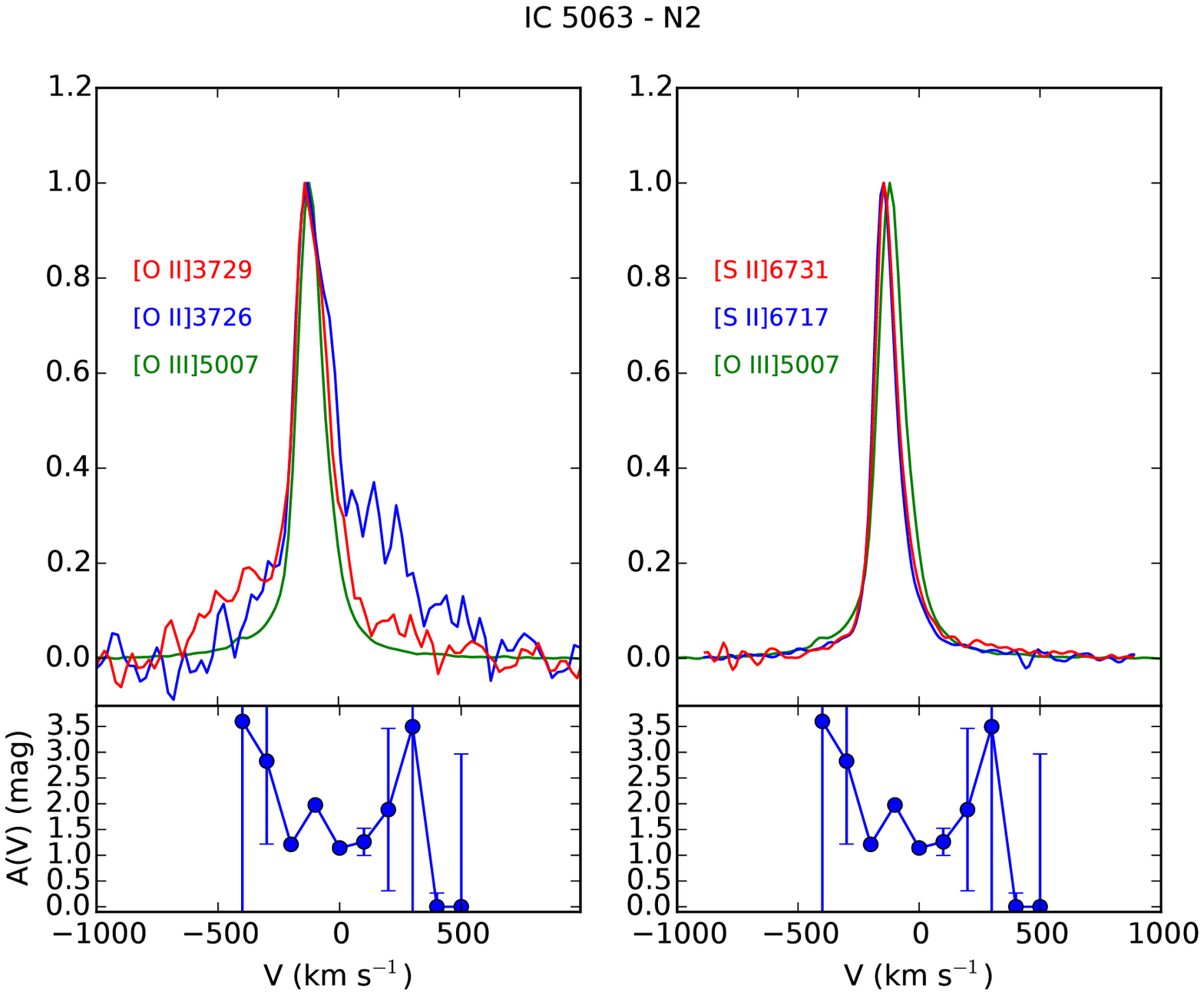}\\
\caption[]{Comparison of the emission lines of the N2 region of IC\,5063. The behaviour of the extinction coefficient as a function of velocity is shown under each plot.}
\label{fig:n2l1_I}
\end{figure*}

\begin{figure*}
\centering
\includegraphics[width=0.45\textwidth]{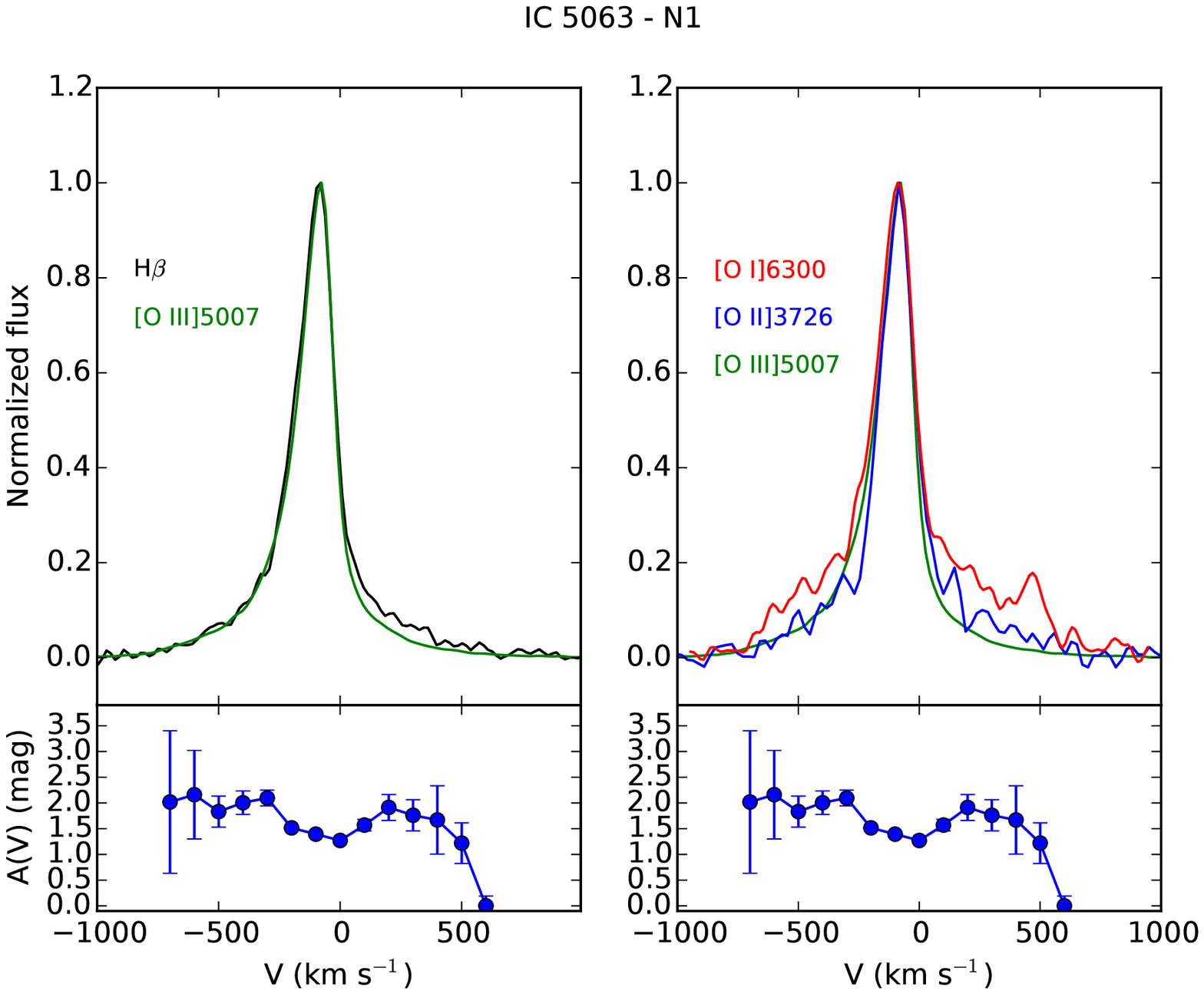} \quad
\includegraphics[width=0.45\textwidth]{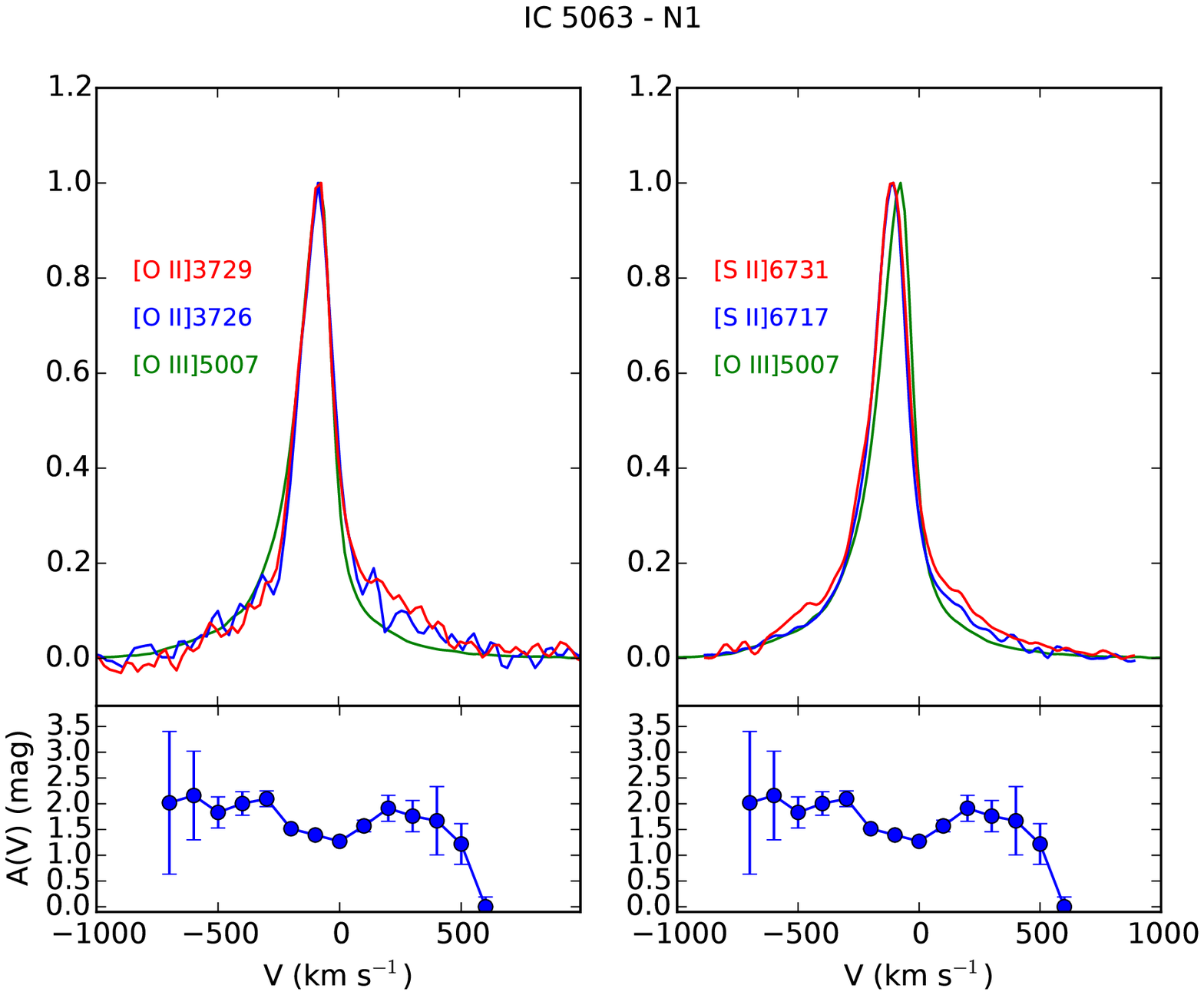}\\
\caption[]{Comparison of the emission lines of the N1 region of IC\,5063. The behaviour of the extinction coefficient as a function of velocity is shown under each plot.}
\label{fig:n1l1_I}
\end{figure*}

\begin{figure*}
\centering
\includegraphics[width=0.45\textwidth]{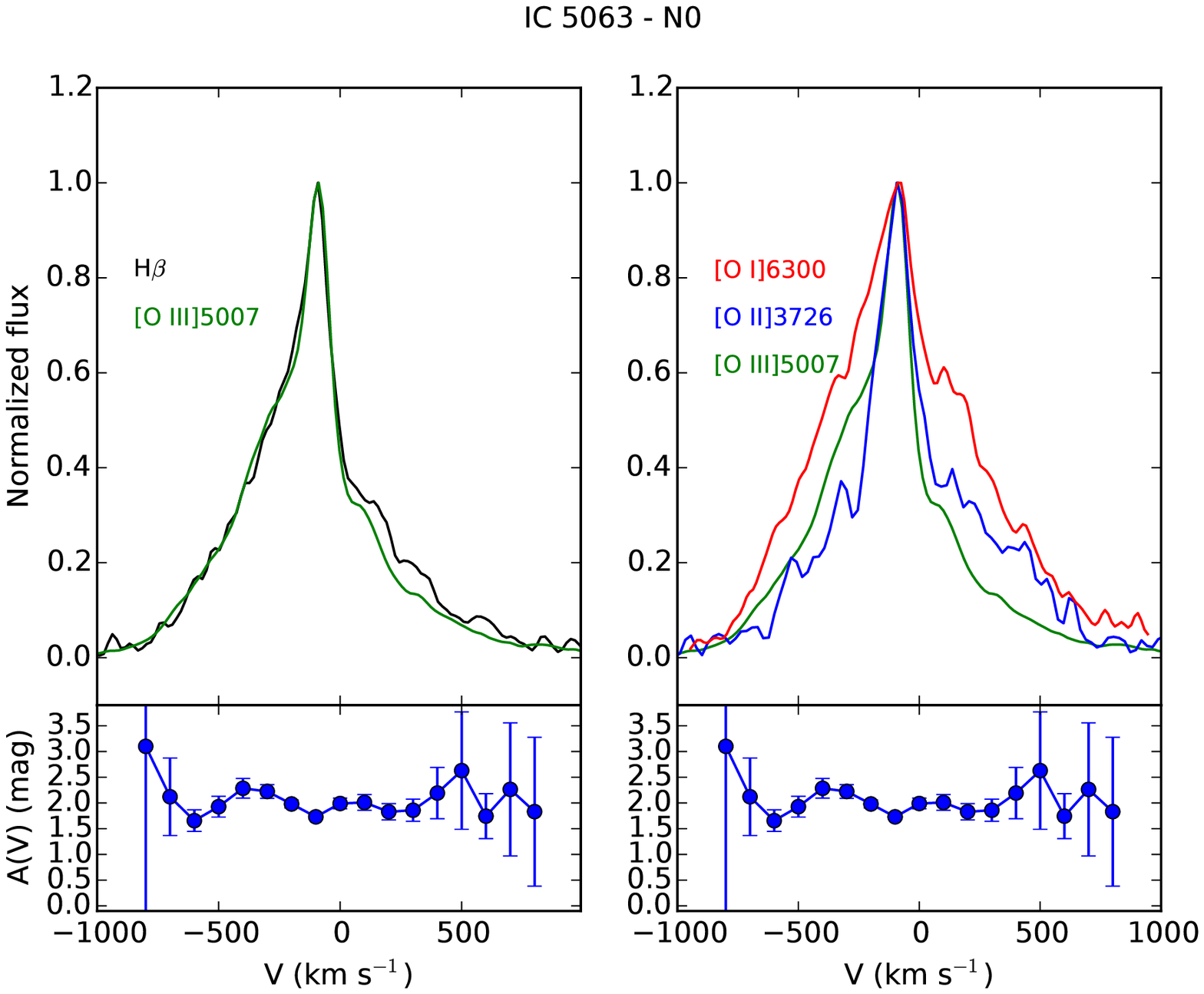} \quad
\includegraphics[width=0.45\textwidth]{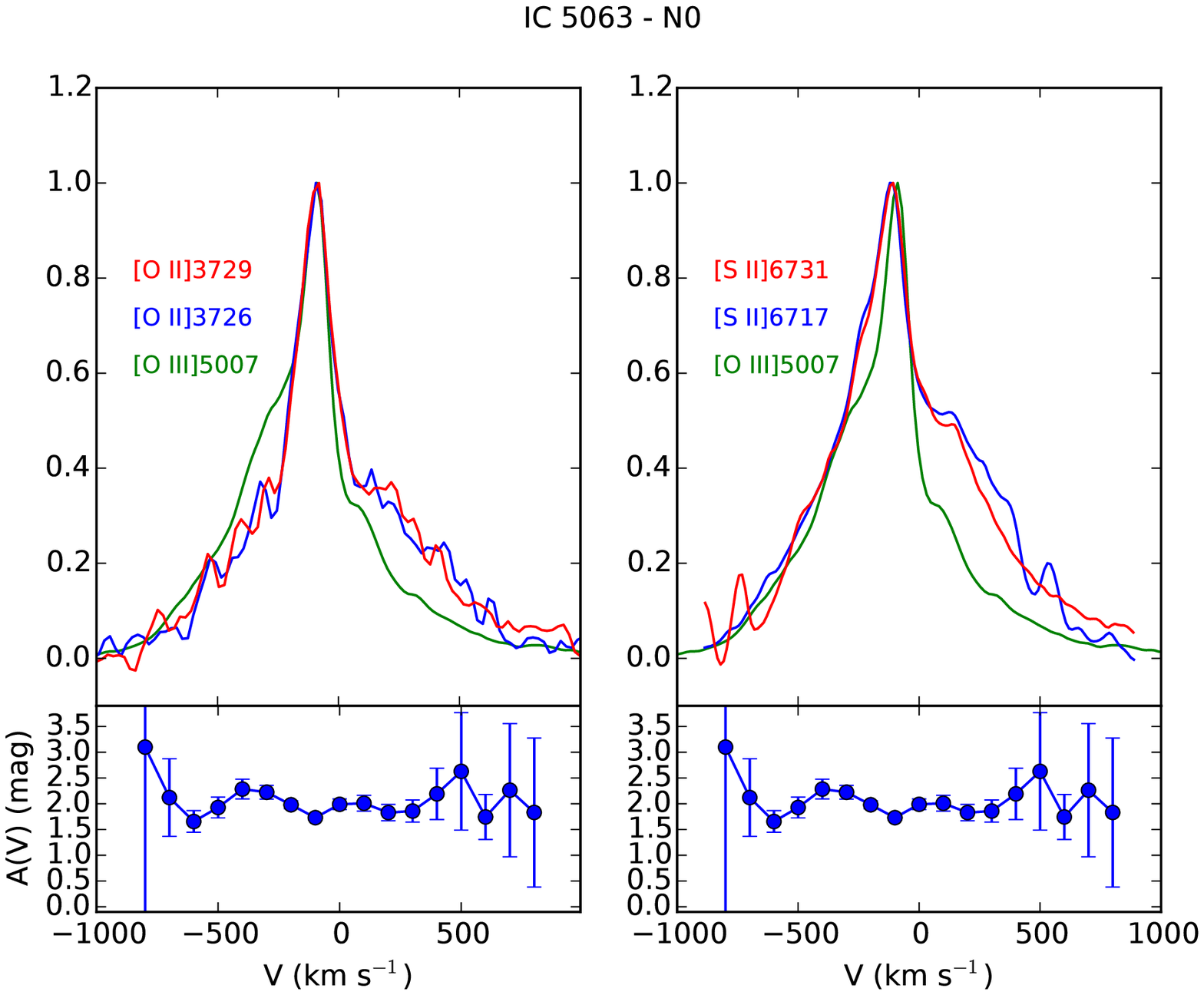}\\
\caption[]{Comparison of the emission lines of the N0 region of IC\,5063. The behaviour of the extinction coefficient as a function of velocity is shown under each plot.}
\label{fig:n0l1_I}
\end{figure*}

\begin{figure*}
\centering
\includegraphics[width=0.45\textwidth]{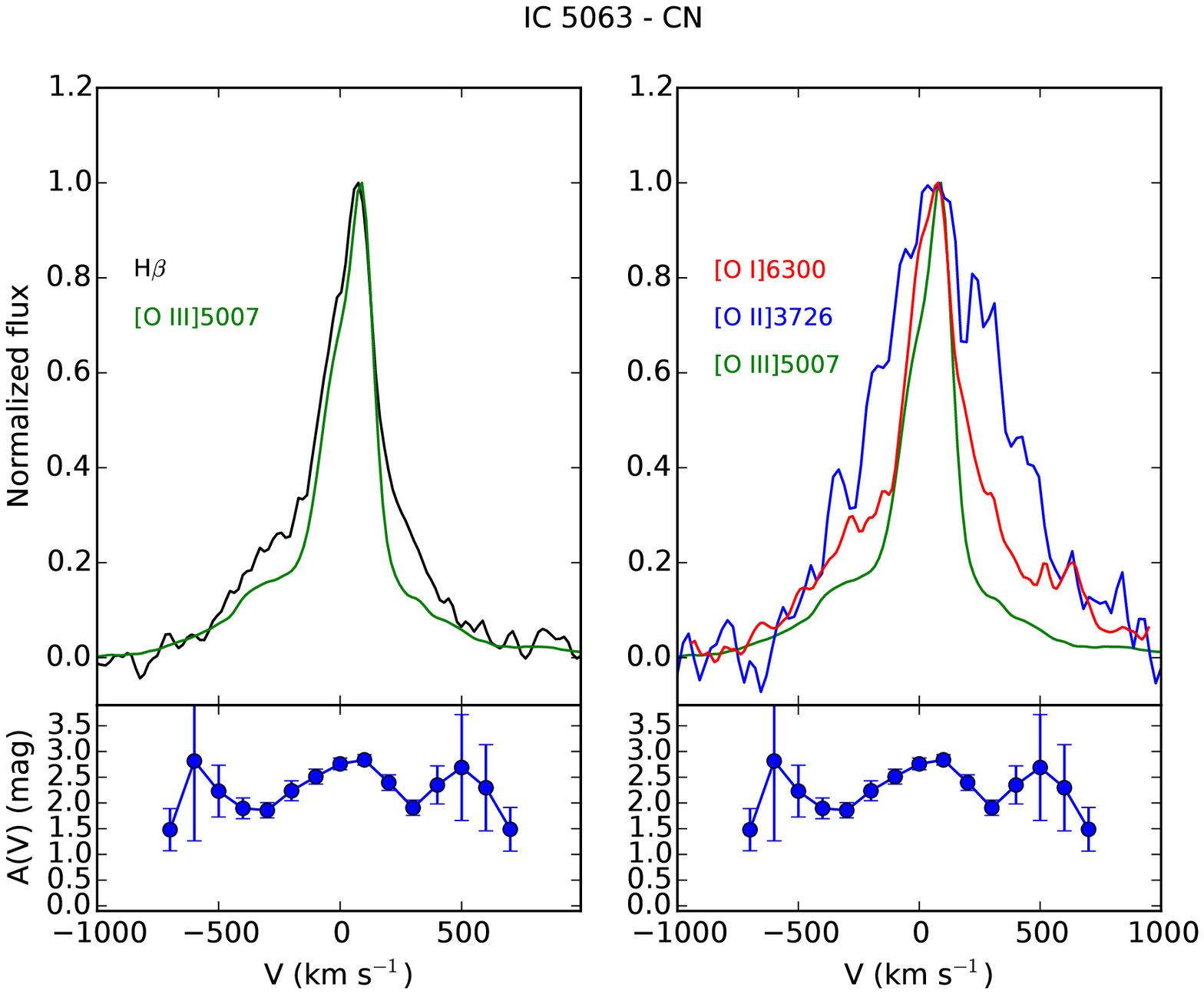} \quad
\includegraphics[width=0.45\textwidth]{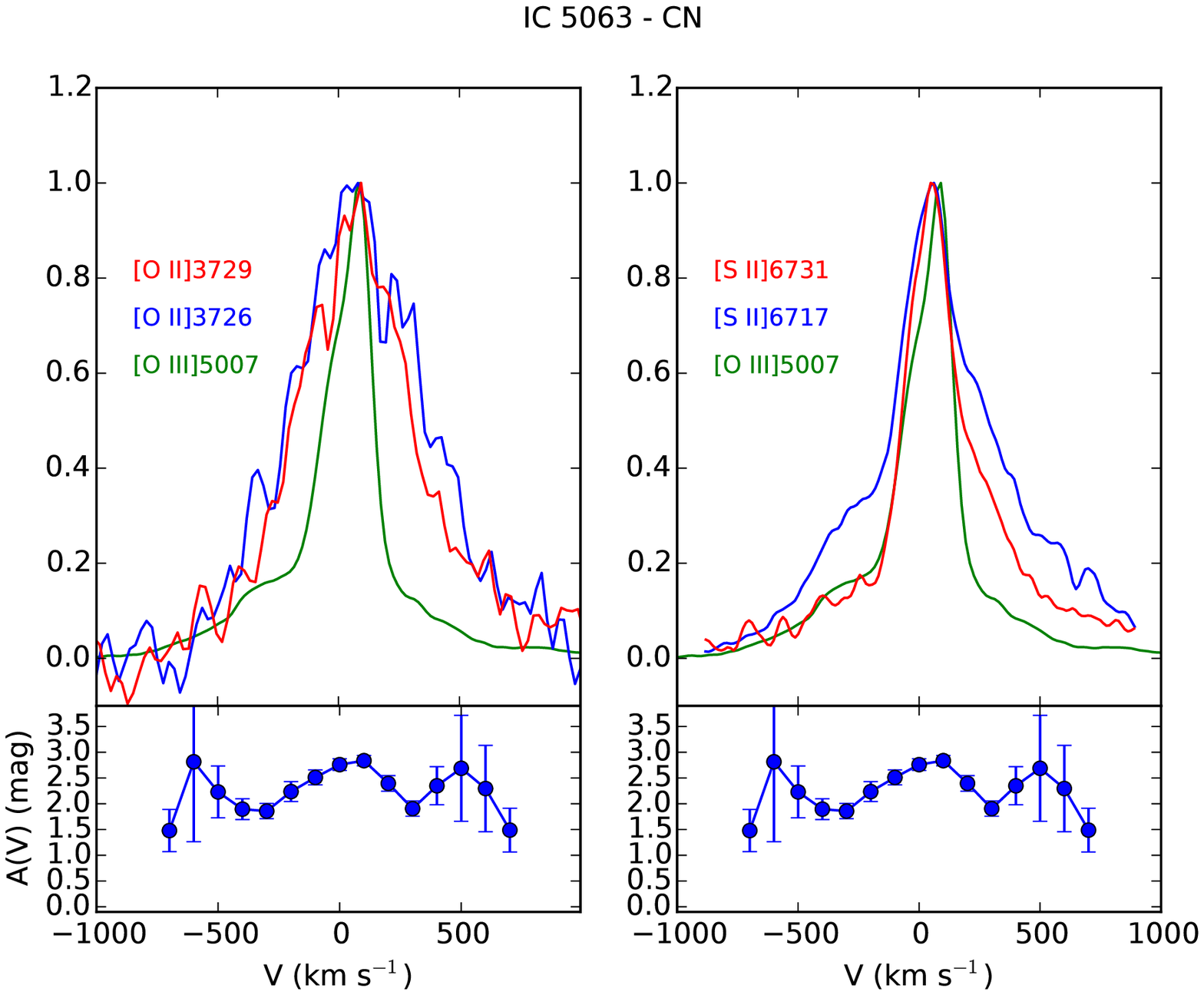}\\
\caption[]{Comparison of the emission lines of the CN region of IC\,5063. The behaviour of the extinction coefficient as a function of velocity is shown under each plot.}
\label{fig:cnl1_I}
\end{figure*}

\begin{figure*}
\centering
\includegraphics[width=0.45\textwidth]{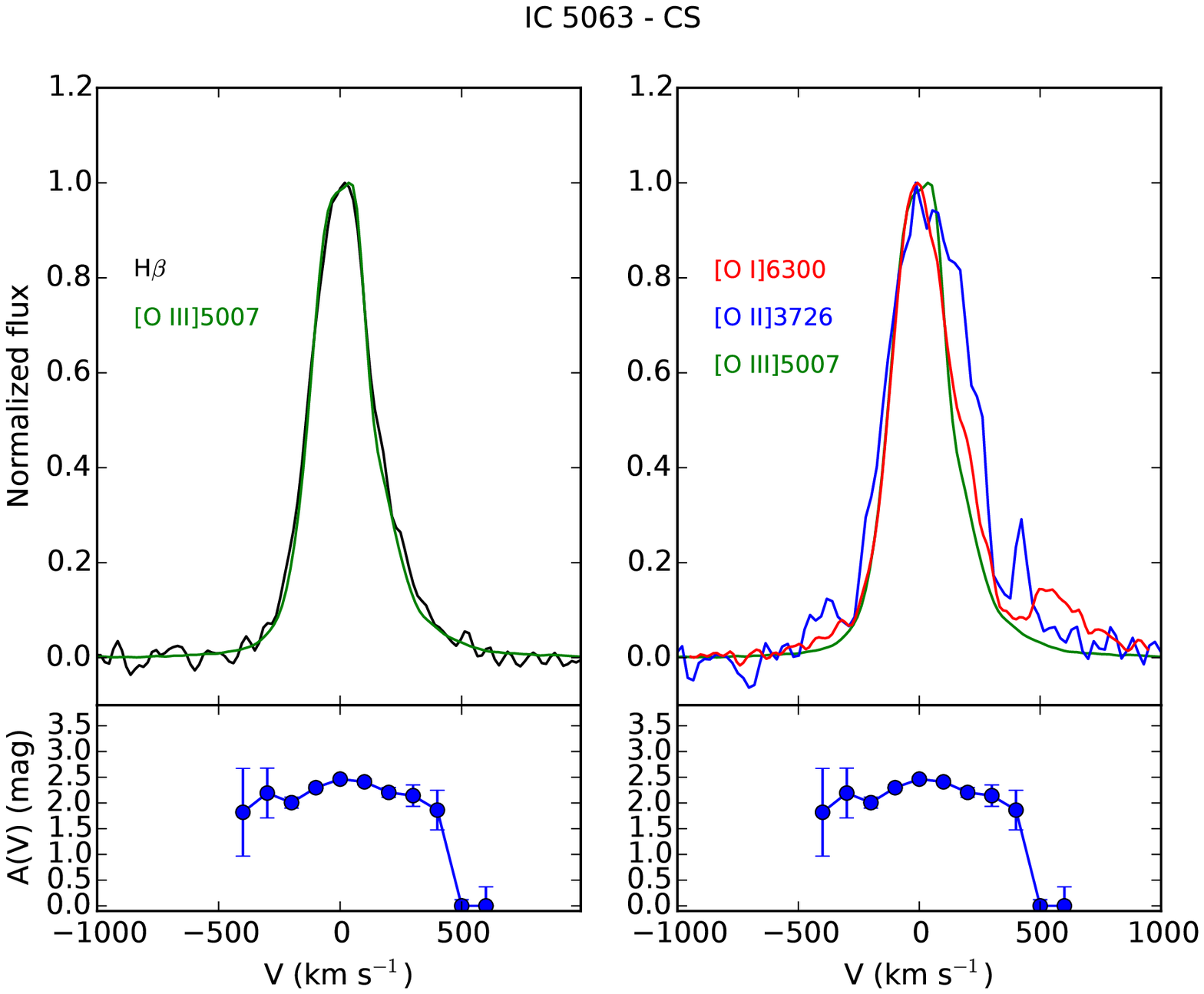} \quad
\includegraphics[width=0.45\textwidth]{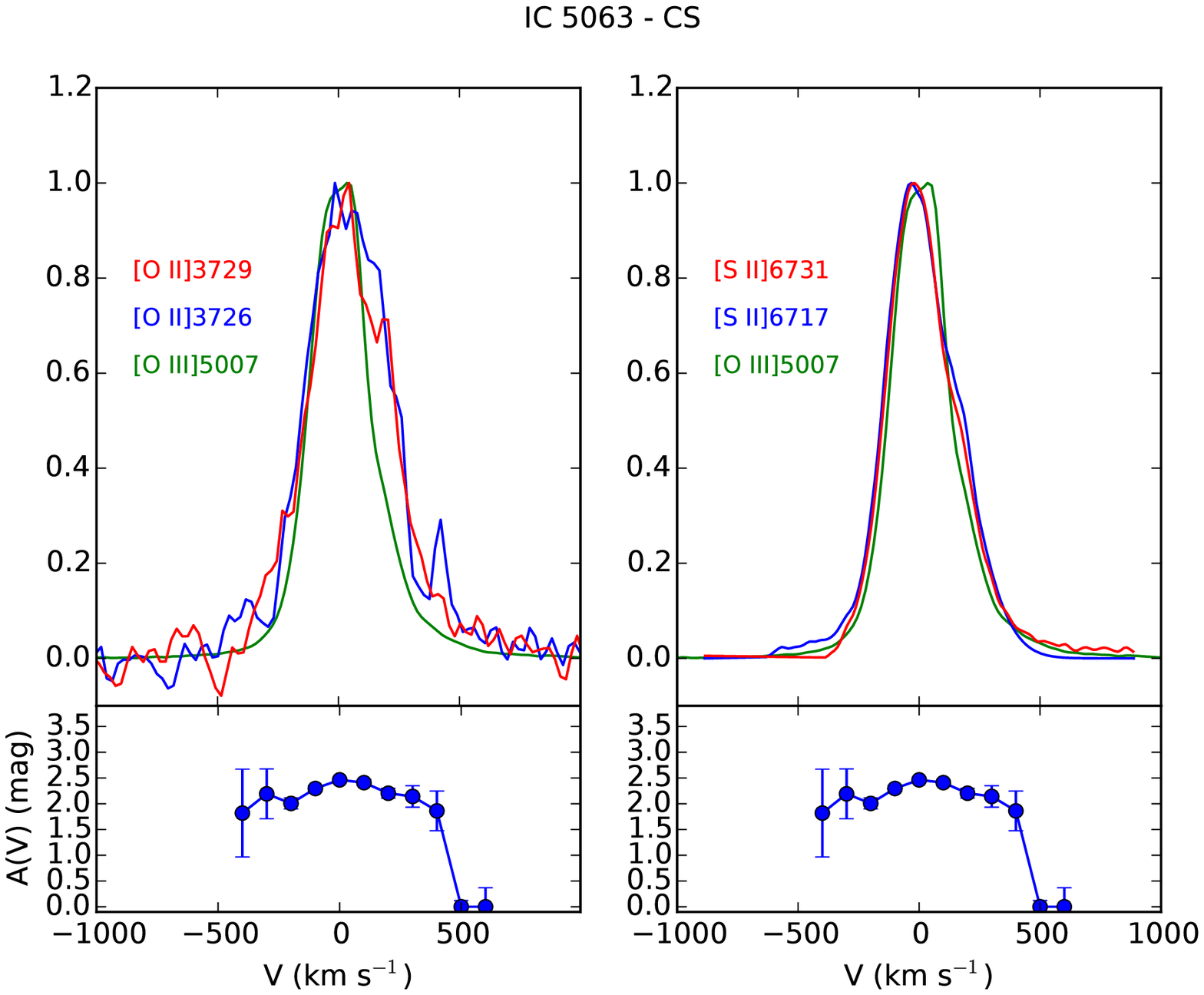}\\
\caption[]{Comparison of the emission lines of the CS region of IC\,5063. The behaviour of the extinction coefficient as a function of velocity is shown under each plot.}
\label{fig:csl1_I}
\end{figure*}

\begin{figure*}
\centering
\includegraphics[width=0.45\textwidth]{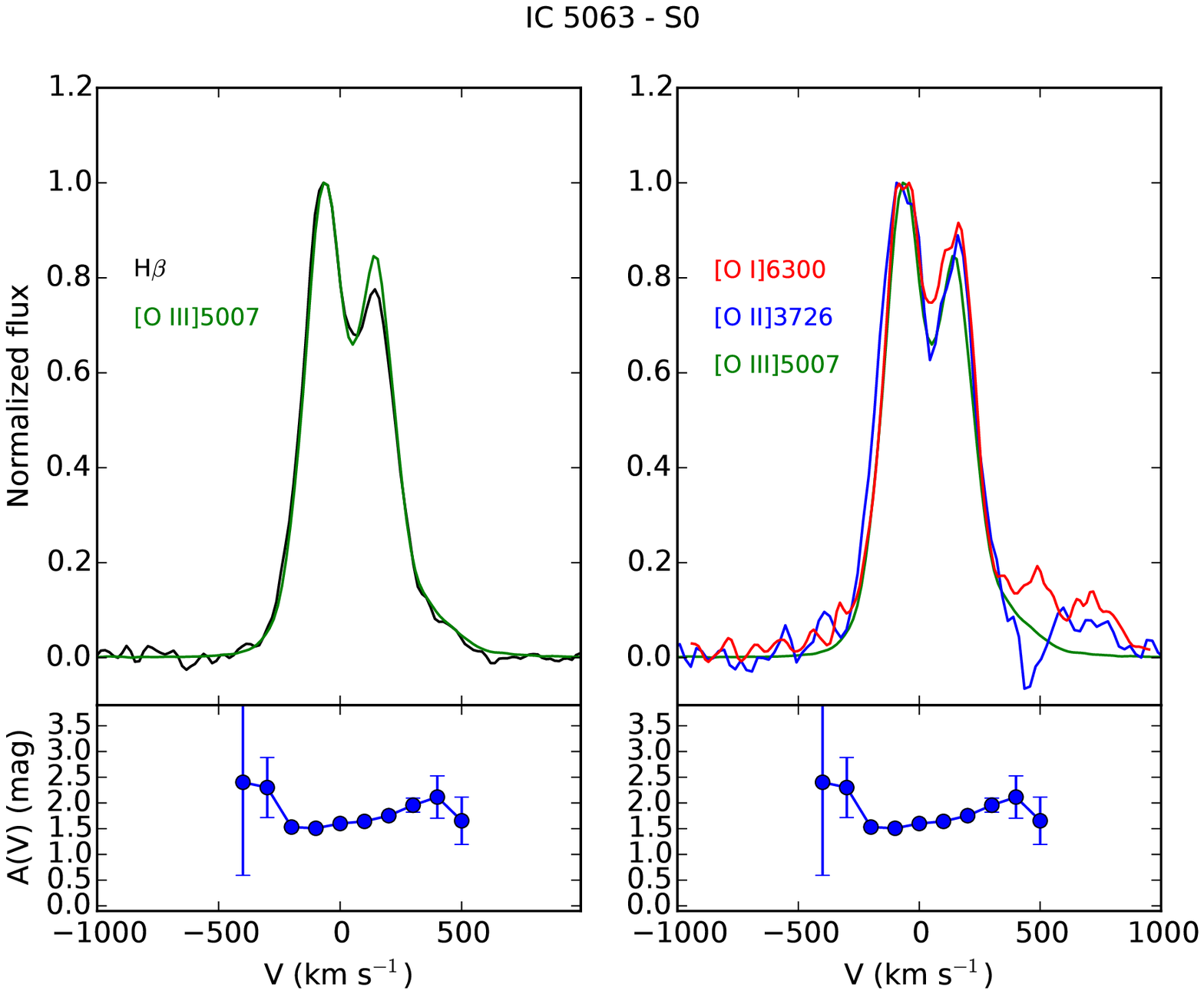} \quad
\includegraphics[width=0.45\textwidth]{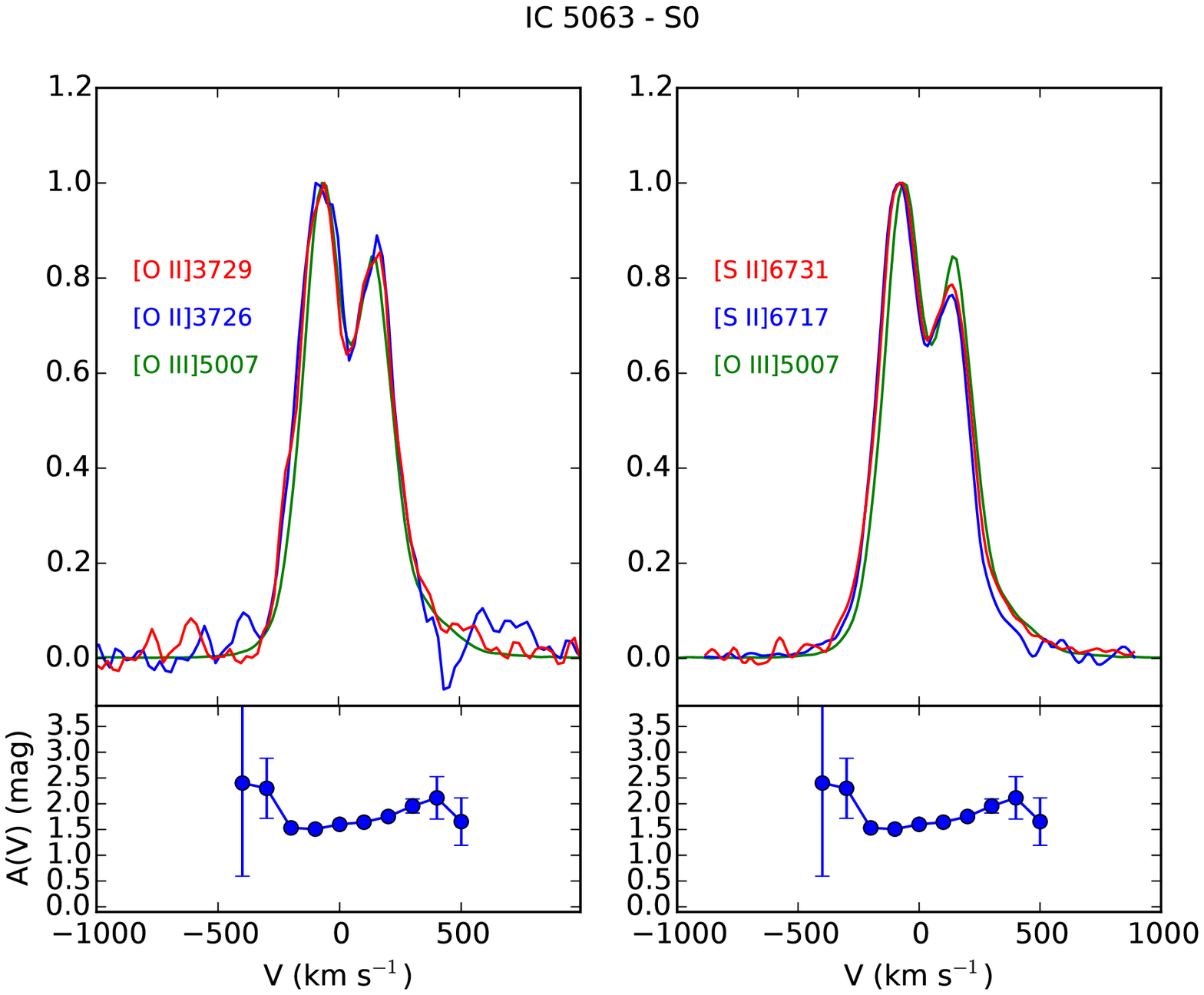}\\
\caption[]{Comparison of the emission lines of the S0 region of IC\,5063. The behaviour of the extinction coefficient as a function of velocity is shown under each plot.}
\label{fig:s0l1_I}
\end{figure*}

\begin{figure*}
\centering
\includegraphics[width=0.45\textwidth]{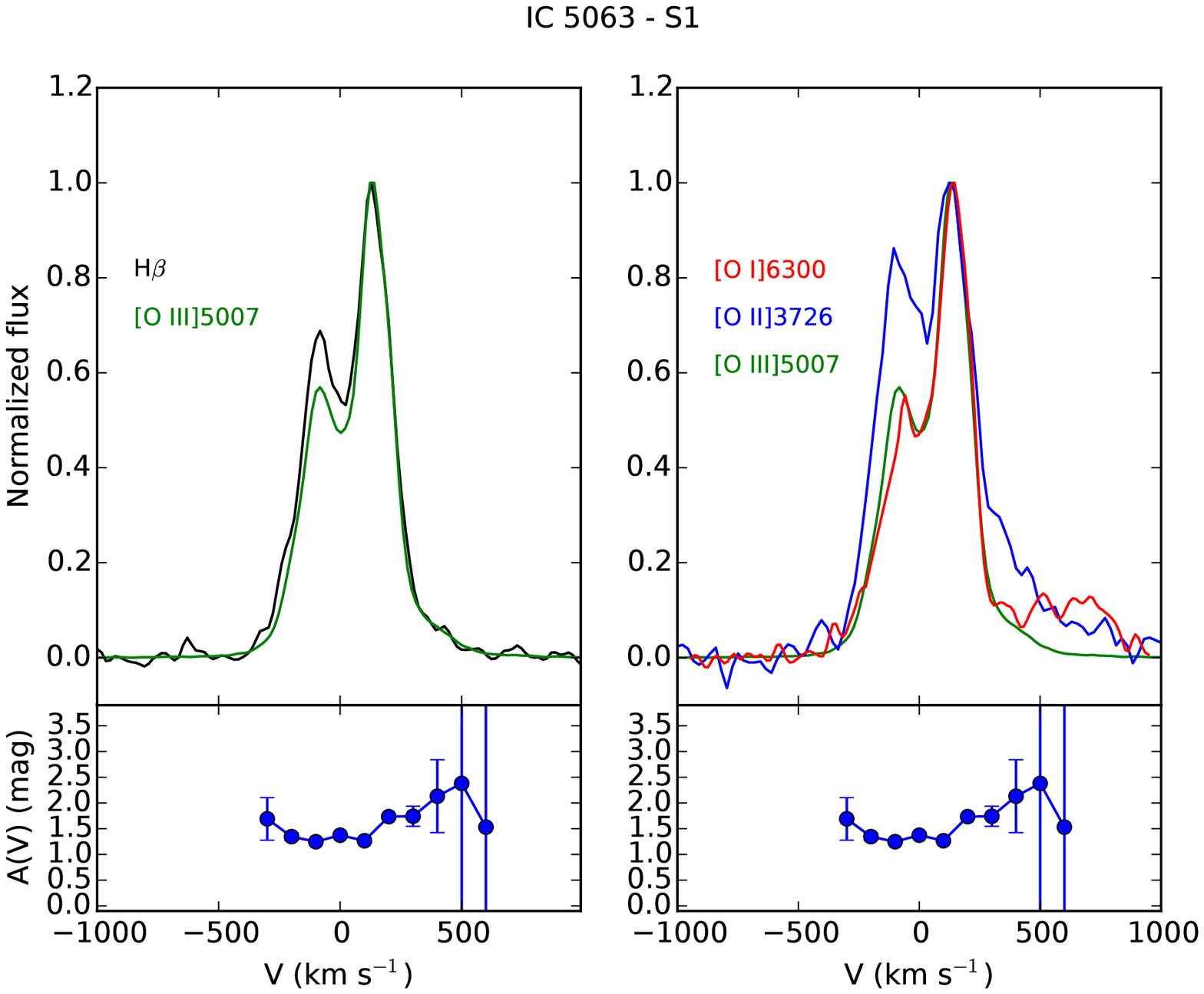} \quad
\includegraphics[width=0.45\textwidth]{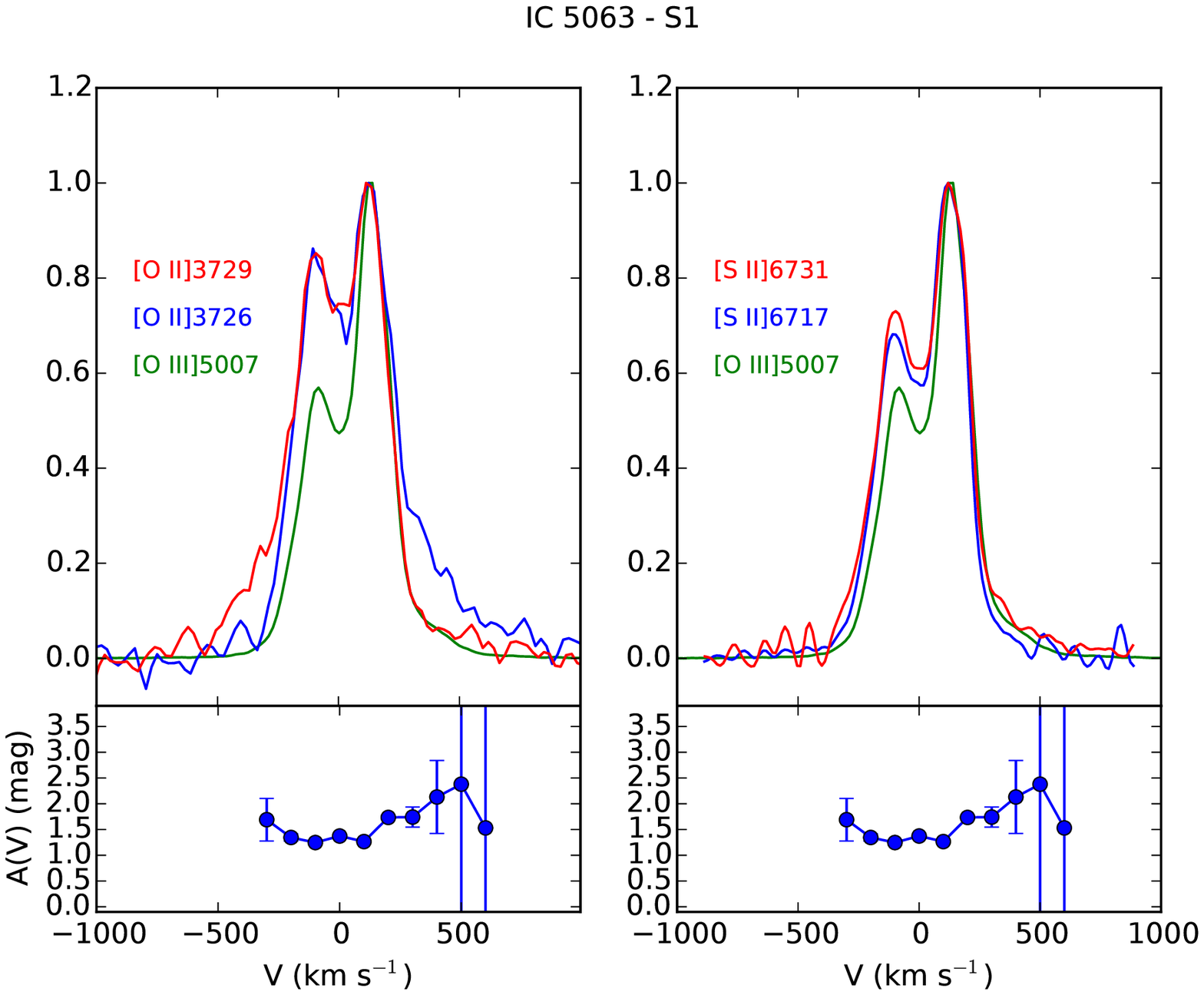}\\
\caption[]{Comparison of the emission lines of the S1 region of IC\,5063. The behaviour of the extinction coefficient as a function of velocity is shown under each plot.}
\label{fig:s1l1_I}
\end{figure*}

\begin{figure*}
\centering
\includegraphics[width=0.45\textwidth]{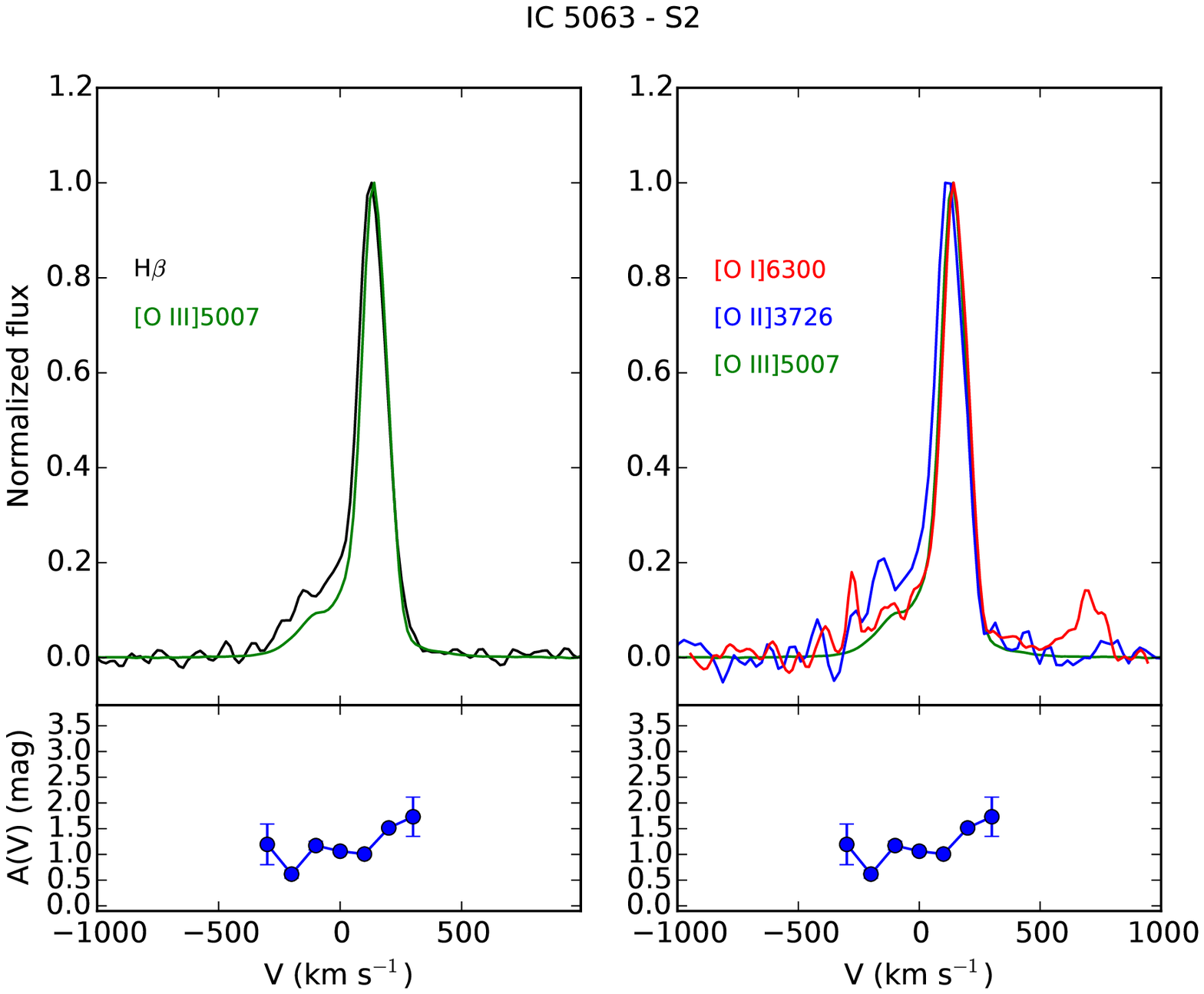} \quad
\includegraphics[width=0.45\textwidth]{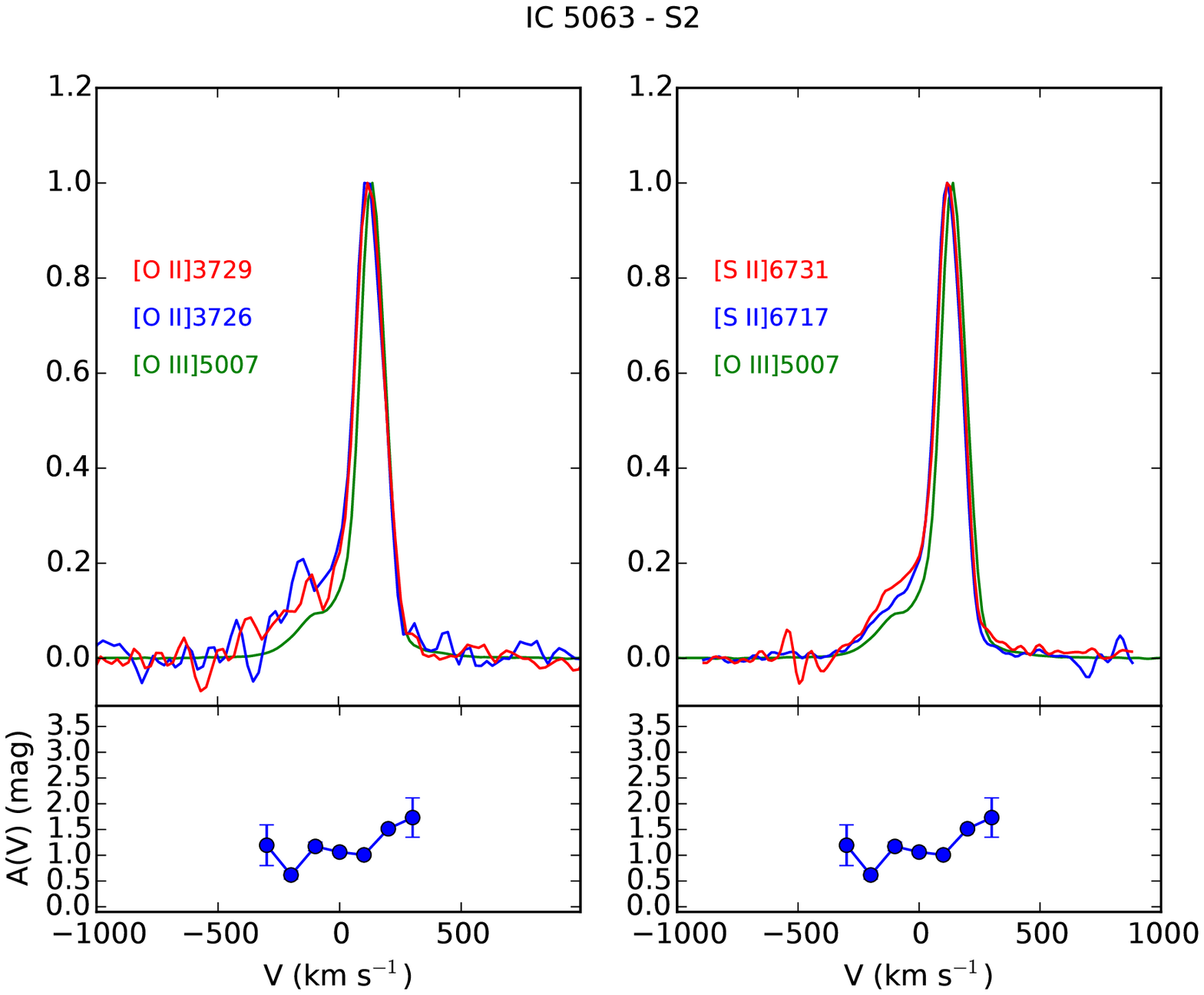}\\
\caption[]{Comparison of the emission lines of the S2 region of IC\,5063. The behaviour of the extinction coefficient as a function of velocity is shown under each plot.}
\label{fig:s2l1_I}
\end{figure*}

\newpage
\subsection{NGC 7212}

\begin{figure*}
\centering
\includegraphics[width=0.45\textwidth]{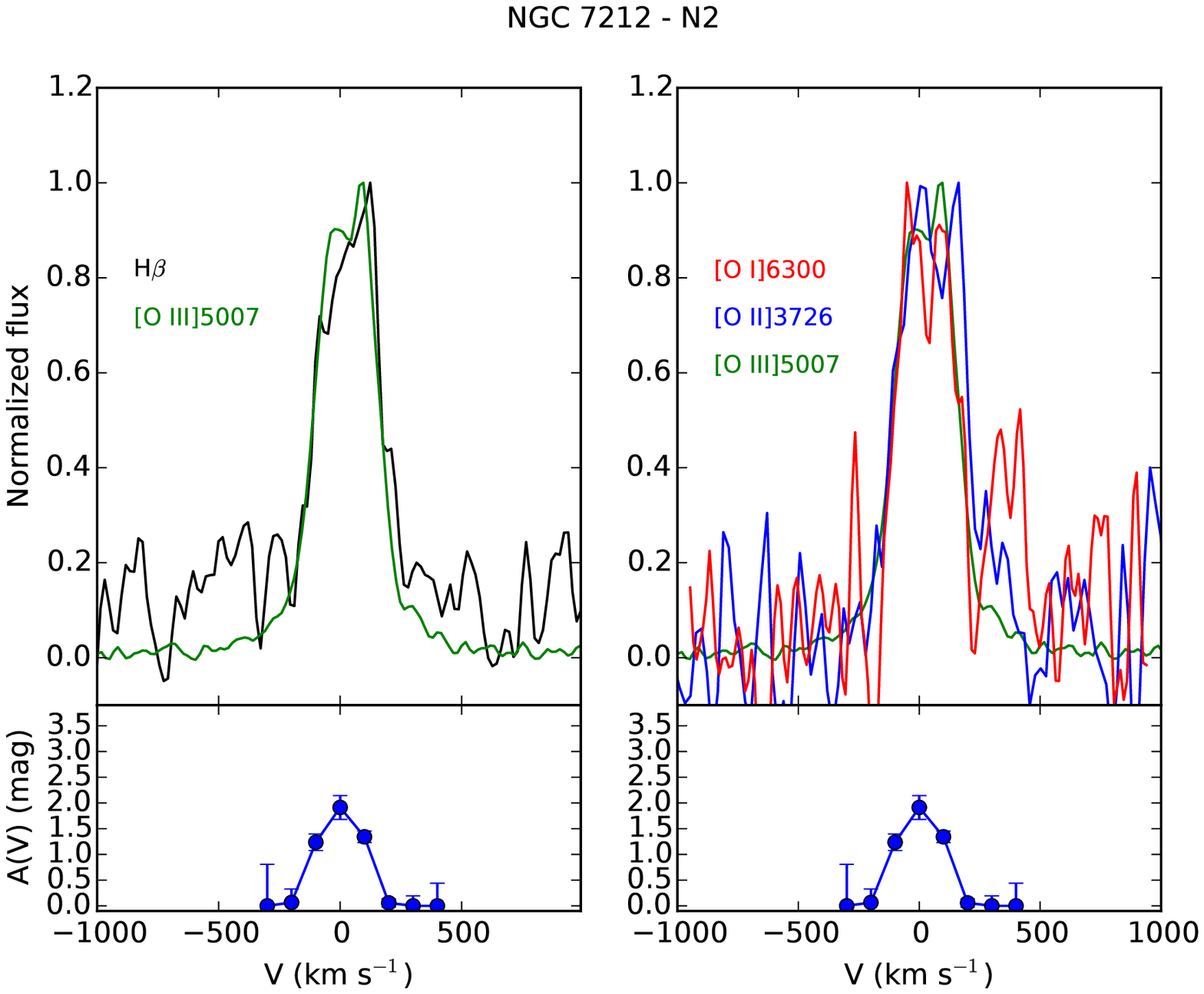} \quad
\includegraphics[width=0.45\textwidth]{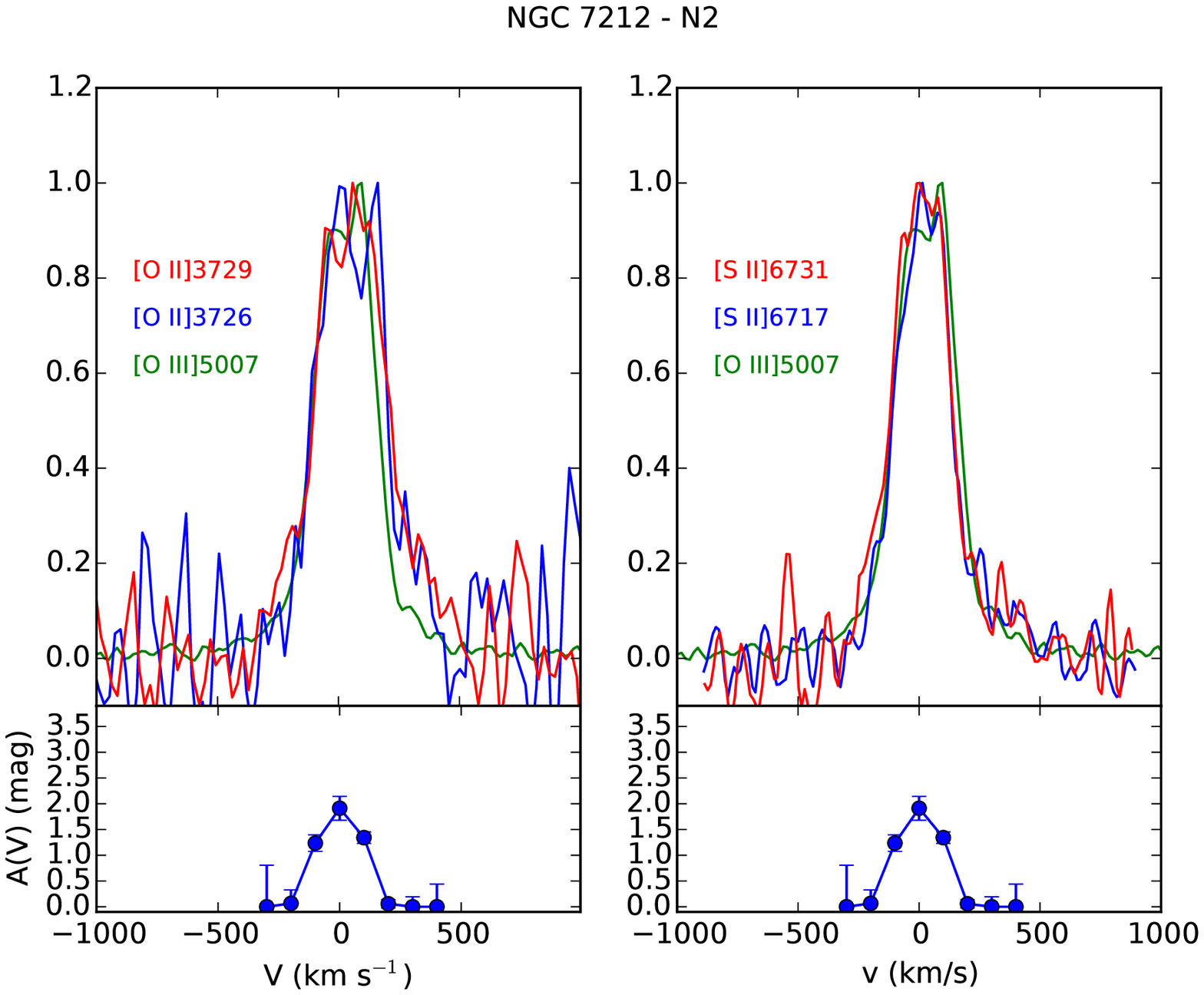}\\
\caption[]{Comparison of the emission lines of the N2 region of NGC\,7212. The behaviour of the extinction coefficient as a function of velocity is shown under each plot.}
\label{fig:n2l1_N}
\end{figure*}

\begin{figure*}
\centering
\includegraphics[width=0.45\textwidth]{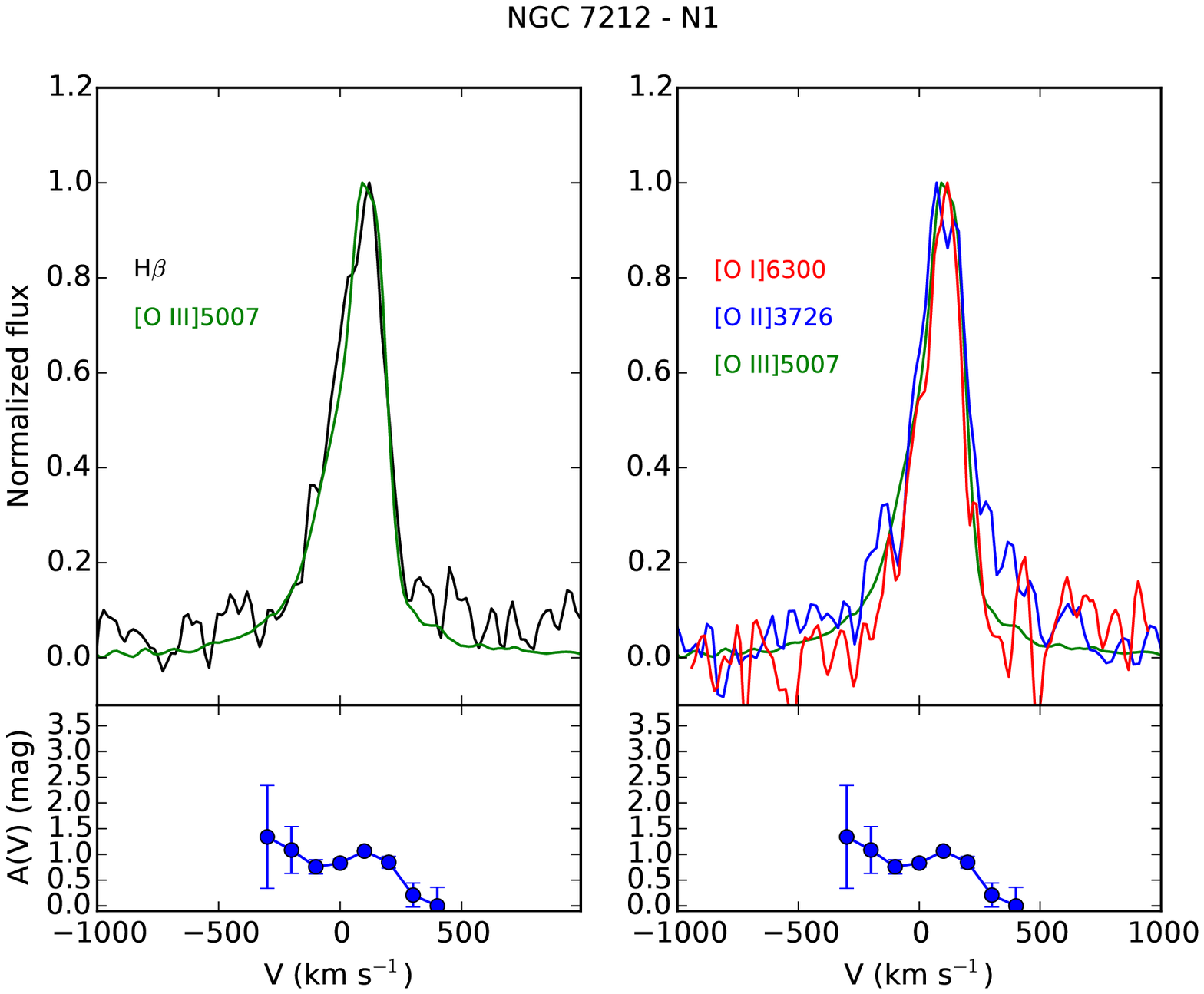} \quad
\includegraphics[width=0.45\textwidth]{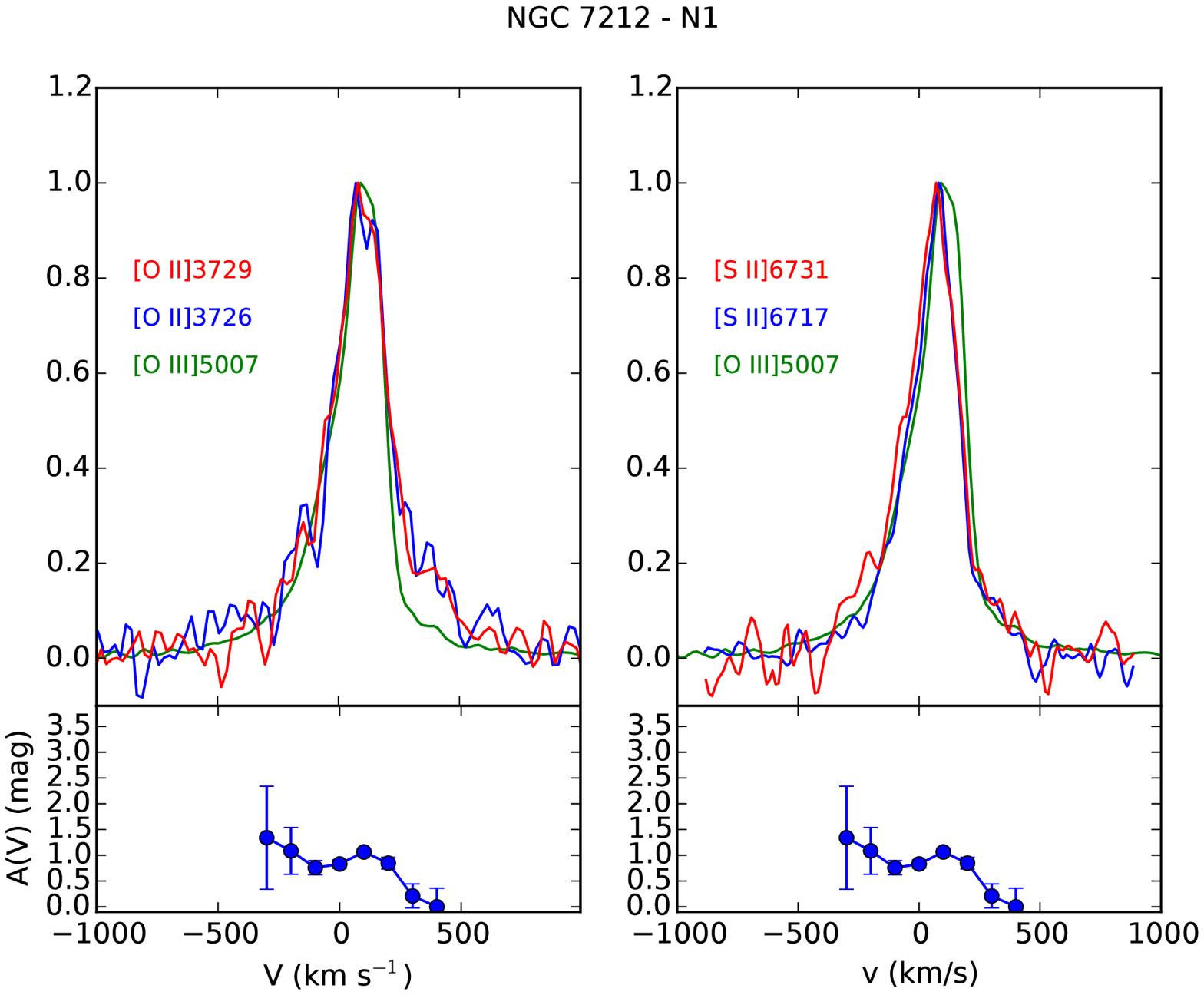}\\
\caption[]{Comparison of the emission lines of the N1 region of NGC\,7212. The behaviour of the extinction coefficient as a function of velocity is shown under each plot.}
\label{fig:n1l1_N}
\end{figure*}

\begin{figure*}
\centering
\includegraphics[width=0.45\textwidth]{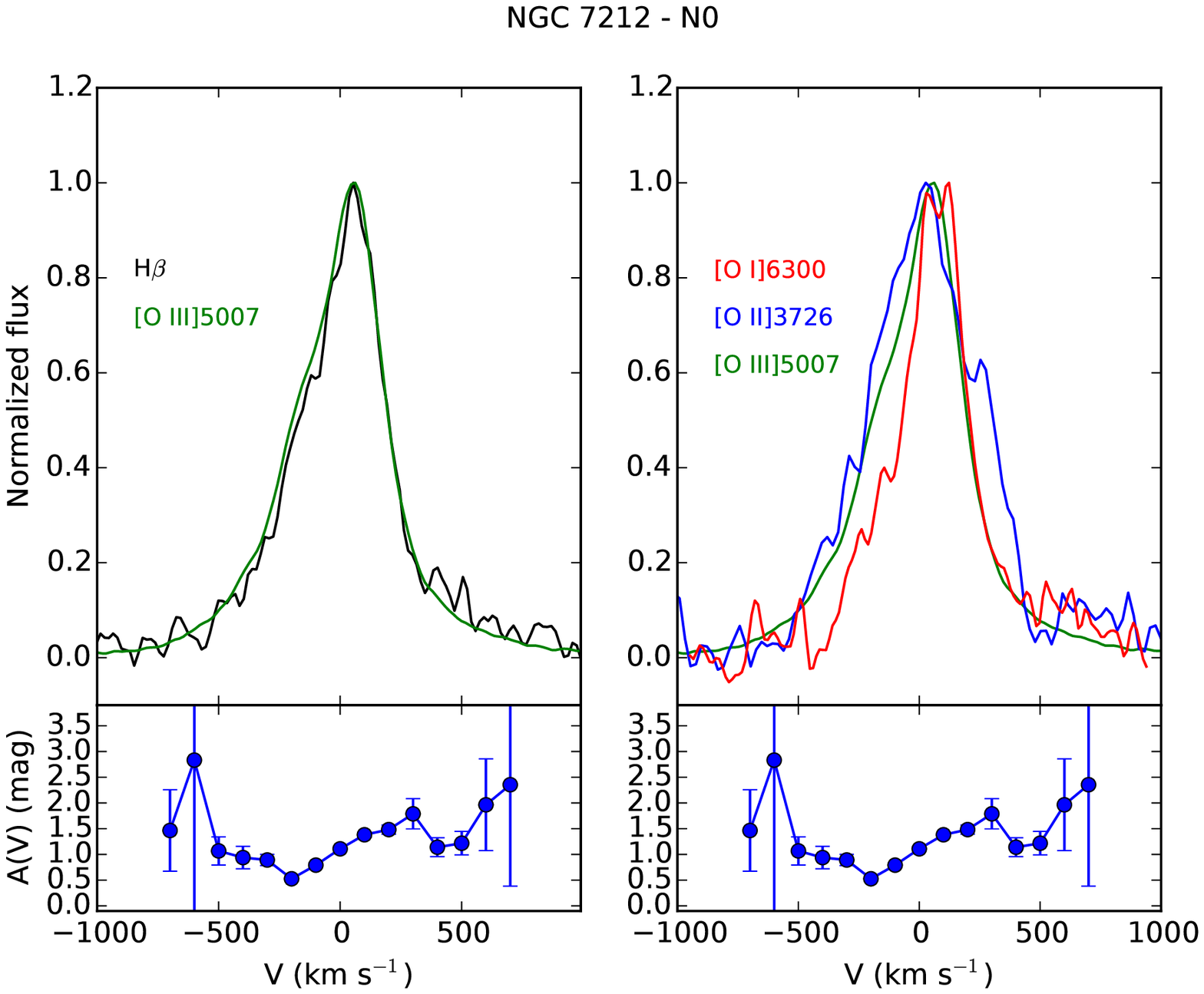} \quad
\includegraphics[width=0.45\textwidth]{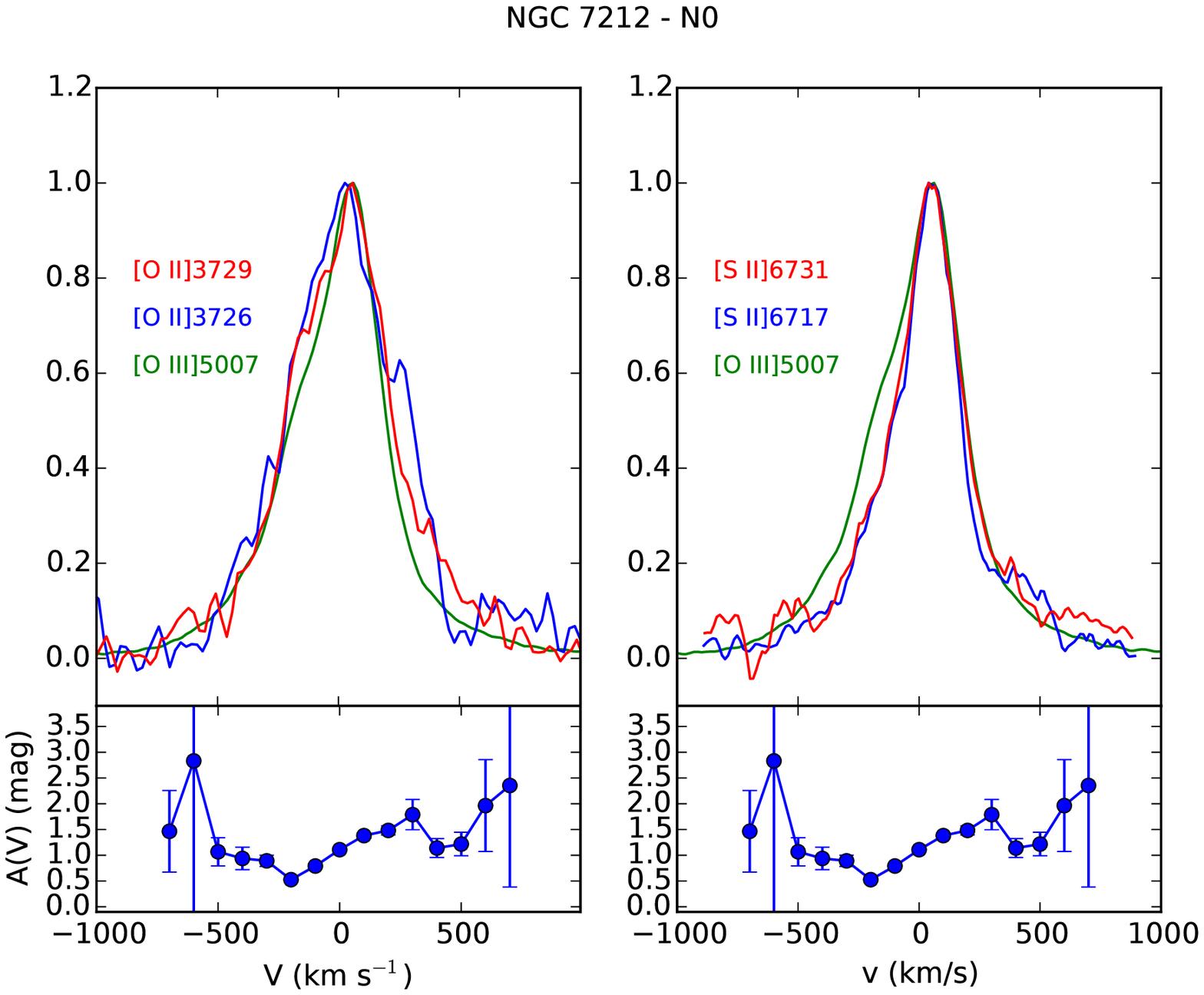}\\
\caption[]{Comparison of the emission lines of the N0 region of NGC\,7212. The behaviour of the extinction coefficient as a function of velocity is shown under each plot.}
\label{fig:n0l1_N}
\end{figure*}

\begin{figure*}
\centering
\includegraphics[width=0.45\textwidth]{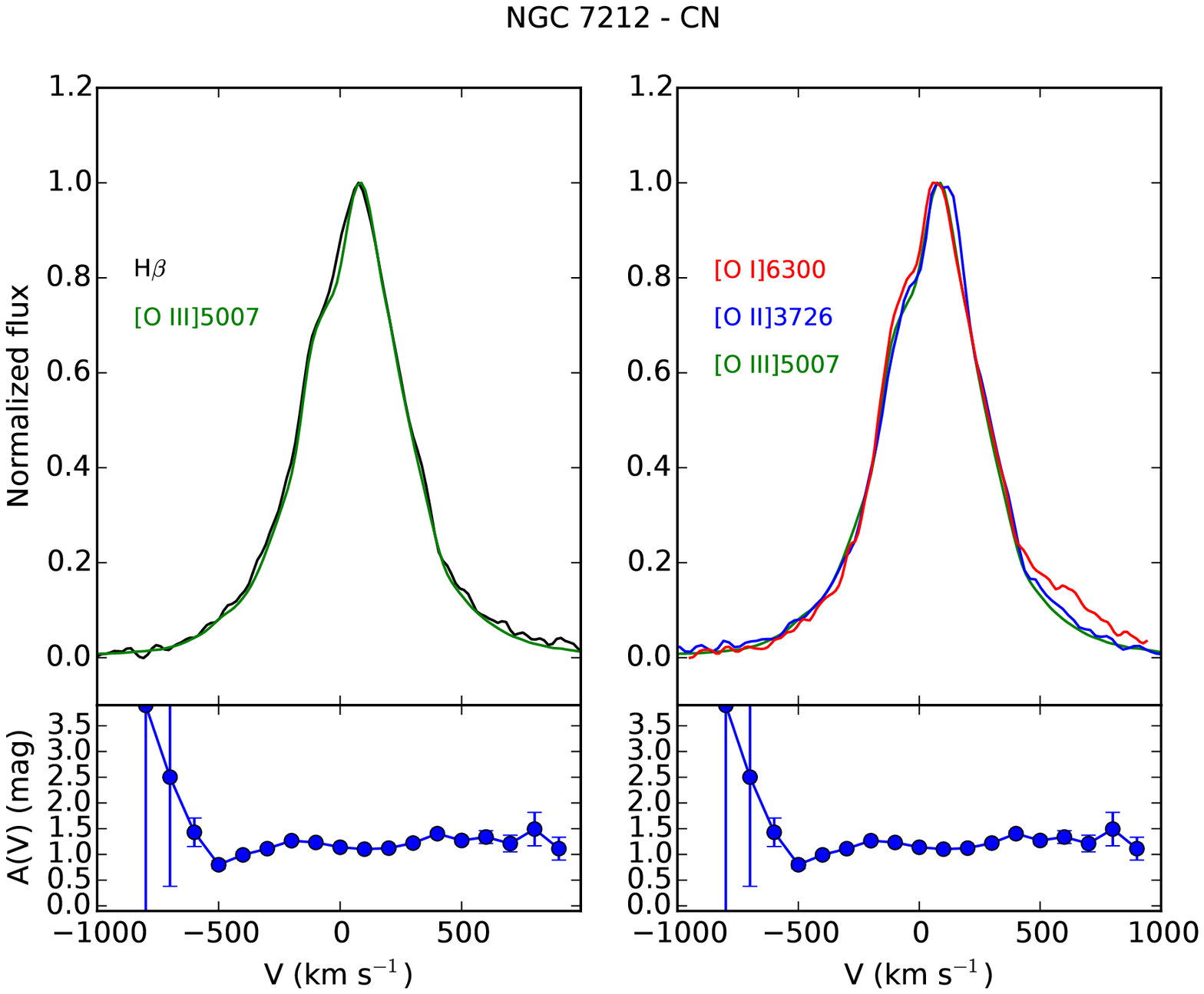} \quad
\includegraphics[width=0.45\textwidth]{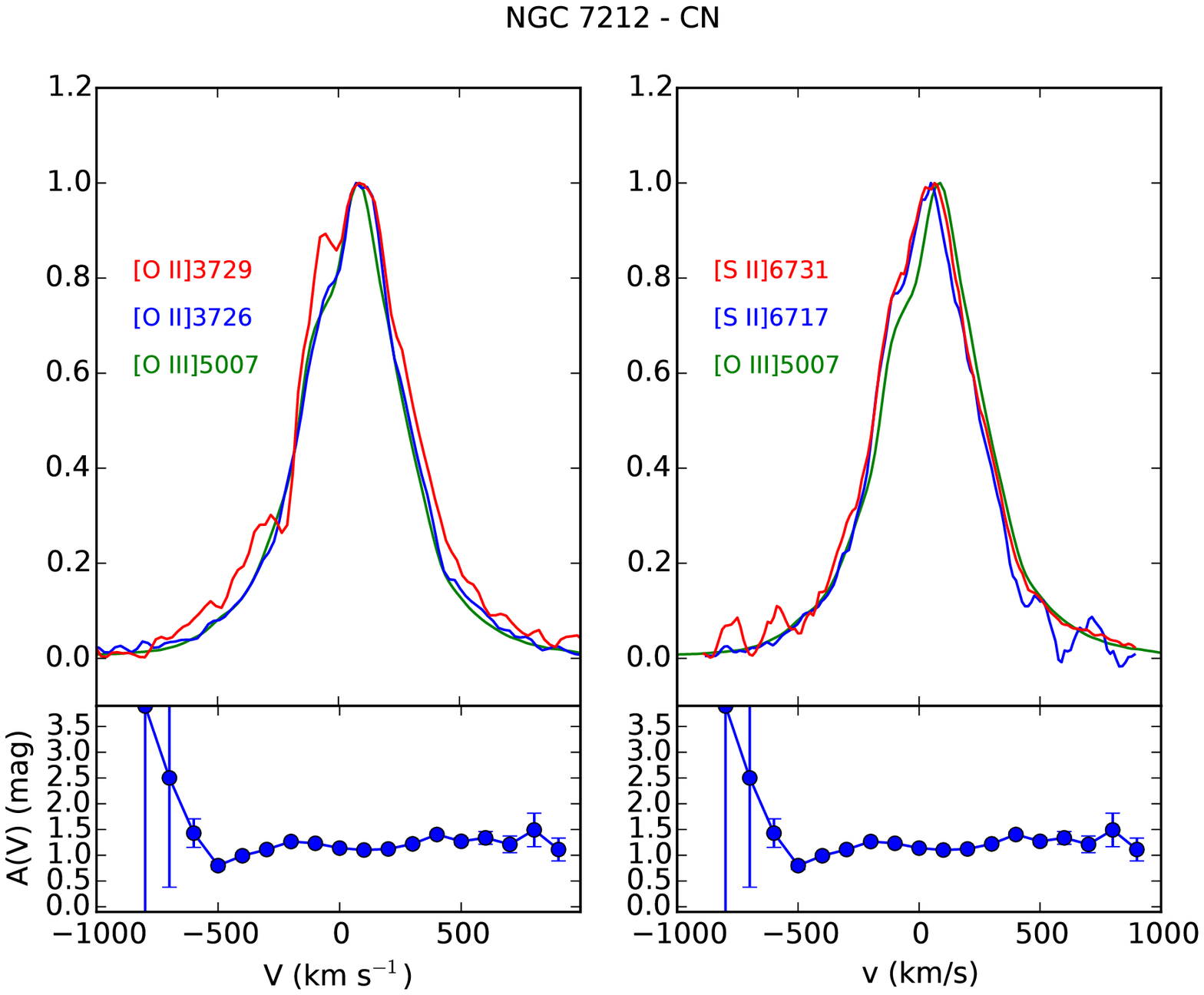}\\
\caption[]{Comparison of the emission lines of the CN region of NGC\,7212. The behaviour of the extinction coefficient as a function of velocity is shown under each plot.}
\label{fig:cnl1_N}
\end{figure*}

\begin{figure*}
\centering
\includegraphics[width=0.45\textwidth]{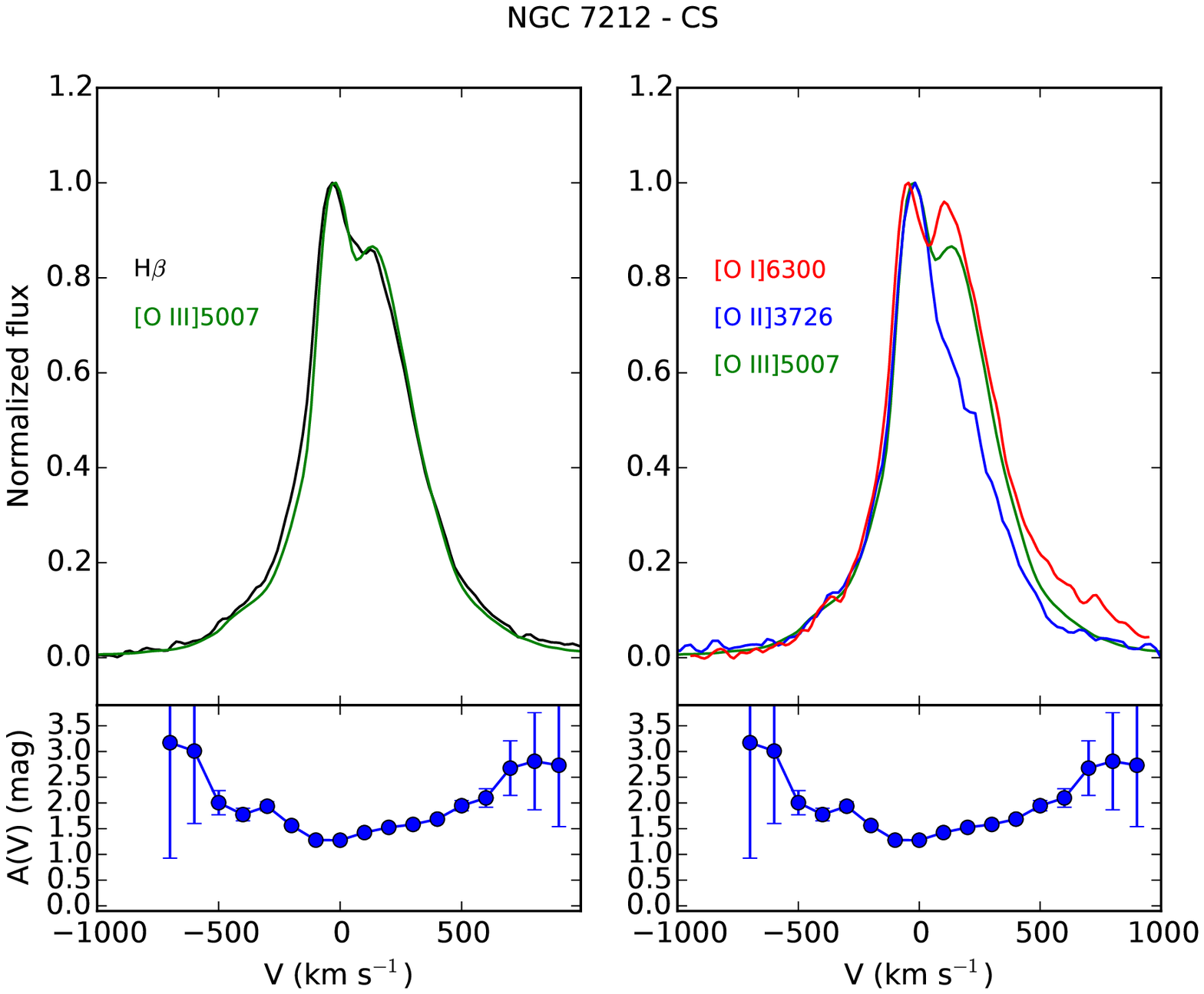} \quad
\includegraphics[width=0.45\textwidth]{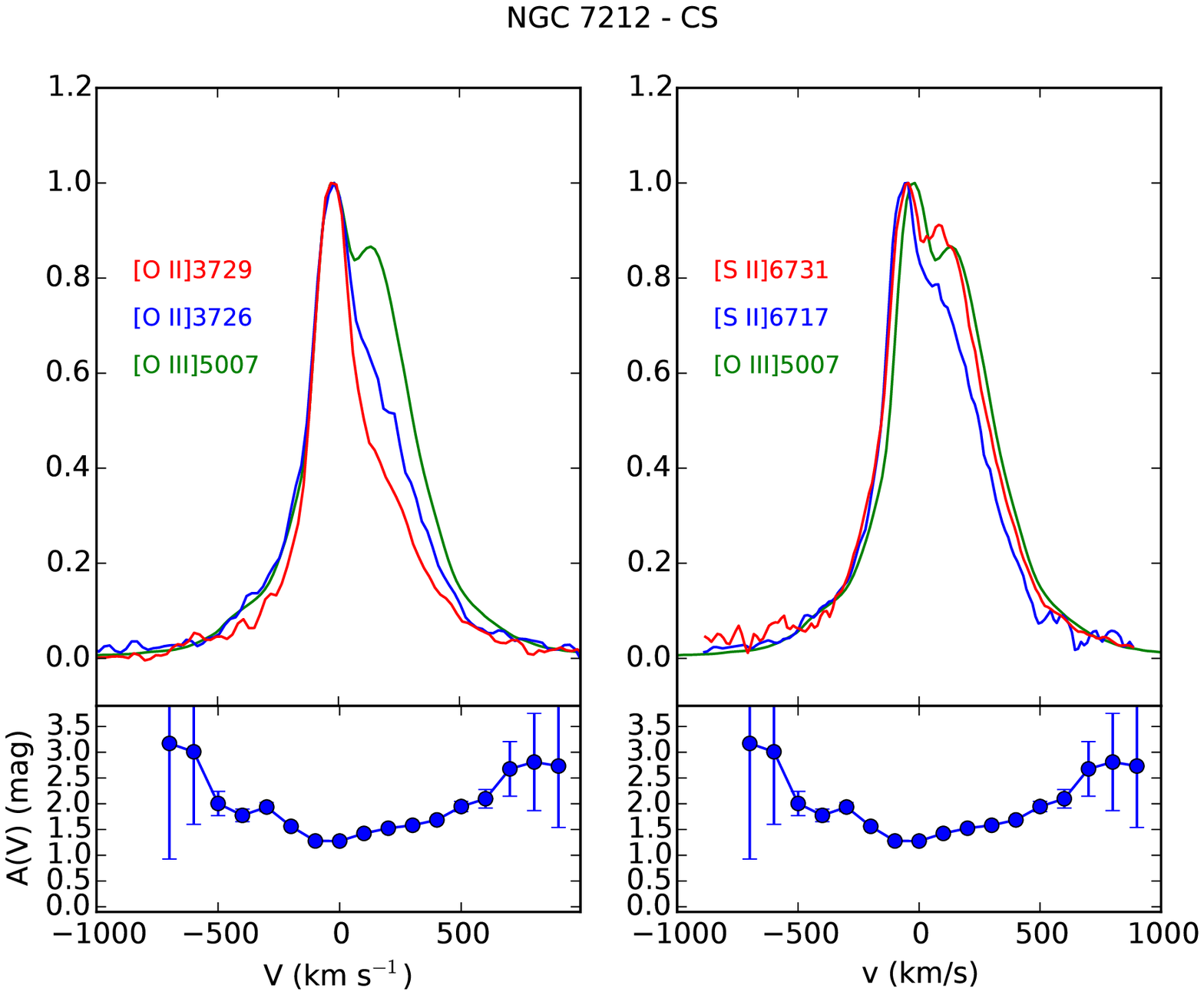}\\
\caption[]{Comparison of the emission lines of the CS region of NGC\,7212. The behaviour of the extinction coefficient as a function of velocity is shown under each plot.}
\label{fig:csl1_N}
\end{figure*}

\begin{figure*}
\centering
\includegraphics[width=0.45\textwidth]{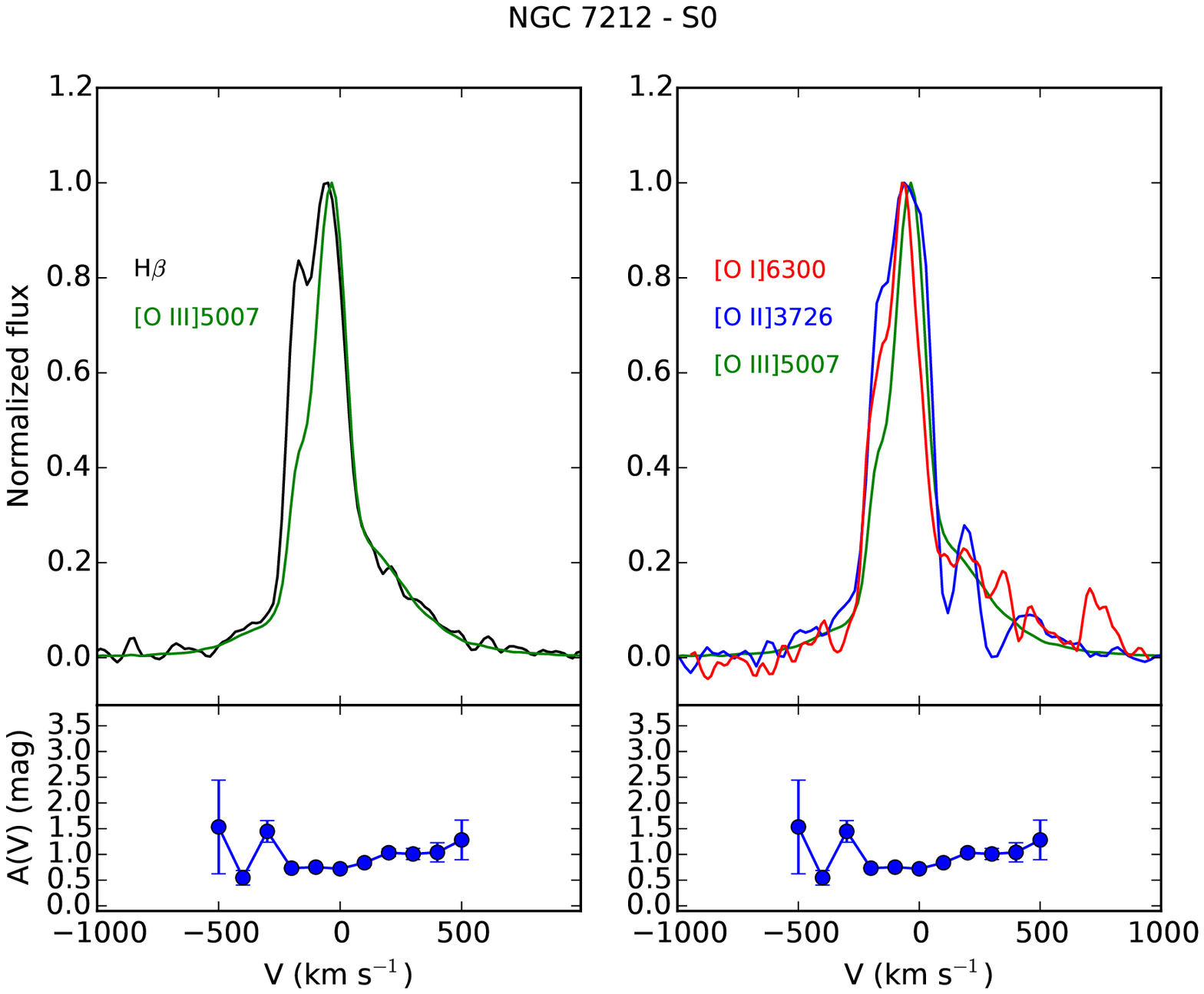} \quad
\includegraphics[width=0.45\textwidth]{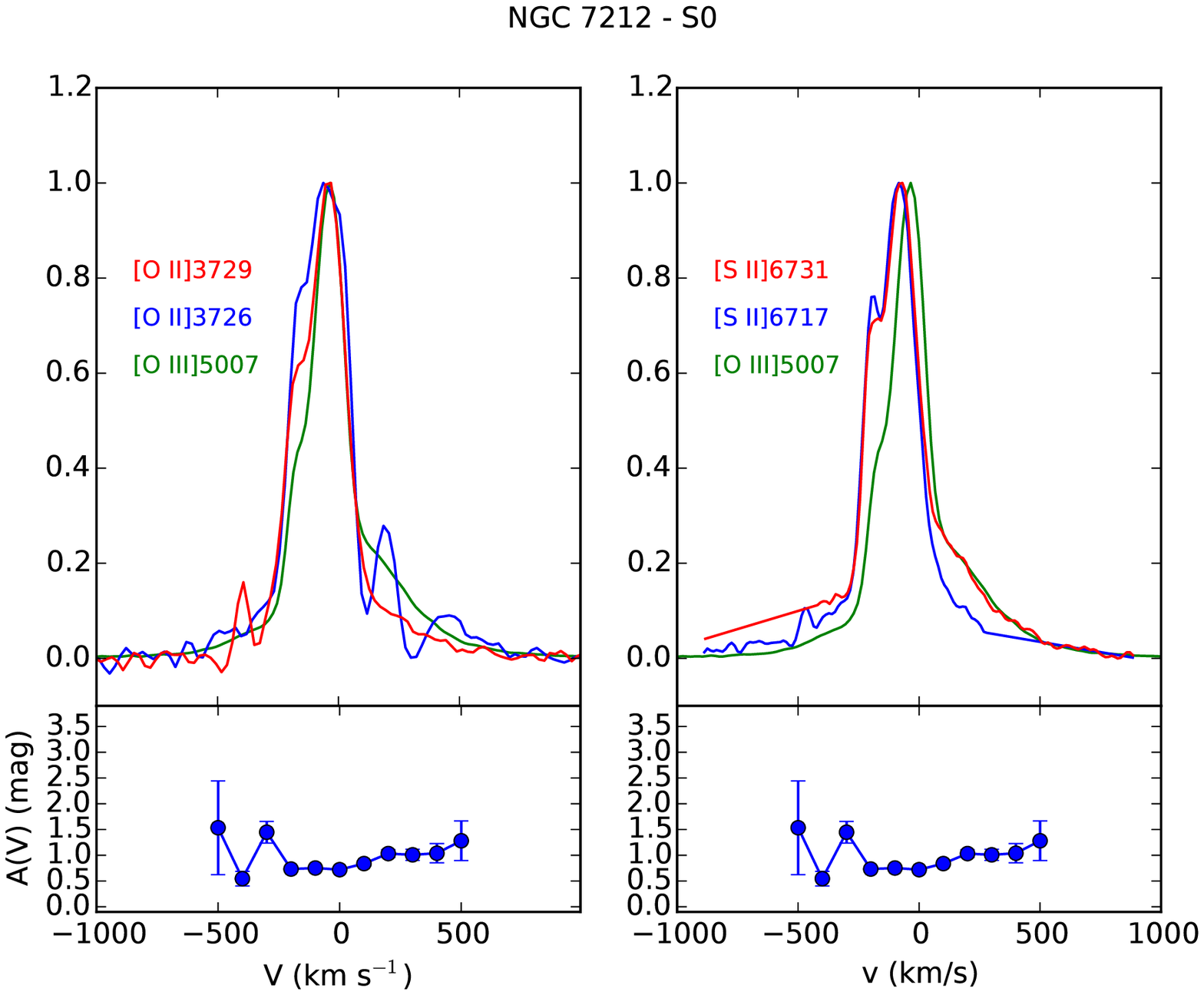}\\
\caption[]{Comparison of the emission lines of the S0 region of NGC\,7212. The behaviour of the extinction coefficient as a function of velocity is shown under each plot.}
\label{fig:s0l1_N}
\end{figure*}

\begin{figure*}
\centering
\includegraphics[width=0.45\textwidth]{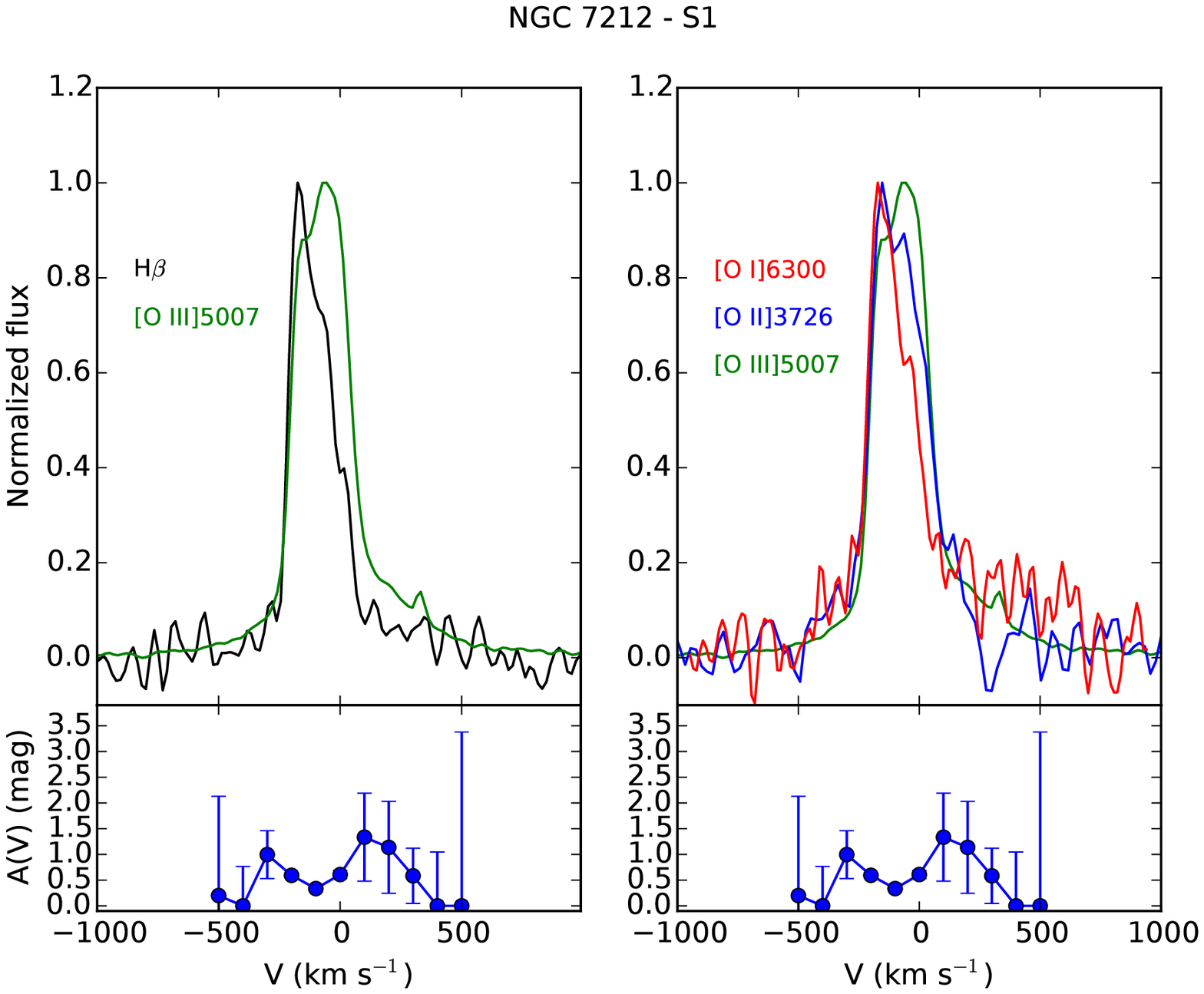} \quad
\includegraphics[width=0.45\textwidth]{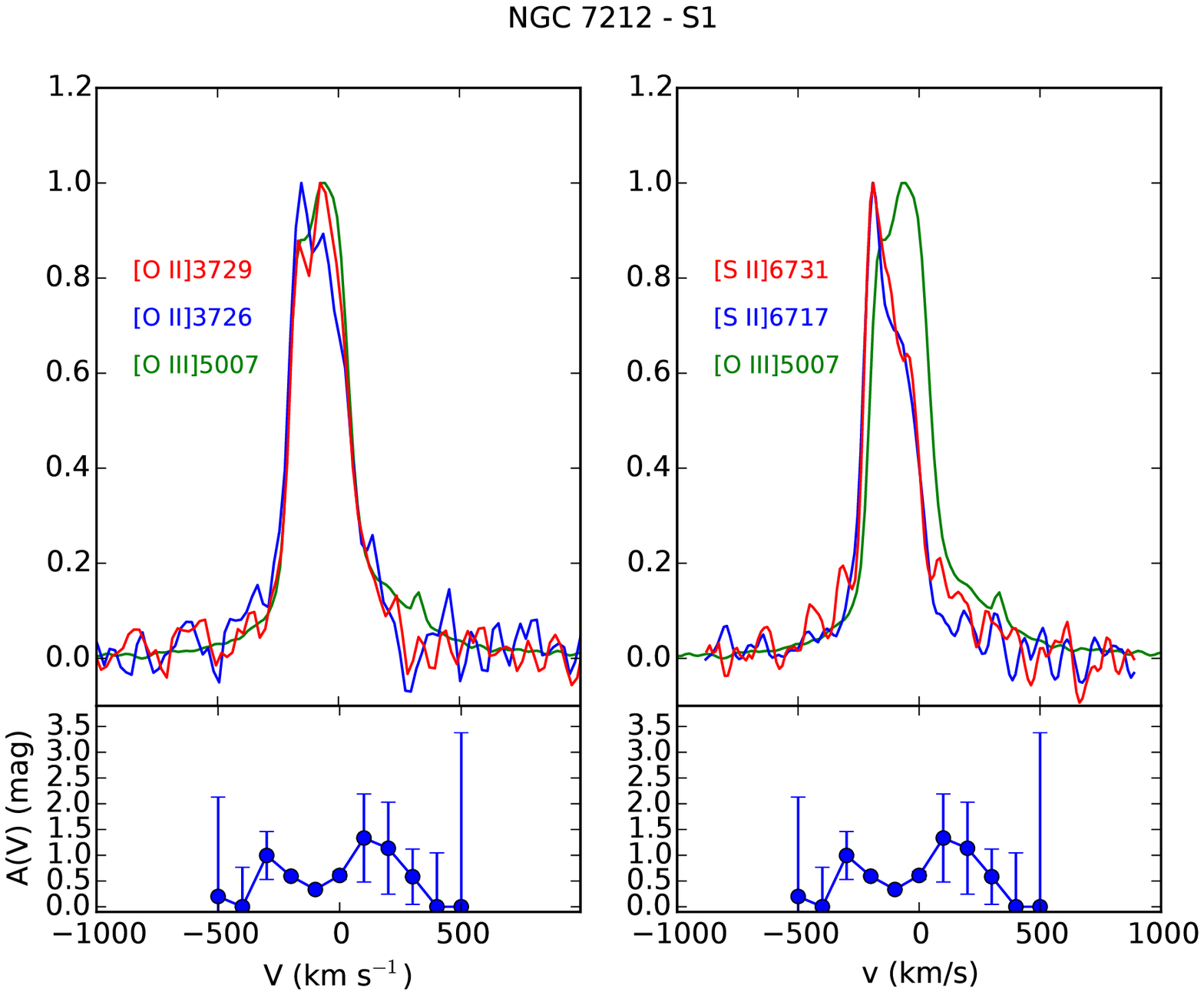}\\
\caption[]{Comparison of the emission lines of the S1 region of NGC\,7212. The behaviour of the extinction coefficient as a function of velocity is shown under each plot.}
\label{fig:s1l1_N}
\end{figure*}

\section{Diagnostic Diagrams}
\label{sec:Adiag}

\begin{figure*} 
\centering
\subfloat[][]{\includegraphics[width=0.43\textwidth]{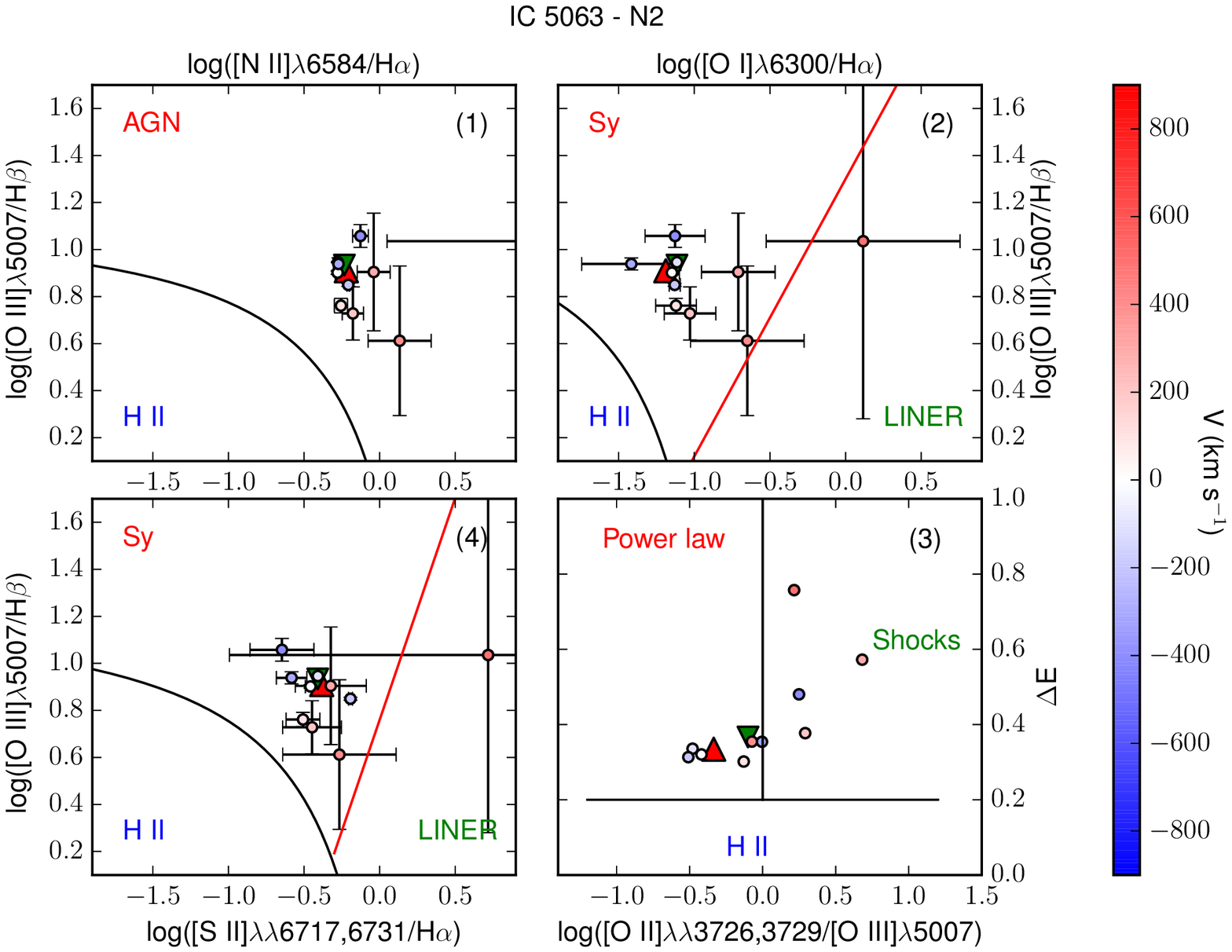}} \quad
\subfloat[][]{\includegraphics[width=0.43\textwidth]{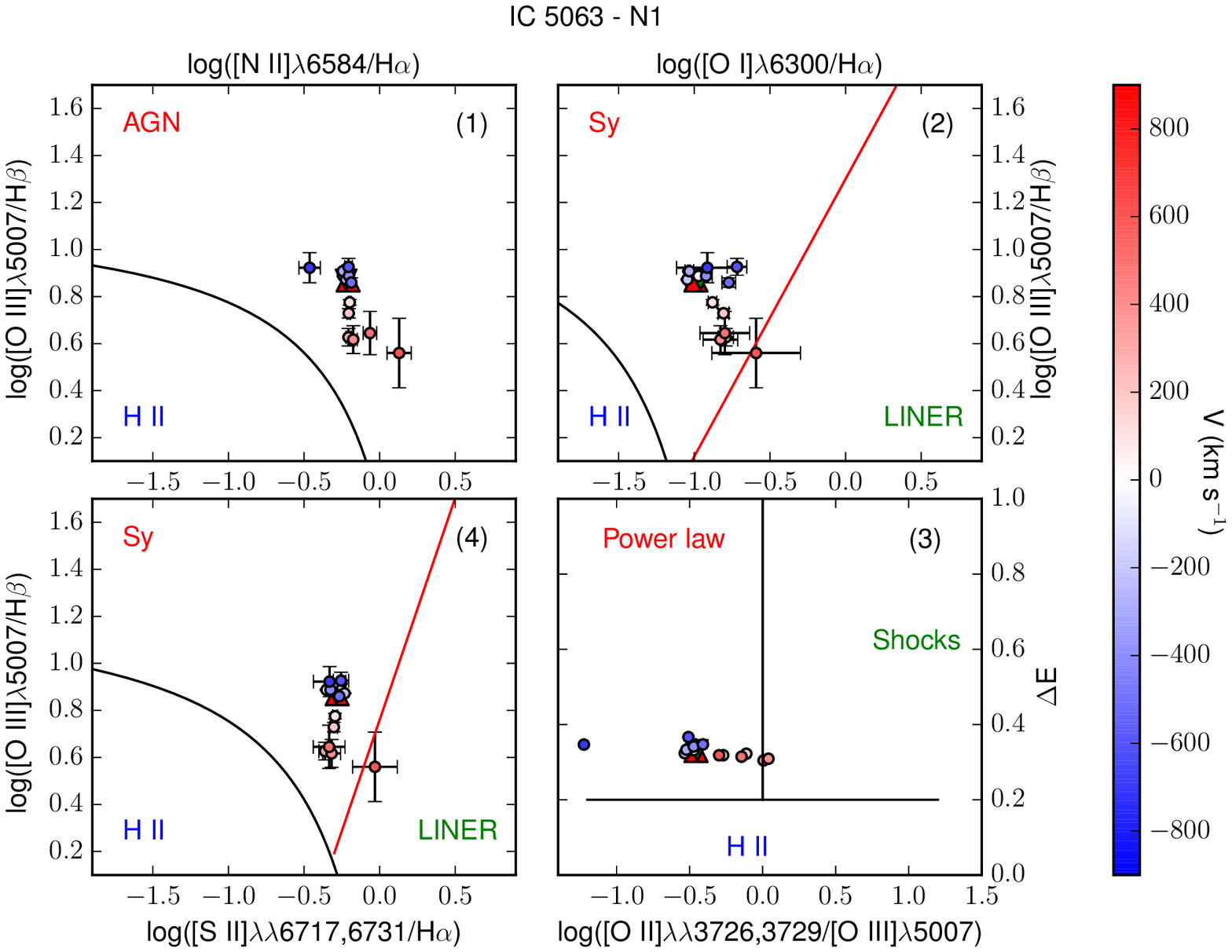}}\\
\subfloat[][]{\includegraphics[width=0.43\textwidth]{IC5063_n0_diag.eps}} \quad
\subfloat[][]{\includegraphics[width=0.43\textwidth]{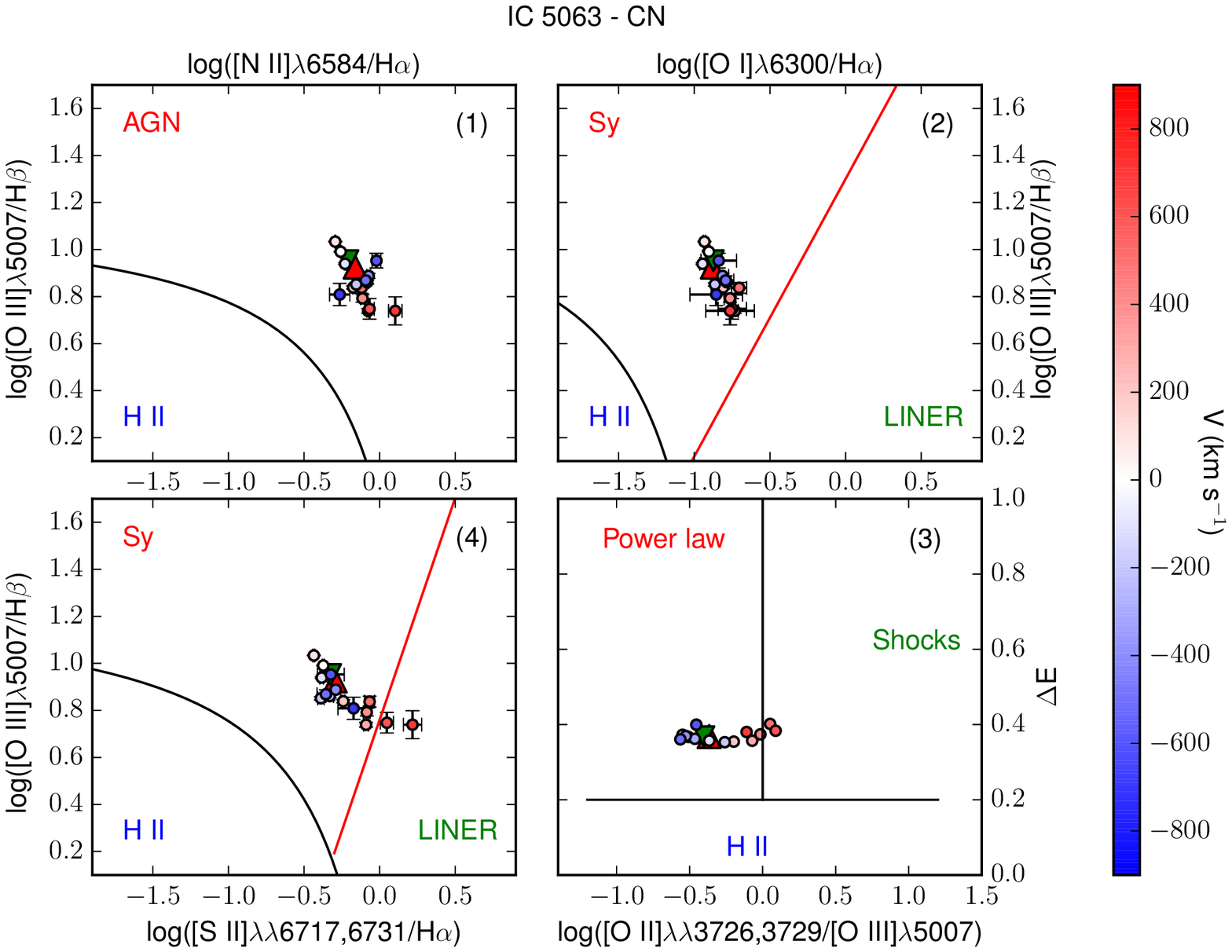}}\\
\subfloat[][]{\includegraphics[width=0.43\textwidth]{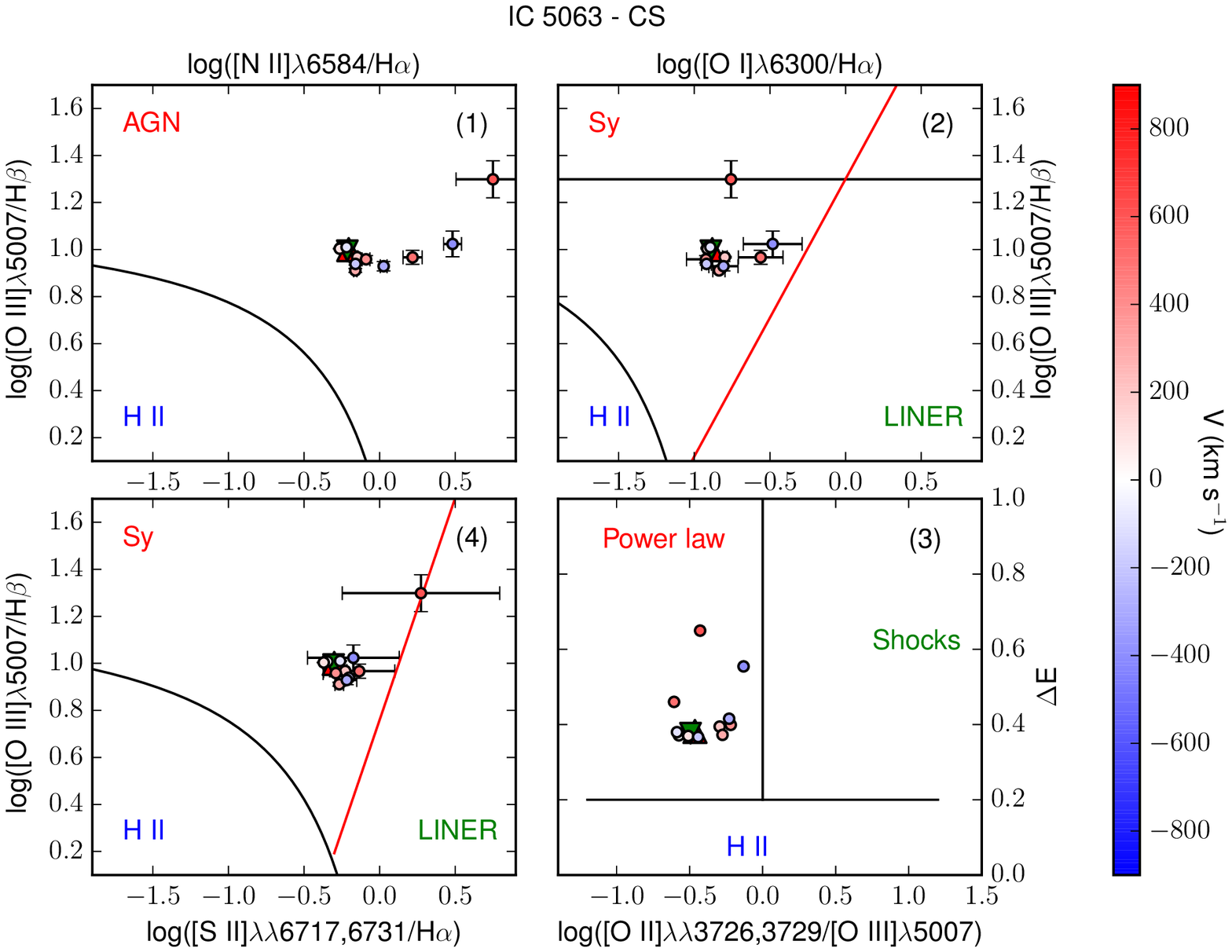}} \quad
\subfloat[][]{\includegraphics[width=0.43\textwidth]{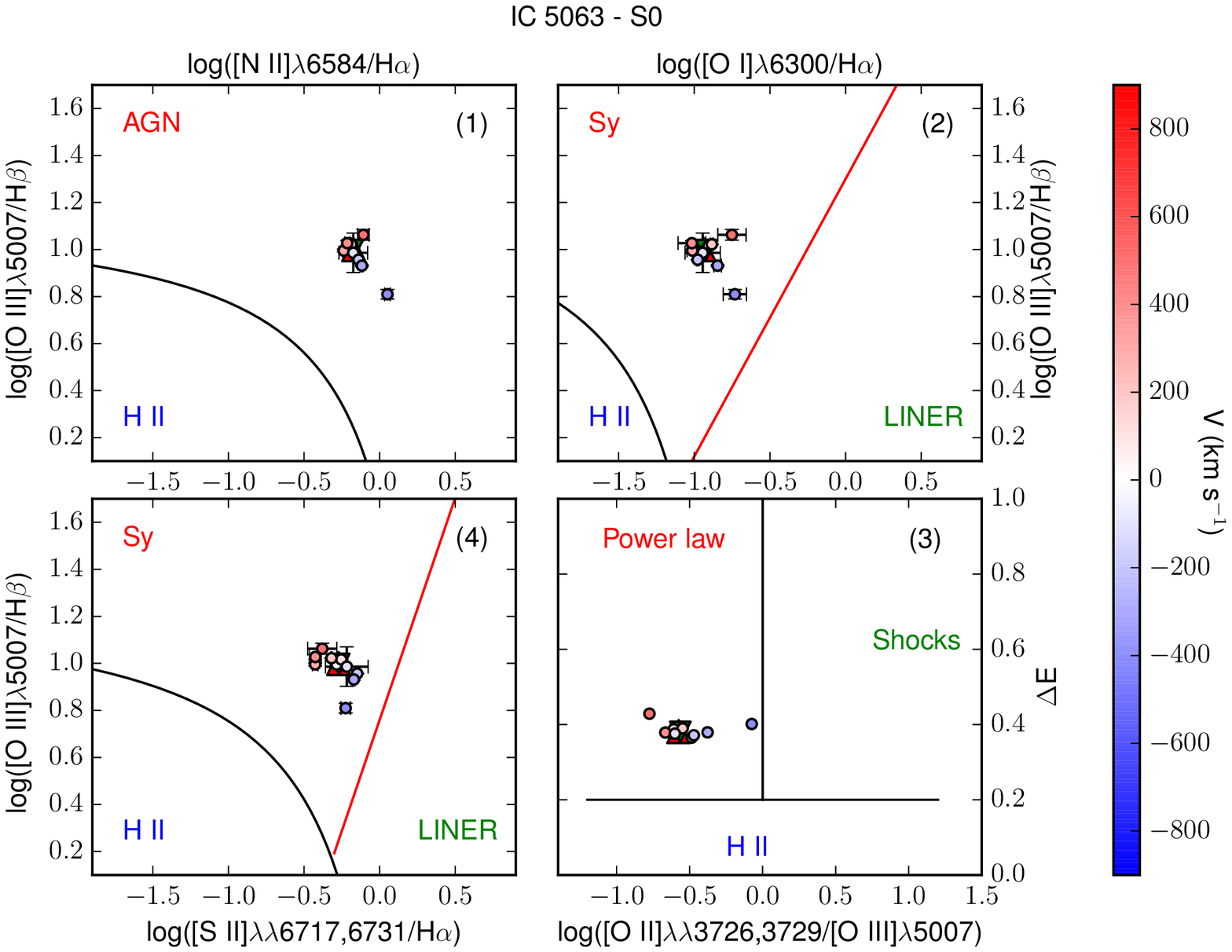}}\\
\caption[]{Diagnostic diagrams of the (a) N2, (b) N1, (c) N0, (d) CN, (e) CS and (f) S0 regions of IC\,5063.  In each plot we show, from the top left panel clockwise: (1) $\log([\ion{O}{III}] \Hb)$ vs $\log([\ion{N}{II}]/ \Ha)$, (2) $\log([\ion{O}{III}]/ \Hb)$ vs $\log([\ion{O}{I}]/ \Ha)$, (3) $\Delta E$ vs $\log([\ion{O}{II}]/[\ion{O}{III}])$, (4) $\log([\ion{O}{III}]/ \Hb)$ vs $\log([\ion{S}{II}]/ \Ha)$ \citep{Baldwin81, Veilleux87}. The colorbar shows the velocity of each bin. The black curves in (1), (2), (4) divide power-law ionized regions (top) and HII regions (bottom). The red lines divide Seyfert-like regions (left) and LINER-like regions (right) \citep{Kewley06}. The black lines in (3) divide HII regions (bottom), power-law ionized region (left) and shock ionized regions (right) \citep{Baldwin81}.  }
\label{fig:diag_css0}
\end{figure*}

\begin{figure*} 
\centering
\subfloat[][]{\includegraphics[width=0.43\textwidth]{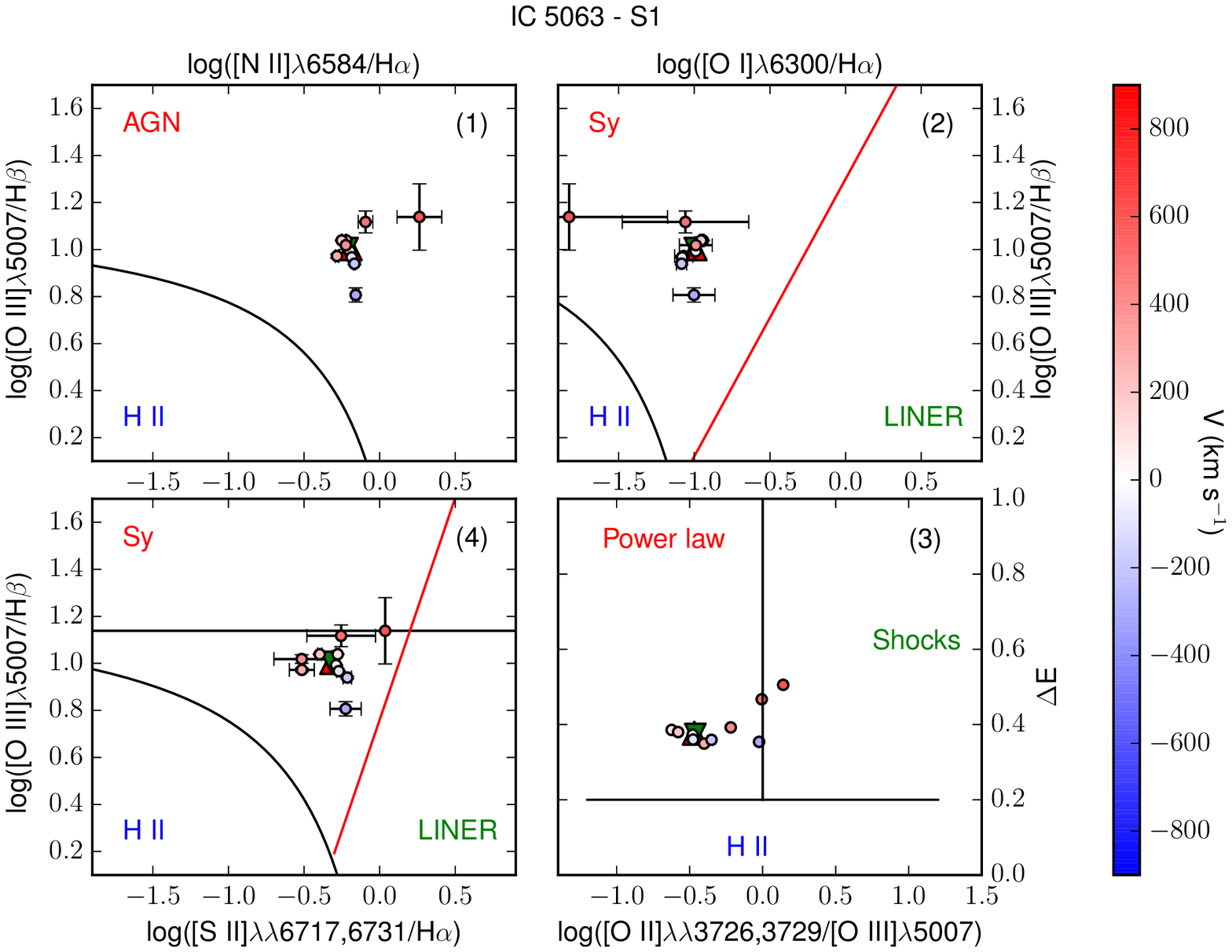}} \quad
\subfloat[][]{\includegraphics[width=0.43\textwidth]{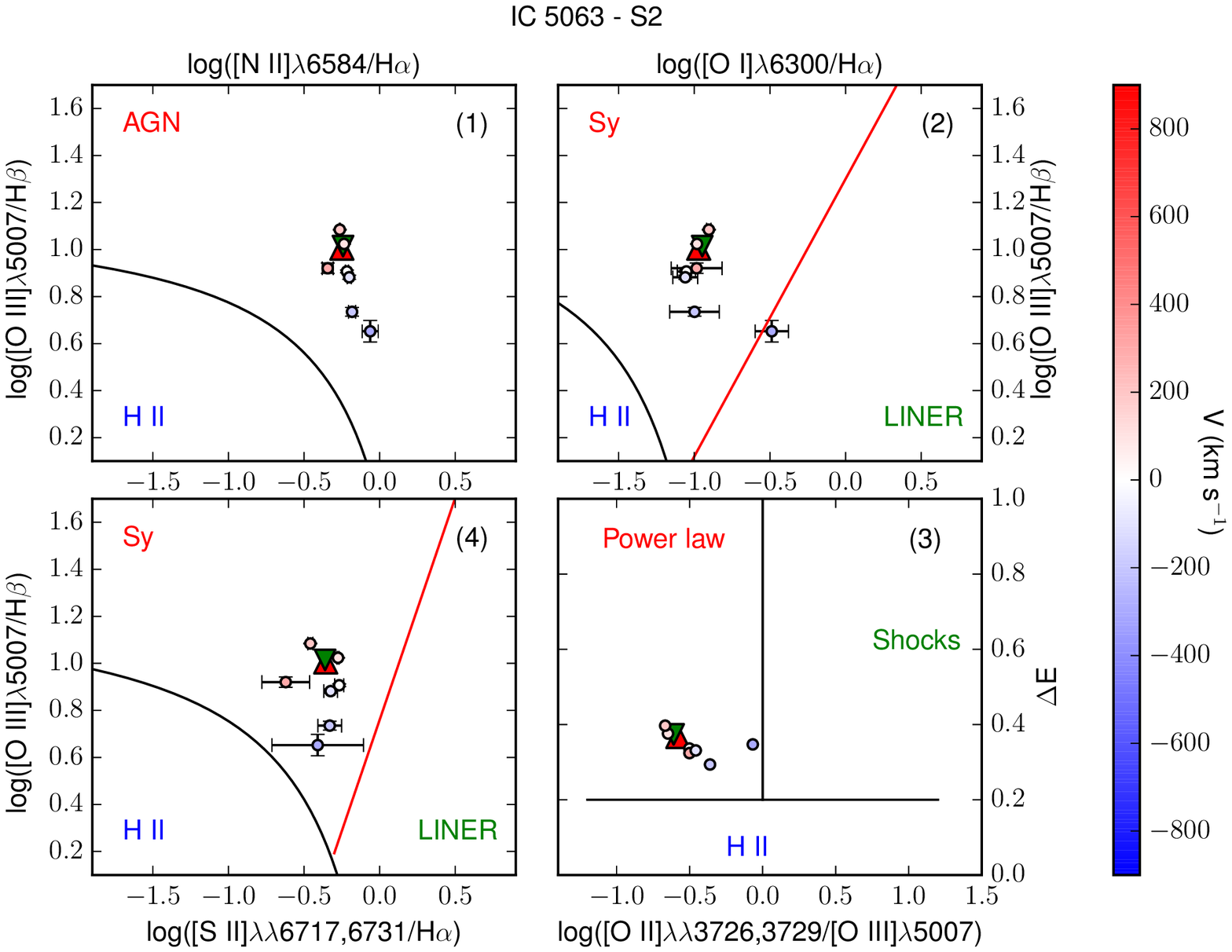}}\\
\subfloat[][]{\includegraphics[width=0.43\textwidth]{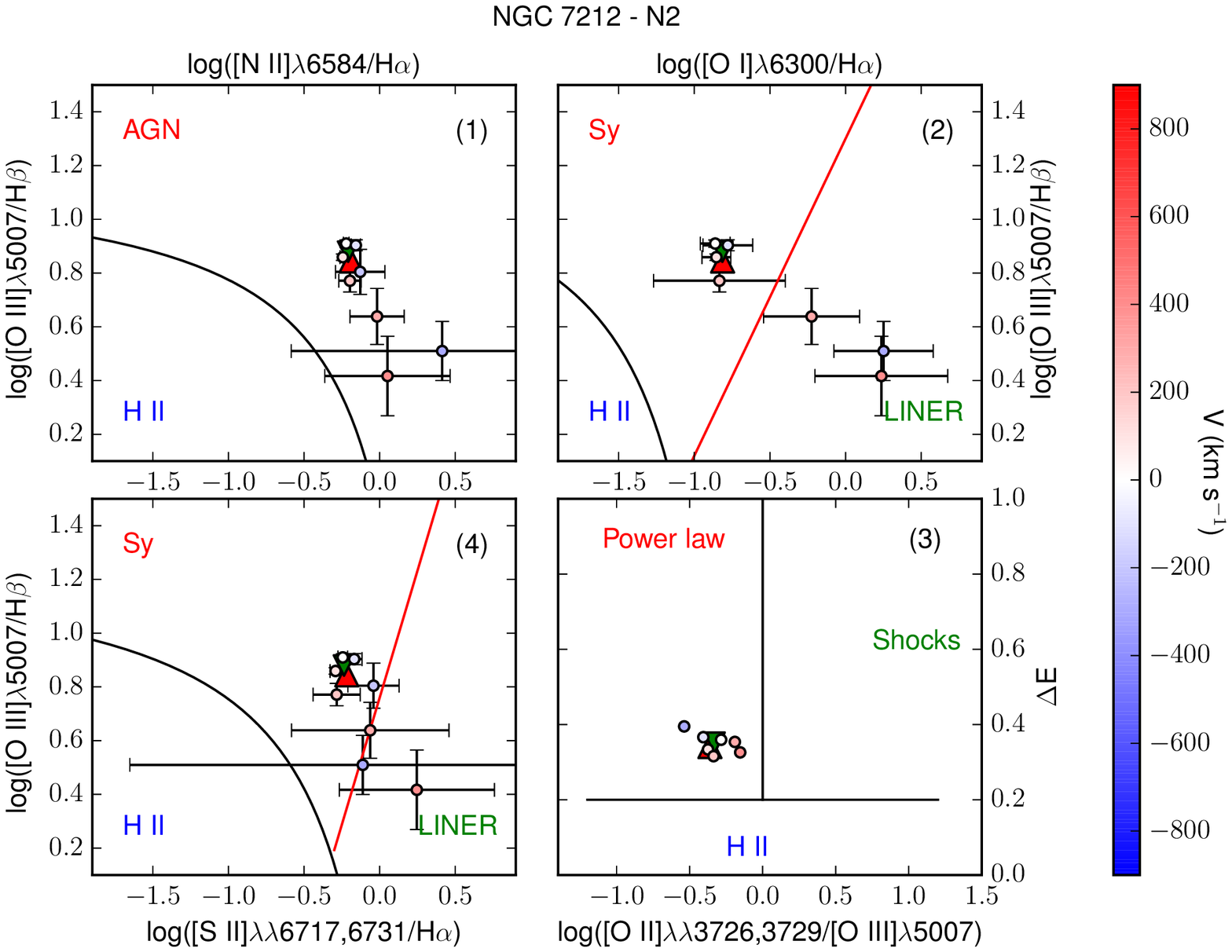}} \quad
\subfloat[][]{\includegraphics[width=0.43\textwidth]{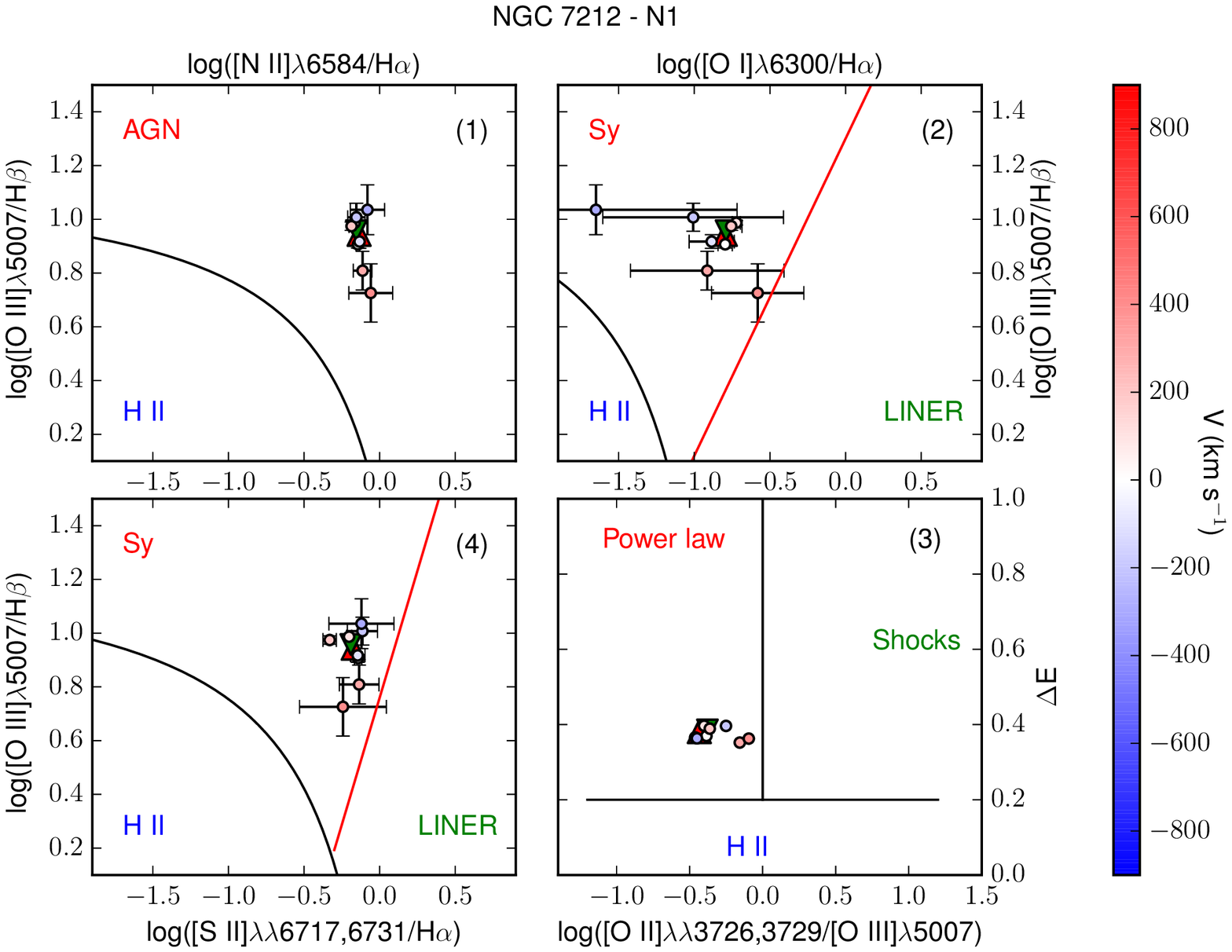}}\\
\subfloat[][]{\includegraphics[width=0.43\textwidth]{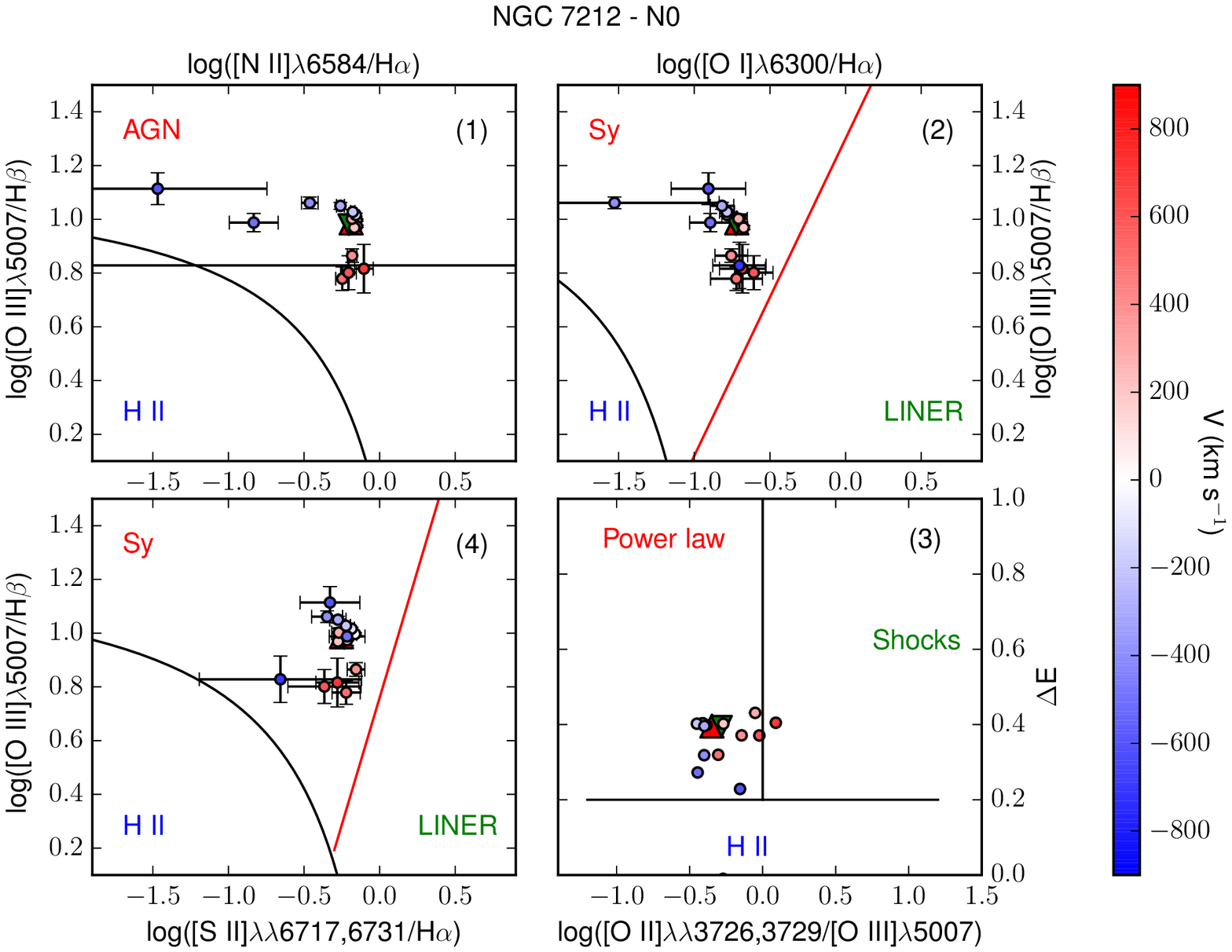}} \quad
\subfloat[][]{\includegraphics[width=0.43\textwidth]{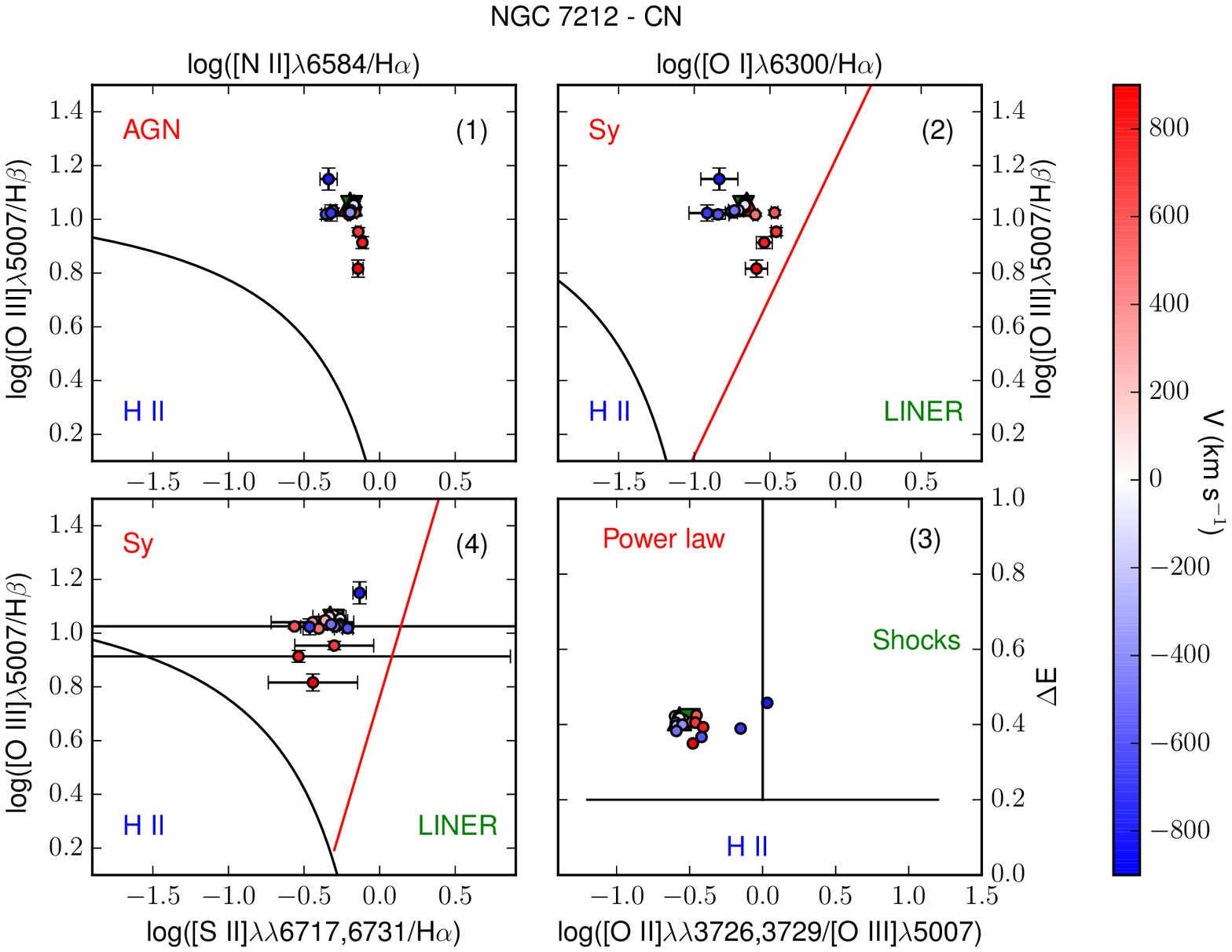}}\\
\caption[]{Diagnostic diagrams of the (a) S1 and (b) S2 regions of IC\,5063 and (c) N2, (d) N1, (e) N0 and (f) CN  regions of NGC\,7212.  In each plot we show, from the top left panel clockwise: (1) $\log([\ion{O}{III}] \Hb)$ vs $\log([\ion{N}{II}]/ \Ha)$, (2) $\log([\ion{O}{III}]/ \Hb)$ vs $\log([\ion{O}{I}]/ \Ha)$, (3) $\Delta E$ vs $\log([\ion{O}{II}]/[\ion{O}{III}])$, (4) $\log([\ion{O}{III}]/ \Hb)$ vs $\log([\ion{S}{II}]/ \Ha)$ \citep{Baldwin81, Veilleux87}. The colorbar shows the velocity of each bin. The black curves in (1), (2), (4) divide power-law ionized regions (top) and HII regions (bottom). The red lines divide Seyfert-like regions (left) and LINER-like regions (right) \citep{Kewley06}. The black lines in (3) divide HII regions (bottom), power-law ionized region (left) and shock ionized regions (right) \citep{Baldwin81}.   }
\label{fig:diag_n0cnN}
\end{figure*}

\begin{figure*}
\centering
\subfloat[][]{\includegraphics[width=0.43\textwidth]{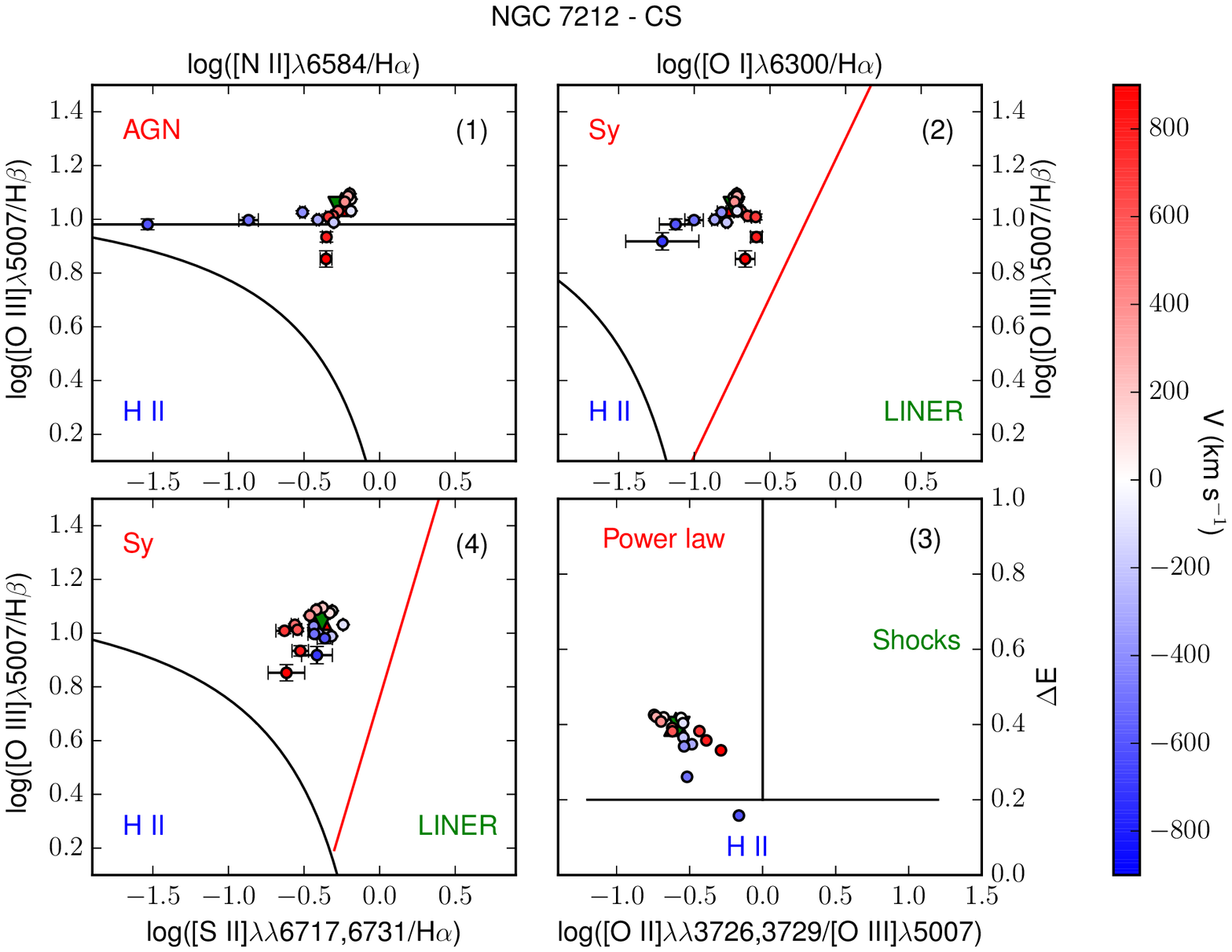}} \quad
\subfloat[][]{\includegraphics[width=0.43\textwidth]{NGC7212_s0_diag.eps}}\\
\subfloat[][]{\includegraphics[width=0.43\textwidth]{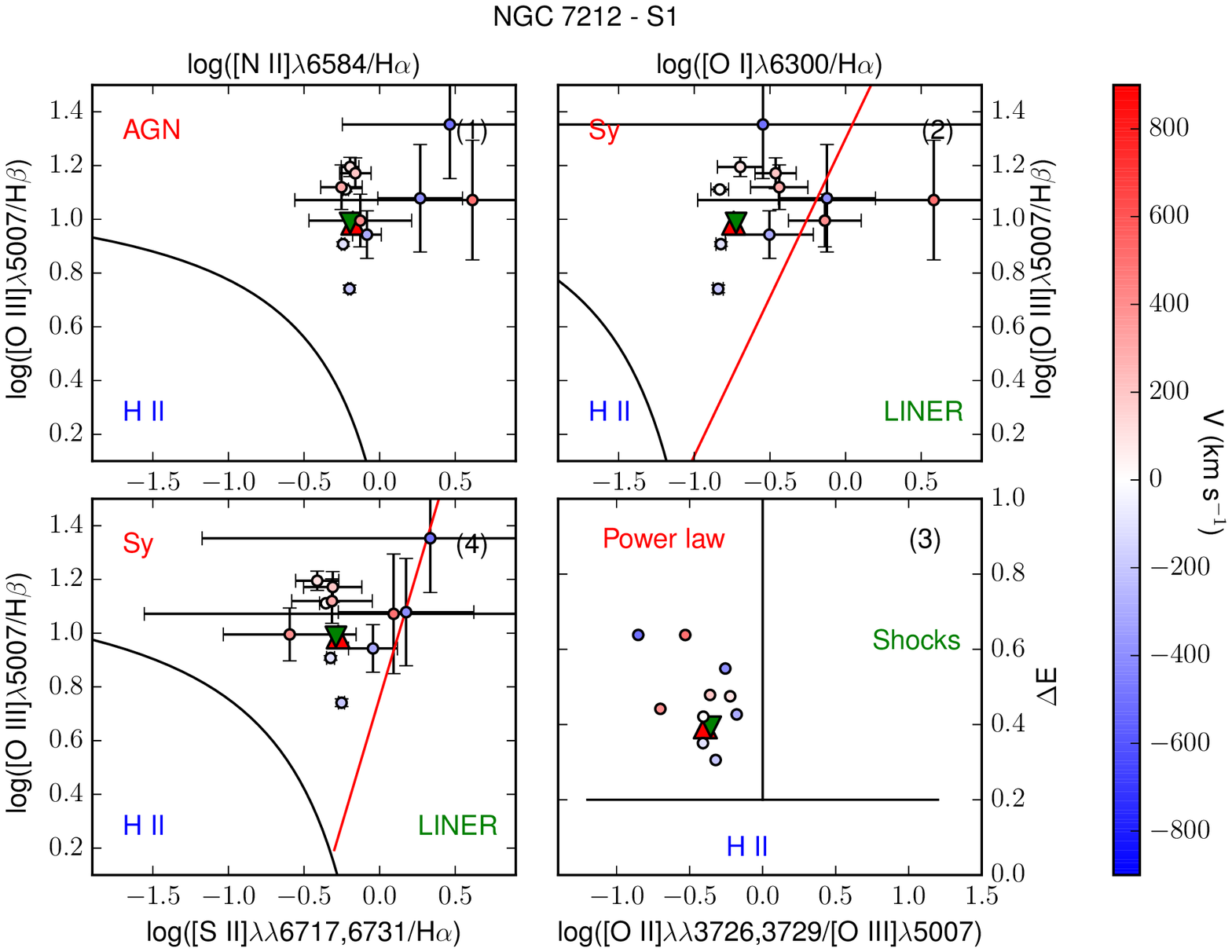}}\\
\caption[]{Diagnostic diagrams of the (a) CS, (b) S0 and (c) S1 regions of NGC\,7212. In each plot we show, from the top left panel clockwise: (1) $\log([\ion{O}{III}] \Hb)$ vs $\log([\ion{N}{II}]/ \Ha)$, (2) $\log([\ion{O}{III}]/ \Hb)$ vs $\log([\ion{O}{I}]/ \Ha)$, (3) $\Delta E$ vs $\log([\ion{O}{II}]/[\ion{O}{III}])$, (4) $\log([\ion{O}{III}]/ \Hb)$ vs $\log([\ion{S}{II}]/ \Ha)$ \citep{Baldwin81, Veilleux87}. The colorbar shows the velocity of each bin. The black curves in (1), (2), (4) divide power-law ionized regions (top) and HII regions (bottom). The red lines divide Seyfert-like regions (left) and LINER-like regions (right) \citep{Kewley06}. The black lines in (3) divide HII regions (bottom), power-law ionized region (left) and shock ionized regions (right) \citep{Baldwin81}. }
\label{fig:diag_s1N}
\end{figure*}


\bsp	
\label{lastpage}
\end{document}